\documentclass{article}
\usepackage{cite}
\usepackage{graphicx}

\begin{document}

\newcommand{\bear}{\begin{eqnarray}}
\newcommand{\eear}{\end{eqnarray}}
\newcommand{\be}{\begin{equation}}
\newcommand{\ee}{\end{equation}}
\newcommand{\beqn}{\begin{eqnarray}}
\newcommand{\eeqn}{\end{eqnarray}}
\newcommand{\beqnn}{\begin{eqnarray*}}
\newcommand{\eeqnn}{\end{eqnarray*}}

\def\vep{\varepsilon}
\def\vf{\varphi}

\newcommand{\ds}{\displaystyle}
\newcommand{\bea}{\begin{eqnarray}}
\newcommand{\eea}{\end{eqnarray}}
\newcommand{\ba}{\begin{array}}
\newcommand{\ea}{\end{array}}
\newcommand{\arcsinh}{\mathop{\rm arcsinh}\nolimits}
\newcommand{\arctanh}{\mathop{\rm arctanh}\nolimits}
\newcommand{\bc}{\begin{center}}
\newcommand{\ec}{\end{center}}

\newcommand{\hr}{\hat\rho}
\def\a{\alpha}
\def\b{\beta}
\def\g{\gamma}
\def\l{\lambda}
\def\s{\sigma}
\def\vf{\varphi}
\def\ep{\epsilon}
\def\vep{\varepsilon}

\newcommand{\rot}{\operatorname{rot}}
\newcommand{\PG}[1]{\ensuremath{\left[#1\right]}}
\newcommand{\PC}[1]{\ensuremath{\left(#1\right)}}
\newcommand{\PN}[1]{\ensuremath{\left|#1\right|}}
\newcommand{\BK}[1]{\ensuremath{\left\langle#1\right\rangle}}
\newcommand{\mat}[1]{\mbox{\boldmath{$#1$}}}
\newcommand{\mc}{\mathcal}
\def\vf{\varphi}
\def\h{\hat}
\def\b{\beta}

\def\h{\hat}
\def\T{\Theta}
\def\A{\Alpha}
\def\up{\Upsilon}

\newcommand{\PH}[1]{\ensuremath{\left\{#1\right\}}}

\begin{center} {\Large \bf
Energy and magnetic moment of a quantum charged particle in time dependent 
magnetic and electric fields of circular and plane solenoids}

\end{center}

\begin{center} {\bf
V.V.~Dodonov$^{1*}$ and M.B.~Horovits$^{1,2}$
}\end{center}
 
\begin{center}

{\it $^1$
Institute of Physics and International Center for Physics,
University of Brasilia, \\
70910-900 Brasilia, Federal District, Brazil}

$^*$E-mail: vdodonov@unb.br\\ 

$^2$ {Instituto Federal de Bras\'{\i}lia, Campus Estrutural, Bras\'{\i}lia 71255-200, DF, Brasil }

\end{center}

\abstract{
We consider a quantum spinless nonrelativistic charged particle moving in the $xy$ plane under the
action of a time-dependent magnetic field, described by means of the linear
vector potential  ${\bf A}=B(t)\PG{-y(1+\alpha),x(1-\alpha)}/2$, with two fixed
values of the gauge parameter $\alpha$: $\alpha=0$ (the circular gauge) and $\alpha =1$ (the Landau gauge). 
Although the magnetic field is the same in all the cases,
the systems with different values of the gauge parameter are not
equivalent for nonstationary magnetic fields due to different structures of
induced electric fields, whose lines of force are circles for $\alpha=0$ and
straight lines for $\alpha=1$. We derive general formulas for the time-dependent mean values of the energy and magnetic moment,
as well as for their variances,
for an arbitrary function $B(t)$. They are expressed in terms of solutions to the classical equation of motion
$\ddot\vep +\omega_{\alpha}^2(t) \vep=0$, with $\omega_1=2\omega_0$.  Explicit results are found in the cases of 
the sudden jump of magnetic field, the parametric resonance, the adiabatic evolution, and for several specific functions $B(t)$, 
when solutions can be expressed in terms of elementary or hypergeometric functions.
These examples show that the evolution of the mentioned mean values can be rather different for the two gauges, 
if the evolution is not adiabatic. It appears that the adiabatic approximation fails when the magnetic field goes to zero.
Moreover, the sudden jump approximation can fail in this case, as well.
The case of slowly varying field changing its sign seems especially interesting. In all the cases, fluctuations of the
magnetic moment are very strong, frequently exceeding the square of the mean value.
}


\tableofcontents

\section{Introduction}      

The motion of a quantum charged particle in a uniform stationary magnetic field 
attracted attention of many authors since the first years of quantum mechanics
\cite{Kennard,Darwin,Fock28,Land30,Page30,Darwin31}. 
For a nonrelativistic spinless particle of mass $m$ and charge $e$, moving in the $xy$~plane 
perpendicular to the magnetic field ${\bf{B}}=(0,0,B)=\mbox{rot}{\bf A}$, the problem is reduced to solving the 
Schr\"odinger equation with Hamiltonian
 (in the Gauss system of units)
\be
\hat{H} = \mat{\hat\pi}^2/(2m), \qquad \mat{\pi}=\mat{p}-e\mat{A}/c.
\label{Ham}
\ee
It is well known that the same vector ${\bf{B}}$ can be obtained from the whole family of
linear vector potentials of the form
\be
\mat{A}= B\PG{-y(1+\alpha),x(1-\alpha)}/2.
\label{pv}
\ee
Two choices of the gauge parameter $\alpha$ are frequently considered in the literature:
$\alpha=0$ (the so called circular or symmetric gauge) and $\alpha=1$ (the Landau gauge).
The solutions to the stationary Schr\"odinger equation (with $B=const$) have different forms for the two gauges:
they are expressed in terms of the Laguerre polynomials for $\alpha=0$ \cite{Fock28} and in terms of the Hermite polynomials
for $|\alpha| = 1$ \cite{Land30}. Nonetheless, the physical consequences, such as the mean energy 
in the thermodynamic equilibrium state or equilibrium magnetization,
are identical. Therefore, one could think that the concrete choice of the gauge is mainly a matter of taste.
However, this is true for {\em time-independent\/} magnetic fields only.

The Schr\"odinger equation with Hamiltonian (\ref{Ham}) and a general function $B(t)$ 
was solved exactly for the first time  in papers \cite{LR,MMT69,MMT70} for the circular gauge and  \cite{DMM72}
for the Landau gauge. 
It was shown that quantum solutions are determined completely by the solution
of the {\em classical equation of motion\/} for the oscillator with a time-dependent frequency,
\be
\ddot\vep +\omega_{\alpha}^2(t) \vep=0.
\label{eqvep}
\ee
In the case of circular gauge, one should put in (\ref{eqvep}) the {\em Larmor frequency\/} 
$\omega_0(t) \equiv \omega(t)=  eB(t)/(2mc)$, whereas
  the {\em cyclotron frequency\/} $\omega_1(t) \equiv \Omega(t) = eB(t)/(mc)$ should be used
 in the case of the Landau gauge.
In particular, the authors of papers \cite{LR,MMT69,MMT70,DMM72} constructed
generalizations of the energy eigenstates (as eigenstates of the {\em quadratic\/} operators - integrals of motion),
which look similar to the time-independent eigenstates, with the functions $\vep(t)$ in the arguments of the Laguerre or
Hermite polynomials. In papers \cite{MMT69,MMT70,DMM72}, generalized coherent states were constructed as eigenstates
of the {\em linear\/} integrals of motion. These states were used to calculate the propagators,
transition amplitudes and transition probabilities between energy levels corresponding to the initial and final asymptotic 
magnetic fields. 
Several other aspects of the problem, where solutions to Equation (\ref{eqvep}) were used, were considered later, e.g., in papers 
\cite{Aga80,Abdalla88,Abdalla88a,Jan89,Dutra91,Baseia92,BasMiMo,DMMPR95,DelMi,Yuce03,Abdal07,Menouar10,%
ManZheb12,Zhebrak13,Menouar15,Aguiar16} 
for $\alpha=0$. 
In particular,  the problem of squeezing in the time dependent
magnetic field with $\alpha=0$ was considered in \cite{Jan89,Baseia92,BasMiMo,DelMi,Aguiar16} 
with respect to the canonical pairs of variables.
The tomographic approach was used in papers \cite{ManZheb12,Zhebrak13}.
Informational aspects of the problem motivated the authors of \cite{Aguiar16}. 
A few papers  \cite{Ying99,Choi03} were devoted to the case of $\alpha=1$.
Note, however, that no {\em explicit\/} solutions to Equation (\ref{eqvep}) with $\omega_{\alpha}(t) \neq const$ 
were considered in all the cited papers.

It was mentioned already in paper \cite{DMM72}, 
that the physical consequences are {\em different\/} for the two gauges in the time-dependent magnetic fields.
The difference can be clearly seen, if one compares explicit expressions for the propagators and transition amplitudes 
for the two gauges given in \cite{MMT70} and \cite{DMM72}.
Other manifestations of the
gauge nonequivalence in the case of time-dependent magnetic fields were
observed in studies \cite{GSQ,DH18}, devoted to the problem of generation of squeezed
states of charged particles in  magnetic fields, with respect to relative and guiding center coordinates.
Clearly, the origin of the gauge nonequivalence is in {\em different spatial distributions of
the induced electric field\/} $\mat{E}(\mat{r},t) = -\partial{\mat{A}}(\mat{r},t)/\partial(ct)$,
whose lines of force are circles for $\alpha=0$ (the circular solenoid) and straight lines for $\alpha=1$
(the plane solenoid). 

The goal of our paper is to compare the 
{\em explicit evolution\/} of such physical quantities as the mean energy
and mean magnetic moment, as well as their variances, for two different physical systems, characterized by two 
different gauge parameters of the time-dependent vector potential with the same magnetic field $B(t)$. 
It appears that none of these two quantities were calculated for time-dependent magnetic fields in all known papers 
\cite{LR,MMT69,MMT70,Aga80,Abdalla88,Abdalla88a,Jan89,Dutra91,Baseia92,BasMiMo,DMMPR95,DelMi,Yuce03,Abdal07,Menouar10,ManZheb12,Zhebrak13,%
Menouar15,Aguiar16,DMM72,Ying99,Choi03}.
We consider several concrete functions $B(t)$ admitting exact explicit solutions of Equation (\ref{eqvep}). 
No one of these functions was considered in connection with the problem under study until now
(except for the obvious case of the constant magnetic field).
Using the explicit solutions, we can establish, in particular, 
conditions of validity of two frequently used approximations: adiabatic and ``sudden jump'' ones.
While two physical situations are different, we believe that their treatment in the frames of a single paper
is justified, because the starting point for the analysis of the two cases is the same Equation (\ref{eqvep})
(although with scaled frequencies). However, the final results are different. Why? It seems that the circular gauge
is so symmetric that it ``hides'' in some sense the presence of the circular induced electric field, as soon as all
final expressions contain the functions $\vep(t)$ and $\dot\vep(t)$ only.
On the other hand, this symmetry is broken for the Landau gauge, where an additional solution, satisfying an inhomogenious oscillator
equation, appears necessary. The inhomogeneous term is proportional to the additional constant of the motion which,
in turn, exists due to the unidirectional structure of the induced electric field. Explicit examples considered in this paper
demonstrate how this difference in the geometry of induced electric fields influences the energy and magnetic moment.

Our plan is as follows.
In Section \ref{sec-main}, we remind the definitions of the main quantities characterizing the motion of a charged particle in the
magnetic field, such as energy, angular momentum and magnetic moment, 
emphasizing the role of the relative and the center of orbit coordinates. Also, we analyze the dynamical equations
for the canonical and ``geometrical'' variables and discuss the choice of initial conditions. 
The details of evolution are considered separately in Sections \ref{sec-circ}--\ref{sec-tanh} for the circular gauge and \ref{sec-Land}--\ref{sec-Landexp} for the Landau gauge.
In particular, in Section \ref{sec-circ} we provide general expressions for the mean values and fluctuations
of the energy and magnetic moment in terms of solutions to Equation (\ref{eqvep}). Three simple approximate solutions are
considered in that section: the adiabatic evolution, the sudden jump of the magnetic field, and 
the parametric resonance.
In Section \ref{sec-part}, we analyze three concrete functions $B(t)$, which permit us to find explicit exact solutions 
to Equation (\ref{eqvep}) in terms of elementary functions. Four other examples, when exact solutions can be written in
terms of the confluent hypergeometric, Gauss hypergeometric, cylindrical and Legendre functions, 
are analyzed in Sections \ref{sec-hyper} and \ref{sec-tanh}.
Section \ref{sec-Land} is devoted to general relations for the Landau gauge, with the same special cases 
as in Section \ref{sec-circ}. Two special cases, when explicit exact solutions can be written in terms of elementary functions,
are considered in Section \ref{sec-Landexp}.
Section \ref{sec-fin} contains a discussion of results.
Some details of calculations are given in Appendices \ref{ap-detcirc}--\ref{ap-SOm-as}.
Appendices \ref{ap-deg} and \ref{ap-gauge} are devoted to the interesting questions arising in connection with our study:
the existence (and sense) of the Landau levels in the time-dependent magnetic field and the non-equivalence of different 
time-dependent gauges.

\section{Basic definitions and equations}
\label{sec-main}

\subsection{Main physical quantities}

The energy operator coincides with Hamiltonian (\ref{Ham}) in the stationary case.
However, it is useful to write it in a different form, using the concept of relative and center of orbit coordinates.
For this purpose, we remember that Hamiltonian (\ref{Ham}) admits two linear
integrals of motion,
\be
\hat{x}_c=\hat{x}+ {\mat{\hat{\pi}}_{y}}/(m\Omega) = (1+\alpha)x/2 +\hat{p}_y/(m\Omega) , 
\qquad
\hat{y}_c=\hat{y} - {\mat{\hat{\pi}}_{x}}/(m\Omega) = (1-\alpha)y/2 -\hat{p}_x/(m\Omega) ,
\label{centroXY}
\ee
provided the magnetic field $B$ does not depend on time. 
Operators (\ref{centroXY})  describe nothing but the
coordinates of the center of a circle, which the particle rotates around with the cyclotron frequency 
$\Omega= e{B}/(mc)$. 
The importance of these integrals of motion was emphasized by many authors during decades 
\cite{Land30,JL49,Dulock66,MM69,FeKa70,Tam71,32,33,34,Kowalski05,Mielnik11,Dodcohmag18,Champel19,Waka20,Kita,Fletcher21}.
Equivalent integrals of motion, obtained by the multiplication of $x_c$ and $y_c$ by $m\Omega$, were considered under
the name ``pseudomomentum'' in papers \cite{32,Konst16}.

The second pair of physical observables consists of two
relative coordinates, 
\[
\hat{x}_r=\hat{x}-\hat{x}_c= - \hat{\mat{\pi}}_y/({m\Omega}) = (1-\alpha)x/2 -\hat{p}_y/(m\Omega),
\]
\[
 \hat{y}_r=\hat{y}-\hat{y}_c= \hat{\mat{\pi}}_x/({m\Omega}) = (1+\alpha)y/2 +\hat{p}_x/(m\Omega).
\]
Then, Hamiltonian (\ref{Ham}) with $B =const$ can be written as
\be
\hat{H} = m\Omega^2\left( \hat{x}_r^2 +\hat{y}_r^2\right)/2 .
\label{Ham0}
\ee
Due to the commutation relations
\be
\left[\hat\pi_x, \hat\pi_y\right] = -im\Omega\hbar, \quad
\PG{\hat{x}_r,\hat{y}_r}= \PG{\hat{y}_c, \hat{x}_c}= {i\hbar}/(m\Omega),
\quad
 \PG{\hat{x}_r,\hat{x}_c}=\PG{\hat{x}_r,\hat{y}_c}=\PG{\hat{y}_r,\hat{x}_c}=\PG{\hat{y}_r,\hat{y}_c}=0,
\label{comut}
\ee
the eigenvalues of operator (\ref{Ham0})  assume discrete values $\hbar\Omega(n+1/2)$.
Moreover, these eigenvalues have infinite degeneracy \cite{Land30}, because they do not depend on the mean values 
of operators $\hat{x}_c$ and $\hat{y}_c$ (or their functions).
These results are well known, of course.

In addition to the energy, there exists another quadratic
integral of motion, which can be considered as the generalized angular momentum
(the same formulas hold for the classical variables and quantum operators):
\be
L = x\pi_y - y\pi_x + \frac{m\Omega}{2} \left(x^2 + y^2\right)
= x p_y - y p_x +  \frac{m\Omega}{2} \alpha\left(x^2 -y^2\right)
=  \frac{m\Omega}{2}\left(x_c^2 +y_c^2 -x_r^2 -y_r^2\right).
\label{L3}
\ee
It coincides formally with the canonical angular momentum
$L_{can} =x p_y - y p_x $
in the only case of ``circular'' gauge of the vector potential. 
It follows from (\ref{L3}) that the ``kinetic'' angular momentum, defined as 
\be
L_{kin} \equiv x\pi_y - y\pi_x
= - m\Omega\left(x_r^2 +y_r^2 + x_c x_r + y_c y_r \right),
\label{Lkin}
\ee
 is not a conserved quantity, and it can vary with time in the generic case \cite{JL49,Li99,Barnett14,Green15,Waka18,vanEnk20},
 except for the special cases of energy eigenstates or their statistical mixtures.   
On the other hand, the ``intrinsic'' angular momentum
\be
J = x_r \pi_y - y_r \pi_x = -m\Omega\left(x_r^2 +y_r^2\right) = -2H/\Omega
\label{J}
\ee
is conserved for the constant magnetic field.
While operators (\ref{Ham0}) and (\ref{L3}) commute, one cannot expect that
the mean value of $L$ can be preserved for {\em time-dependent\/} functions $\Omega(t)$,
unless $\alpha=0$,
because $d\langle \hat{L}\rangle/dt = \frac12 m\alpha\langle x^2 -y^2\rangle d\Omega /dt$.

To introduce the magnetic moment operator, we use the  definition of the classical magnetic moment \cite{LL,Jackson}
\be
{\bf M} = \frac1{2c}\int dV \left[{\bf r}\times{\bf j}\right].
\label{Mclas}
\ee
Then, using the expression for the quantum probability current density, 
\be
{\bf j} = \frac{ie\hbar}{2m}\left(\psi \nabla\psi^* - \psi^* \nabla\psi\right) - \frac{e^2}{mc}{\bf A} \psi^*\psi,
\ee
one can write the right-hand side of (\ref{Mclas})  as the mean value of operator
\be
\hat{\cal M} = \frac{e}{2mc}\hat{L}_{kin}.
\label{M2}
\ee
A formula equivalent to (\ref{M2}) was justified (for $\alpha=0$) in \cite{Felder76,March85}, 
using the thermodynamical approach. Another proof of the definition (\ref{M2}) for an arbitrary gauge was given
in \cite{Stewart00} (see also \cite{Friar81,Ishi99,Bliokh12,Bliokh17,Waka21}).

\subsection{Equations describing the time evolution}

As soon as we are interested in the evolution of the mean energy and mean magnetic moment, we have to calculate the mean values
of various products of operators $\hat{x}_{r,c}$ and $\hat{y}_{r,c}$ as functions of time.
At first glance, one could use the simplest form of the Ehrenfest equation for the mean values of some operator,
$d\langle\hat{O}\rangle/dt = (i/\hbar)\langle[\hat{H},\hat{O}]\rangle$. Then, the commutator (\ref{comut}) yields
$d\langle\hat{\pi}_x\rangle/dt = -\Omega \langle\hat{\pi}_y\rangle$ and  
$d\langle\hat{\pi}_y\rangle/dt = \Omega \langle\hat{\pi}_x\rangle$, without any dependence on the gauge parameter $\alpha$.  
But this is true only for the time-independent frequency $\Omega$. In the general case, one has to use the complete
Ehrenfest equation, 
$d\langle\hat{O}\rangle/dt = (i/\hbar)\langle[\hat{H},\hat{O}]\rangle +\langle \partial\hat{O}/\partial t\rangle$,
taking into account that the operator $\mat{\hat\pi}$ in (\ref{Ham}) contains the explicit time dependence through the 
vector potential (\ref{pv}) with a time-dependent function $B(t)$.
However, the equation for $d\langle\mat{\hat\pi}\rangle/dt$ contains the derivative $d\Omega/dt$ in addition to $\Omega(t)$.
For this reason, we prefer to start from the equations for the mean values of the {\em canonical operators\/}, since
these operators do not contain time-dependent functions in their definitions.
Omitting the symbol of quantum mechanical averaging $\langle\cdots\rangle$, we obtain the following equations (formally 
coinciding with the  equations for classical variables due to the linearity):
\be
\dot{x} = p_x/m +\omega(t)(1+\alpha)y, 
\qquad
 \dot{y} = p_y/m -\omega(t)(1-\alpha)x,
 \label{dotx}
 \ee
 \be
\dot{p}_x = \omega(t)(1-\alpha)p_y -m\omega^2(t)(1-\alpha)^2 x,  
\qquad
\dot{p}_y = -\omega(t)(1+\alpha)p_x -m\omega^2(t)(1+\alpha)^2 y, 
\label{dotpy}
\ee
where $\omega(t) = eB(t)/(2mc)$ is the  Larmor frequency.
It is convenient to introduce the vector ${\bf Q} = (x,y,p_x,p_y)$ (whose components are either mean values of quantum operators
or classical variables). Then, solutions to the system (\ref{dotx})-(\ref{dotpy}) can be written in the compact form as
\be
{\bf Q}(t) = \mat{\Lambda}_Q(t){\bf Q}(0),
\label{LamQ}
\ee
where $\mat{\Lambda}_Q(t)$ is some $4\times4$ matrix.
Also, it is convenient to introduce the $4\times4$ symmetrical
 matrix $\mat{\sigma}=\Vert\sigma_{ij}\Vert$, consisting of
 all symmetrical second order moments 
$\sigma_{ij}=\langle\hat{Q}_i \hat{Q}_j + \hat{Q}_j\hat{Q}_i\rangle/2$. 
 Then, it is known (see, e.g., \cite{book3}) that the linear transformation (\ref{LamQ}) results in the following 
relation between the matrices $\mat{\sigma}(t)$ and $\mat{\sigma}(0)$:
\be
\mat{\sigma}(t)=\mat{\Lambda}_Q(t) \mat{\sigma}(0) \tilde{\mat{\Lambda}}_Q(t),
\label{sigLamsig}
\ee
 where $\tilde{\mat{\Lambda}}_Q$ means the transposed matrix. 
 From the physical point of view,
it is convenient to use the  matrices corresponding to the ``geometrical'' coordinates, combined in the vector
${\bf q} = (x_r,y_r,x_c,y_c)$, instead of vector ${\bf Q}(t)$. Knowing the transformation
${\bf q} = U{\bf Q}$ with
\[
U = \frac12\left\Vert
\begin{array}{c c c c}
1-\alpha	&	0	& 0 & -r^{-1} \\
0 & 1+\alpha & r^{-1} & 0 \\
1+\alpha	&	0	& 0 & r^{-1} \\
0 & 1-\alpha & -r^{-1} & 0 
\end{array}
\right\Vert ,
\quad
U^{-1} = \left\Vert
\begin{array}{c c c c}
1	&	0	& 1 & 0 \\
0 & 1 & 0 & 1 \\
0	&	r(1-\alpha)	& 0 & -r(1+\alpha) \\
-r(1+\alpha) & 0 & r(1-\alpha) & 0 
\end{array}
\right\Vert , 
\]
where $r=m\omega$, we arrive at the final expression
\be
\mat{\sigma}_q(t)=\mat{\Lambda}_q(t) \mat{\sigma}_q(0) \tilde{\mat{\Lambda}}_q(t),
\label{sigLamfin}
\ee
where
\be
\mat{\Lambda}_q(t)= U(t)\mat{\Lambda}_Q(t)U^{-1}(0).
\label{LamUQU}
\ee
Here, matrix $U(t)$ contains the current Larmor frequency $\omega(t)$, whereas $U(0)$
contains the initial frequency $\omega(0)$.

In general, the $4\times4$ symmetric matrix $\mat{\sigma}_q(0)$ can have 10 independent elements (obeying some restrictions due to the
uncertainty relations). Therefore, it is difficult to analyze the problem for the most general initial states.
We consider the most natural situation, when the initial state is the thermodynamic equilibrium state,
corresponding to the inverse temperature $\beta$. Then we have the matrix with four non-negative parameters \cite{DH-jump},
\be
\mat{\sigma}_q(0)= G\left\Vert
\begin{array}{cccc }
1 & 0	&	-\rho & 0	 \\
0	&	1 & 0 & -\rho \\
-\rho & 0 & s\Upsilon & 0 \\
0 & -\rho & 0 & \Upsilon/s
\end{array}
\right\Vert , 
\label{sig0}
\ee
\be
G = \frac{ \hbar {\cal C}}{4m\omega_i}, \quad {\cal C} =\coth(\hbar\omega_i\beta) \ge 1, \quad
\rho = \frac{\tanh(\hbar\omega_i\beta)}{\hbar\omega_i\beta} \le 1,
\quad \Upsilon =\frac{\tanh(\hbar\omega_i\beta)}{\tanh(\hbar\beta \nu)} \ge 1.
\label{Geq}
\ee
Actually, matrix (\ref{sig0}) corresponds to the equilibrium state of the charged particle, confined by means of
a weak parabolic potential, so that $\nu$ is some effective frequency of this potential,
satisfying the restriction $\nu \ll \omega$.
The real coefficient $s$ characterizes the degree of anisotropy of the potential ($s=1$ in the isotropic case).
The initial mean values of the energy and magnetic moment are as follows,
\be
{\cal E}_i = \hbar\omega_i{\cal C}, \qquad
{\cal M}_{i} = \mu_B{\cal C}(\rho-1)= \mu_B \left[(\hbar\omega_i\beta)^{-1} - \coth(\hbar\omega_i\beta)\right],
\label{EiMi}
\ee
where $\mu_B = e\hbar/(2mc)$ is the Bohr magneton. We see that the nonzero value of parameter $\rho$ is necessary to
ensure the famous Landau--Darwin formula (\ref{EiMi}) for the diamagnetism of a free charged particle \cite{Land30,Darwin31} 
in all temperature regimes.
We shall pay especial attention to two limit cases.
In the high temperature limit, $\hbar\beta\omega_i \ll 1$, we have $\rho \approx 1$ and $\Upsilon \gg 1$. 
On the other hand, in the extreme low temperature limit, $\hbar\beta\nu \gg 1$,
we have $\Upsilon = 1$, $\rho = 0$, and  $G = {\hbar}/(2m\Omega_i)$.
We assume that the direction of the initial magnetic field (or the $z$-axis) is chosen in such a way that $\omega_i >0$.

Under real conditions, the particle moves inside some container with an effective radius $R$. Hence, the approximations and
results of this study have sense under the restrictions 
\be
\mbox{Tr}(\sigma_q) \ll R^2, \qquad 2G\left( 1 + s_0 \Upsilon\right) \ll R^2, \qquad 2s_0 = s+s^{-1}.
\label{impcond}
\ee
At zero temperature, we have the restriction on the magnetic field $B \gg \hbar c/(|e|R^2)$. 
Note that the particle mass does not enter this inequality. 
For $R \sim 1\,$cm, the restriction is very weak: $B \gg 10^{-7}\,$G.
Remember that $\Omega \approx 10^{11}\,$s$^{-1}$ for electrons in the field $B\approx 6\times 10^3\,$G.
Then, the low-temperature limit means that $T\ll 1\,$K. On the other hand, the high-temperature limit is more adequate 
for ions, whose cyclotron frequency $\Omega$ is several ($3$ to $5$) orders of magnitude smaller than the electron frequency. 

It is convenient to split the $4\times4$ matrices $\mat{\sigma}_q(t)$ and $\mat{\Lambda}_q(t)$ into $2\times2$ blocks:
\be
\mat{\sigma}_q(t) =  G\left\Vert
\begin{array}{c c}
\sigma_r & \sigma_{rc} \\
\tilde\sigma_{rc} & \sigma_c
\end{array}
\right\Vert,
\qquad
\mat{\Lambda}_q(t) =  \left\Vert
\begin{array}{c c}
\lambda_1 & \lambda_2 \\
\lambda_3 & \lambda_4
\end{array}
\right\Vert.
\label{Lamq-0}
\ee
Matrices $G\sigma_r$ and $G\sigma_c$ describe fluctuations of the relative and guiding center coordinates, respectively.
Matrix $G\sigma_{rc}$ describes correlations between these two subsystems. Note that initial fluctuations of the guiding center
coordinates are stronger than those of relative coordinates, especially if $\Upsilon \gg 1$.

Using formula (\ref{sig0}), we can write the blocks of $\mat{\sigma}_q(t)$ as follows,
\be
\sigma_r = \lambda_1 \tilde\lambda_1 + \Upsilon \lambda_2 S \tilde\lambda_2 
- \rho\left(\lambda_2 \tilde\lambda_1 + \lambda_1 \tilde\lambda_2\right),
\qquad
\sigma_c = \lambda_3 \tilde\lambda_3 + \Upsilon \lambda_4 S \tilde\lambda_4 
- \rho\left(\lambda_4 \tilde\lambda_3 + \lambda_3 \tilde\lambda_4\right),
\label{sigr}
\ee
\be
\sigma_{rc} = \lambda_1 \tilde\lambda_3 + \Upsilon \lambda_2 S \tilde\lambda_4 
- \rho\left(\lambda_2 \tilde\lambda_3 + \lambda_1 \tilde\lambda_4\right),
\label{sigrc}
\ee
where $S=\mbox{diag}(s,s^{-1})$ is the diagonal matrix.

We suppose that the confining potential is removed at the time instant $t=0$, and the system starts to evolve in accordance
with Hamiltonian (\ref{Ham}).
Then, the main tool for calculating mean values of the energy and magnetic moment is the
transformation matrix $\mat{\Lambda}_q$. In turn, it is determined by the solutions to the set of four linear differential
equations with time-dependent coefficients (\ref{dotx})-(\ref{dotpy}). 
This set can be reduced to a single second order differential equation in two special cases: $\alpha=0$ and $\alpha=1$ 
(or $\alpha=-1$). These cases are studied separately in Sections \ref{sec-circ}--\ref{sec-tanh}
and \ref{sec-Land}--\ref{sec-Landexp}.

\section{The circular gauge: general}
\label{sec-circ}

For $\alpha=0$, it is convenient  to introduce the complex variables
$z=x+iy$ and $p=p_x + i p_y$ \cite{LR,MMT69,MMT70}. They obey the equations
\[
\dot{z} = p/m - i\omega(t) z, \quad
\dot{p} = -i\omega(t) p -m\omega^2(t) z.
\]
Writing
\[
z=\Phi \tilde{z}, \quad p = \Phi \tilde{p}, \quad \Phi = \exp\left[ -i\int_0^t \omega(\tau)d\tau\right],
\]
we get the equations
\be
\dot{\tilde{z}} = \tilde{p}/m, \quad \dot{\tilde{p}} = -m\omega^2(t) \tilde{z},
\label{dottilz}
\ee
whose consequence is (\ref{eqvep})  with $\alpha=0$ for $\tilde{z}(t)$.
We fix the pair of independent complex solutions $\vep(t)$ and $\vep^*(t)$,
imposing the condition on the Wronskian \cite{MMT69,MMT70}
\be
\dot\vep \vep^* -\dot{\vep}^* \vep =2i \quad \mbox{or} \;\; \mbox{Im}\left(\dot\vep \vep^*\right)=1.
\label{incondvep}
\ee
We assume that $\omega(t)=\omega_{i} = const >0$ for $t\le 0$ and 
$\vep(t)= \omega_{i}^{-1/2}\exp\left(i\omega_{i} t\right)$ for $t \le 0$.
 This means that we choose the initial conditions 
\be
\vep(0)= \omega_{i}^{-1/2}, \quad \dot\vep(0)= i\omega_{i}^{1/2}.
\label{invep}
\ee
Solutions to equations (\ref{dottilz}) are linear combinations,
\[
\tilde{z}(t) = C_1\vep(t) + C_2\vep^*(t), \quad \tilde{p}(t) = m\left[ C_1\dot\vep(t) + C_2\dot\vep^*(t)\right],
\]
where constant coefficients $C_{1,2}$ are determined by the initial conditions.
Thus we arrive at formulas
\[
z(t) = \omega_{i}^{1/2}\Phi(t)\left[z(0)\mbox{Re}(\vep) +p(0)\mbox{Im}(\vep)/(m\omega_{i})\right],
\]
\[
p(t) = m\omega_{i}^{1/2}\Phi(t)\left[z(0)\mbox{Re}(\dot\vep) +p(0)\mbox{Im}(\dot\vep)/(m\omega_{i})\right].
\]
Further details of calculations and explicit forms of blocks (\ref{sigr})-(\ref{sigrc}) of matrix 
$\Lambda_{q}(t)$ (\ref{Lamq-0}) are given in Appendix \ref{ap-detcirc}.

Mean values of the energy and magnetic moment depend on traces of matrices $\sigma_r$ and $\sigma_{rc}$. These traces have the
following explicit forms:
\be
\mbox{Tr}(G\sigma_r) = \frac{G\omega_i}{2\omega^2(t)} \left[ |F_-|^2 + s_0\Upsilon |F_+|^2 - 2\rho \mbox{Re}(F_- F_+) \right],
\label{TrGsigr}
\ee
\be
\mbox{Tr}(G\sigma_{rc}) = \frac{G\omega_i}{2\omega^2(t)} \left[ (1 + s_0\Upsilon)\mbox{Re}\left(F_{-} F_{+}^*\right) - 
\rho \mbox{Re}\left(F_{-}^2 + F_{+}^2\right) \right],
\label{TrGsigrc}
\ee
where
\be
F_{\pm}(t) = \omega(t)\vep(t) \pm i \dot\vep(t), \qquad
2s_0 = s+ s^{-1}.
\label{def-Fpm}
\ee
Note that the traces (\ref{TrGsigr}) and (\ref{TrGsigrc}) are invariant with respect to the transformation $s \to s^{-1}$.
Two important special cases will be analyzed in more details in the subsequent sections.

1) The adiabatic regime:
\be
\vep(t) \approx [\omega(t)]^{-1/2}\exp[i\vf(t)], \quad \dot\vep(t) \approx i[\omega(t)]^{1/2}\exp[i\vf(t)], 
\qquad \vf(t) = \int_0^t \omega(x)dx,
\label{vep-ad}
\ee
\be
F_{-}(t) \approx 2[\omega(t)]^{1/2}\exp[i\vf(t)], \quad F_{+}(t) \approx 0. 
\label{vepFadiab}
\ee
In this case, matrix $\mat{\sigma}_q(t)$ assumes the form
\be
\mat{\sigma}_q^{(ad)}(t) = \frac{G\omega_i}{\omega(t)}
\left\Vert
\begin{array}{c c c c}
1 & 0 & -\rho\cos(2\vf) & 0 \\
0 & 1 & 0 & -\rho\cos(2\vf) \\
-\rho\cos(2\vf) & 0 & \Upsilon & 0 \\
0 & -\rho\cos(2\vf) & 0 & \Upsilon
\end{array} 
\right\Vert.
\label{sig-ad}
\ee

2) The asymptotic regime, when 
 the frequency $\omega(t)$ assumes a constant value $\omega_{f}$ after some time interval $T$ 
(or asymptotically as $t\to\infty$). In this case, one can write the solution $\vep(t)$ for $t>T$ as
\be
\vep(t) = |\omega_{f}|^{-1/2}\left[u_{+} e^{i|\omega_{f}|t} + u_{-} e^{-i|\omega_{f}|t}\right],
\label{uvsol}
\ee
where constant complex coefficients $u_{\pm}$  obey the condition
\be
|u_{+}|^2 - |u_{-}|^2 =1,
\label{uvcond}
\ee
which is the consequence of (\ref{incondvep}). Then, we have for $t>T$,
\be
F_{\pm}(t) = 2|\omega_{f}|^{1/2} u_{\mp} e^{\mp i|\omega_{f}|t} .
\ee

\subsection{Evolution of the mean energy}

Equations (\ref{Ham0})  and (\ref{TrGsigr}) lead to the following 
expressions for the mean energy: 
\be
{\cal E}(t) =  {m\Omega^2(t)} \mbox{Tr}(G\sigma_r)/2
 =  \frac{{\cal E}_i}{4\omega_i} \left[ |F_-|^2 + s_0\Upsilon |F_+|^2 - 2\rho \mbox{Re}(F_- F_+) \right], \qquad
 {\cal E}_i = 4m\omega_i^2 G.
\label{Eqfin}
\ee
In the asymptotic regime (\ref{uvsol}),
the ratio of the final energy to the initial one equals
\be
{\cal E}_{f}/{\cal E}_{i} = \left(|\omega_f|/\omega_i\right)\PG{|u_{+}|^2 +s_0 \Upsilon |u_{-}|^2 -2\rho\mbox{ Re}(u_+u_-)}.
\label{EfEi-u-}
\ee

\subsubsection{Adiabatic regime}
\label{sec-subsub-adiab}

Taking the solution to Equation (\ref{eqvep}) in the form (\ref{vep-ad}),
we have
\[
\dot\vep = \left(i\omega^{1/2} - \frac{\dot\omega}{2\omega^{3/2}}\right)\exp(i\vf), \qquad
\dot\vep \vep^* = i - \frac{\dot\omega}{2\omega^{2}},
\]
so the condition $\mbox{Im}(\dot\vep \vep^*) =1$ is satisfied automatically. 
In this case, 
 ${\cal E}(t) \approx 4 mG\omega_i \omega(t)$,  meaning that the ratio ${\cal E}(t)/ \omega(t)$
is the known adiabatic invariant, which does not depend on parameters $\rho, \Upsilon, s$.
However, this invariant exists for $\omega(t)>0$ only.
Indeed, calculating the second derivative 
of $\vep(t)$, one arrives at the equation
\be
\ddot\vep + \omega^2\vep = \left(\frac{3\dot\omega^2}{4\omega^{5/2}} - \frac{\ddot\omega}{2\omega^{3/2}}\right)\exp(i\vf).
\label{adterms}
\ee
The right-hand side of (\ref{adterms}) can be neglected under the conditions
\be
|\ddot\omega/\omega^3| \ll 1, \quad |\dot\omega/\omega^2| \ll 1.
\label{adcond}
\ee
 If the Larmor frequency $\omega(t)$ changes its sign, slowly passing through the value $\omega=0$, the inequalities (\ref{adcond})
  cannot be guaranteed,
 and the situation can be quite different, as shown in Sections \ref{sec-hyper} and \ref{sec-tanh}.

\subsubsection{Sudden jump of the magnetic field}
\label{sec-jump-gen}

A simple special case is an
 instantaneous jump of the frequency from the value $\omega_i$ at $t<0$ to $\omega_f$ at 
$t>0$. Then we have at $t>0$ the solution (\ref{uvsol}) with
\be
u_{\pm} = \frac{|\omega_{f}| \pm \omega_{i}}{2\sqrt{|\omega_{f}|\omega_{i}}},
\label{upmjump}
\ee
so 
\be
{\cal E}_{f}/{\cal E}_{i} = \PG{\PC{\omega_f^2 + \omega_i^2}\PC{1+ s_0\Upsilon} -2\omega_i |\omega_f|\PC{ s_0\Upsilon -1}
-2\rho\PC{\omega_f^2-\omega_i^2}}/\PC{4\omega_i^2}.
\label{Eqfcirc2}
\ee
Equation (\ref{Eqfcirc2}) is symmetric with respect to the inversion $\omega_f \to -\omega_f$.
In particular, ${\cal E}_{f}/{\cal E}_{i} = 1$ if $\omega_f = -\omega_i$ (the instantaneous inversion of the magnetic field).
Another interesting feature of formula (\ref{Eqfcirc2})  is the non-analyticity of the sudden jump ratio 
${\cal E}_{f}/{\cal E}_{i}$ as function of the final frequency $\omega_f$ at point $\omega_f=0$ if $s_0\Upsilon >1$. 
This discontinuity of the derivative is clearly seen as a cusp in Figure \ref{fig-EfEi-wf}.

Equation (\ref{Eqfcirc2}) predicts that
the mean energy does not go to zero after the instantaneous jump of the frequency to $\omega_f = 0$ 
(contrary to the adiabatic evolution):
\be
{\cal E}_{f}/{\cal E}_{i} = \PG{1 +s_0\Upsilon +2\rho}/4 \ge 1/2.
\label{EfEi-sudden0}
\ee
One can question this result, because
the limit $\omega_f \to 0$ is not justified in the initial equations (\ref{uvsol}) and (\ref{upmjump}).
However, the exact solution to Equation (\ref{eqvep}) with $\omega(t) =0$ at $t>0$, satisfying the initial conditions (\ref{invep}),
has the form 
(this solution was used in Ref. \cite{Hacyan} in connection with the concept of ``quantum sling'')
\be
\vep(t) = \omega_i^{-1/2}(1 + i\omega_i t).
\label{vep-om0}
\ee
 Hence, $F_{\pm}(t) = \mp \omega_i^{1/2}$, so  
Equation (\ref{Eqfin}) results in the same formula (\ref{EfEi-sudden0}).
The minimal value $1/2$ of the right-hand side of Equation (\ref{EfEi-sudden0}) is achieved for zero temperature ($\Upsilon=1$ and $\rho=0$) 
in the isotropic trap ($s_0=1$). 
In this limit, ${\cal E}_{f}/{\cal E}_{i} = (\omega_i^2 + \omega_f^2)/(2\omega_i^2)$.
But the final energy can be much higher than the initial one after instantaneous switching off the magnetic field,
if $\Upsilon \gg 1$ or $s_0\gg 1$ (the high temperature initial state or a strongly anisotropic trap). The approximate
formula in this case reads ${\cal E}_{f}/{\cal E}_{i} \approx s_0\Upsilon(\omega_i - |\omega_f|)^2/(4\omega_i^2)$
(provided the difference $1- |\omega_f|/\omega_i$ is not very small).

The model of  instantaneous jumps of parameters was used by many authors for the analysis of various physical processes
\cite{LR,DelMi,Mielnik11,Parker71,JanYu,Gra,Bechler88,Ma,Lo90,Bas92,JanAd92,DKN93,Olend93,Kiss94,Sas94,Kira95,Titt96,Mend00,%
Minguzzi05,Campos08,Hoffman11,Abah12,Rajab14,DH19,Vicari19,Tiba20,Tiba21}.
 Its validity is analyzed in the next sections. 
In particular, we show in  Section \ref{sec-om1t} that the exact results for $\omega_f =0$ in some cases 
can be quite different from (\ref{EfEi-sudden0}).

\subsubsection{Parametric resonance}

An approximate solution to Equation (\ref{eqvep}) in the form (\ref{uvsol}), with $\omega_{f}=\omega_{i}$ and
{\em slowly time dependent coefficients},
\be
u_{+}(t) = \cosh(\omega_{i}\gamma t), \quad u_{-}(t) = -i\sinh(\omega_{i}\gamma t),
\label{upm-param}
\ee
exists in the parametric resonance case, when the magnetic field is 
harmonically modulated at the {\em twice Larmor frequency\/} 
 (see, e.g., \cite{Dodcohmag18,DKN93,Louis61,Moll67}):
 \be
\omega(t) = \omega_{i}\left[1 + 2\gamma \cos(2\omega_{i} t)\right], \quad |\gamma| \ll 1.
\label{omres}
\ee 
Then, 
\be
{{\cal E}(t)}/{{\cal E}(0)} = \cosh^2(\omega_{i}\gamma t) + s_0\Upsilon \sinh^2(\omega_{i}\gamma t).
\label{Eqres0}
\ee
Note that coefficient $\rho$ does not enter Equation (\ref{Eqres0}), because $\mbox{Re}(u_{+} u_{-})=0$ in the case involved.

\subsection{Energy fluctuations}
\label{sec-enfluc}

The energy fluctuations can be characterized by the variance $\sigma_E =\langle \hat{H}^2\rangle - \langle \hat{H}\rangle^2$,
where
\be
\langle\hat{H}^2\rangle = (2m\omega^2)^2 \langle \hat{x}_r^4 + \hat{y}_r^4 + \hat{x}_r^2\hat{y}_r^2 + \hat{y}_r^2\hat{x}_r^2 \rangle.
\label{H2}
\ee
The fourth order moments in the right-hand side of (\ref{H2}) can be easily calculated for the
initial equilibrium state, because this state is {\em Gaussian}. Moreover, since the Hamiltonian (\ref{Ham})
is {\em quadratic\/} with respect to the canonical variables, it transforms any Gaussian state to another Gaussian state.
Therefore, we can use well known formulas of the classical probability theory 
(with some modifications due to the non-commutativity of the coordinate and momentum operators) 
for average values of the Gaussian distributions (see, e.g., \cite{183vol}). 
Namely, the mean values of {\em symmetrical\/} (or Wigner--Weyl) products \cite{Hill84} of four operators,
 $\hat{A}$, $\hat{B}$, $\hat{C}$ and $\hat{D}$ (with zero mean values), can be expressed as sums of pair products of their 
second order central moments \cite{183vol}:
\beqn
\langle{ABCD}\rangle_W =
\overline{AB}\cdot\overline{CD} + \overline{AC}\cdot\overline{BD} + \overline{AD}\cdot\overline{BC}.
\label{basic}
\eeqn
Here $A,B,C,D$ can be any of variables $x_r, y_r, x_c, y_c$. The  symbol $\langle{ABCD}\rangle_W$ means
the quantum mechanical average value of the sum of all different products of operators $\hat{A},\hat{B},\hat{C},\hat{D}$,
taken in all possible orders, divided by the number of terms in the sum. The second order central moments are defined as
$\overline{AB} \equiv \langle \hat{A}\hat{B}+\hat{B}\hat{A}\rangle/2$. 
Mean values of concrete products of operators in predefined
orders can be expressed in terms of symmetrical mean values with the aid of commutation relations.
The explicit expressions are given in Appendix \ref{ap-42}.
Using that formulas, we obtain 
\[
\sigma_E = \PG{2m\omega^2(t)}^2\PC{2\sigma_{11}^2+2\sigma_{22}^2+\PG{x_r,y_r}^2}.
\]
Comparing this expression with (\ref{sigrcirc}) and (\ref{Eqfin}) in the case of $s=1$, we arrive at a surprisingly simple result
\be
\sigma_E(t) ={\cal E}^2(t)-\PG{\hbar \omega(t)}^2,
\ee
which holds for any values of parameters $\Upsilon$ and $\rho$.
If $s\neq 1$, then the formula for $\sigma_E$ is much more involved (containing trigonometric functions of $\vf$). For this reason,
we do not consider here the case of $s\neq 1$ in connection with the dynamics of fluctuations.
In view of Equation (\ref{EiMi}), the initial level of energy fluctuations is given by the formulas 
\be
\sigma_E(0) = (\hbar\omega_i)^2\left({\cal C}^2 -1\right), \qquad
\sigma_E(0)/{\cal E}_i^2 = 1 -\tanh^2(\hbar\omega_i\beta).
\label{sigE0}
\ee

\subsection{Evolution of the mean magnetic moment}

Equations (\ref{Lkin}), (\ref{M2}), (\ref{TrGsigr}) and (\ref{TrGsigrc}) result in the following explicit expression for
the time dependent mean value of the magnetic moment:
\be
{\cal M}(t)  = -\,\frac{eG}{c}\omega(t)\mbox{Tr}\PC{\sigma_{r}+\sigma_{rc}} =
 -\,\frac{\mu_B {\cal C}}{2}\PH{\omega(t)|\vep|^2 +1 + \Upsilon s_0 \PG{\omega(t)|\vep|^2 -1}
-2\rho \omega(t)\mbox{Re}(\vep^2)}.
\label{meanmag}
\ee
Note that the derivative $\dot\vep(t)$ does not enter the formula for the mean magnetic moment, in contradistinction to the
formula (\ref{Eqfin}) for the mean energy.
In particular,  we have in the zero temperature case (${\cal C} = \Upsilon =1$,  $\rho=0$)
\be
{\cal M}^{(l)}(t) = -(\mu_B {\cal C}/2) \left[\omega(t)|\vep|^2 (1+ s_0) + 1 - s_0 \right].
\label{meanmaglow}
\ee
Parameter $\Upsilon$ {\em almost\/} disappears from Equation (\ref{meanmag}) in the adiabatic case, when $\omega(t)|\vep|^2 -1 \approx 0$:
\be
{\cal M}_{ad}(t)=\mu_B {\cal C}\PG{\rho \cos(2\varphi)-1}, \qquad \vf(t) = \int_0^t \omega(\tau)d\tau.
\label{Madiab}
\ee
According to Equation (\ref{Madiab}),
the mean value of the magnetic moment is the adiabatic invariant for $\rho=0$ and ${\cal C} =1$ only 
(zero temperature initial state).
If $\rho>0$, ${\cal M}_{ad}(t)$ is an oscillating function of time (being always negative).
Note that $|{\cal M}_{ad}(t)|$ can achieve  very big values in the high temperature case, when ${\cal C} \gg 1$.
Also, the parameter $\Upsilon$ will make a contribution when $\Upsilon \gg 1$, due to corrections to the adiabatic approximation.
We return to this issue in Section \ref{sec-fin}.

In non-adiabatic regimes, when the difference $\omega(t)|\vep|^2 -1$ is not close to zero (including all situations with
$\omega \le 0$), Equation (\ref{meanmag}) shows that the contribution of terms containing parameter $\rho$ can be neglected 
(because $\rho$ is close to zero for low temperatures and $\rho \ll s_0\Upsilon$ in the high-temperature case). 
This observation will help us to simplify many formulas.

In the asymptotic regime  (\ref{uvsol}) we obtain 
\be
 |\omega_{f}||\vep|^2=|u_-|^2+|u_+|^2+2 \mbox{Re}(u_-u_+^{*})\cos(2 |\omega_{f}|t)+2 \mbox{Im}(u_-u_+^{*})\sin(2 |\omega_{f}|t),
\label{as-mod} 
\ee
\be
 |\omega_{f}|\mbox{Re}(\vep^2)= \mbox{Re}(u_-^2+u_+^2)\cos(2 |\omega_{f}|t)+ \mbox{Im}(u_-^2-u_+^2)\sin(2 |\omega_{f}|t)+2 \mbox{Re}(u_-u_+).
\label{as-Re} 
\ee
Using the formula $a\cos(x) + b\sin(x) = \sqrt{a^2 + b^2} \sin(x +\phi)$ (where $\phi$ is some phase which is not interesting
for our purposes),
we can rewrite the right-hand side of Equation (\ref{meanmag}) as a sum of constant (averaged over temporal oscillations) and oscillating parts:
\be
{\cal M}(t) = \langle\langle {\cal M}\rangle\rangle + \widetilde{\Delta {\cal M}}\sin(2 |\omega_{f}|t +\phi),
\label{M-av-fluc}
\ee
\be
\langle\langle {\cal M}\rangle\rangle = -\mu_B {\cal C}\sigma\left[|u_{\sigma}|^2 + |u_{-\sigma}|^2 s_0\Upsilon
- 2 \rho \mbox{Re}(u_-u_+)\right], \qquad \sigma = \omega_f/|\omega_f|,
\label{<<M>>}
\ee
\beqn
|\widetilde{\Delta {\cal M}}|  &=& |\mu_B {\cal C}|\left\{ (1 +s_0\Upsilon)^2 |u_{+}u_{-}|^2 - 
2\rho(1 +s_0\Upsilon)\mbox{Re}(u_{+}u_{-})\left(|u_{+}|^2 +|u_{-}|^2\right)
\right. \nonumber \\ && \left.
+\rho^2\left[|u_{+}|^2 +|u_{-}|^2 + 2\mbox{Re}(u_{+}^2u_{-}^2)\right]\right\}^{1/2}.
\label{flucM}
\eeqn
Note that the consequence of identity (\ref{uvcond}) is the relation 
\be
|u_{+}|^2 +|u_{-}|^2 = \sqrt{1 + 4 |u_{+}u_{-}|^2}.
\label{u+u_sqrt}
\ee
The most simple expressions can be written for  $s_0\Upsilon =1$ and $\rho=0$:
\be
\langle\langle {\cal M}\rangle\rangle = -\mu_B {\cal C}\sigma\sqrt{1 + 4 |u_{+}u_{-}|^2}, \qquad
|\widetilde{\Delta {\cal M}}|  = 2|\mu_B {\cal C}u_{+}u_{-}|.
\label{Msimple-low}
\ee
Other simple formulas can be written in the high-temperature case $s_0\Upsilon \gg 1$. If 
$\omega_f <0$, then,
\be
\langle\langle {\cal M}\rangle\rangle \approx \frac12 \mu_B {\cal C}s_0\Upsilon\left(1 +\sqrt{1 + 4 |u_{+}u_{-}|^2}\right), \qquad
|\widetilde{\Delta {\cal M}}|  \approx |\mu_B {\cal C}s_0\Upsilon u_{+}u_{-}|.
\label{Msimple-high}
\ee
We see that the amplitude of oscillations is close to the
average value if $|u_{+}u_{-}| \gg 1$, being always smaller than the average value.
Consequently,  ${\cal M}(t)$ does not change the sign in the asymptotic regime in these two special cases. 

In non-adiabatic regimes, the terms containing parameter $\rho$ can be neglected in Equations (\ref{<<M>>}) and (\ref{flucM}).
In these cases, we have to calculate the coefficient $|u_{-}|^2$ only. In particular, for $\omega_f <0$ we can use 
the following approximate formulas:
\be
\langle\langle {\cal M}\rangle\rangle \approx \mu_B {\cal C}\left[|u_{-}|^2 ( 1+  s_0\Upsilon) + s_0\Upsilon \right],
\qquad
|\widetilde{\Delta {\cal M}}|  = \mu_B {\cal C} (1 +s_0\Upsilon) |u_{-}|\sqrt{1 + |u_{-}|^2}. 
\label{<<M>>simple}
\ee

\subsubsection{The case of sudden jump}

Formulas (\ref{meanmag}), (\ref{as-mod}) and (\ref{as-Re})
can be simplified in the special case of the sudden jump of magnetic field, when coefficients $u_{\pm}$ are real:
see Equation (\ref{upmjump}). Then, for any sign of the final frequency $\omega_f$, we obtain
\be
{\cal M}(t) = -\mu_B{\cal C}\left\{ \frac{\omega_f+\omega_i}{2\omega_i} -\rho \frac{\omega_f}{\omega_i}
+ \Upsilon s_0 \frac{\omega_f -\omega_i}{2\omega_i}
+\sin^2(\omega_f t)
\left[\rho W_{+} \!-\! \frac{W_-}{2}\left(1+\Upsilon s_0 \right)\right]\right\},
\ee
where
$ W_\pm= (\omega_f^2\pm\omega_i^2)/(\omega_i\omega_f)$.
In particular, at zero temperature we have the ratio
\be
R \equiv \frac{|\widetilde{\Delta {\cal M}}|}{|\langle\langle {\cal M}\rangle\rangle|} = 
\left|\frac{\omega_f^2 -\omega_i^2}{\omega_f^2 +\omega_i^2}\right|.
\label{Rjumplow}
\ee
The magnetic moment changes its sign immediately after the jump (at $t=0+$) if 
\[
\omega_f < \omega_* = \omega_i\,\frac{\Upsilon s_0 -1}{\Upsilon s_0 +1 -2\rho}.
\]
Note that $\omega_*$ is only slightly smaller than $\omega_i$ in the high-temperature case ($\Upsilon \gg 1$ and $\rho \approx 1$).
But even in the zero-temperature case ($\Upsilon =1$ and $\rho =0$), $\omega_*$ can be close to $\omega_i$ in strongly anisotropic
initial traps wih $s_0 \gg 1$.
If $\omega_f =0$ exactly, then 
\be
{\cal M}_f/(\mu_B{\cal C})  = (\Upsilon s_0 -1) /{2} = const \ge 0
\label{Mjump-om0}
\ee
 after switching off the field.
The same result follows from Equation (\ref{meanmag}) with $\omega(t)=0$ and $\vep(t) = \omega_i^{-1/2}(1 + i\omega_i t)$ at $t>0$. 
However, if $\omega_f \neq 0$, then the magnetic moment oscillates with frequency $2|\omega_f|$, and the amplitude of oscillations
can be rather high.
For example, for $|\omega_f| \ll \omega_i$ we have
\be
{\cal M}(t)= -\, \frac{\mu_B{\cal C}}{2}\left[1 \!-\! \Upsilon s_0
+ \frac{\omega_i}{\omega_f}\sin^2(\omega_f t)
\PC{1 \!+\! \Upsilon s_0 \!+\! 2\rho}\right].
\label{Mt-omfsmall}
\ee
Due to the fraction ${\omega_i}/{\omega_f}$, the magnetic moment can attain periodically very high negative values 
(i.e., of the same sign as the initial value ${\cal M}_i$)
for $\omega_f >0$
(and positive values for $\omega_f <0$). 
Moreover, ${\cal M}(t)$ changes its sign during the evolution if $\omega_f >0$, because $1 \!-\! \Upsilon s_0 <0$.

Equations (\ref{Mjump-om0}) and (\ref{Mt-omfsmall}) show that the division of the mean magnetic moment in the constant and oscillating parts
(\ref{M-av-fluc})
is questionable for $\omega_f \to 0$, when the period of oscillations becomes extremely large. Indeed, Equation (\ref{Mt-omfsmall})
yields the ratio $R = |\widetilde{\Delta {\cal M}}|/|\langle\langle {\cal M}\rangle\rangle| \approx 1$ if
$|1 \!-\! \Upsilon s_0| \ll |\omega_i/\omega_f|
\PC{1 \!+\! \Upsilon s_0 \!+\! 2\rho}$, while Equation (\ref{Mjump-om0}) yields $\widetilde{\Delta {\cal M}} =0$ if
$\omega_f=0$ exactly.

After the sudden inversion of magnetic field ($\omega_f =-\omega_i$) we have the positive function
\be
{\cal M}(t)/(\mu_B{\cal C}) = \Upsilon s_0 - \rho +2\rho\sin^2(\omega_i t),
\label{M-circ-sudinv}
\ee
which shows that the amplitude of oscillations is very small compared with the average value $\langle\langle {\cal M}\rangle\rangle$ 
for this specific choice of the final frequency.

\subsubsection{Parametric resonance}

In the parametric resonance  case (\ref{upm-param}) we have
\[
{\cal M}_{res}(t)= -\mu_B{\cal C}\PG{\cosh^2(\omega_i\gamma t)+ \Upsilon s_0\sinh^2(\omega_i\gamma t)+ 
\PC{1+\Upsilon s_0}\sinh(2\omega_i\gamma t)\sin(2\omega_i t)/2- \rho\cos(2\omega_i t)}.
\]
This quantity grows with time by the absolute value, but it does not change its initial negative sign, despite strong oscillations
with the frequency $2\omega_i$. In particular, when $\omega_i\gamma t \gg 1$, the ratio $[-{\cal M}_{res}(t)/(\mu_B{\cal C})]$
rapidly oscillates between the maximal value close to 
$\exp(2\omega_i\gamma t)\PC{1+\Upsilon s_0}/2$ and the minimal value which is close to zero.

\subsection{Magnetic moment fluctuations}

The magnetic moment fluctuations can be characterized by the variance 
$\sigma_M \equiv \langle \hat{\cal M}^2\rangle - \langle \hat{\cal M}\rangle^2$.
Using Equations (\ref{Ham0}), (\ref{Lkin}) and (\ref{M2}), we can write 
$\hat{\cal M}^2 = e^2\hat{H}^2 /(mc\Omega)^2 + \hat{K}$, where
\[
\hat{K} = \frac{(e\omega)^2}{c^2}\Big[\hat{x}_c^2\hat{x}_r^2 + \hat{y}_c^2\hat{y}_r^2
+ \hat{x}_c\hat{y}_c \hat{x}_r \hat{y}_r + \hat{y}_c\hat{x}_c  \hat{y}_r \hat{x}_r
+ \hat{x}_c \left( \hat{y}_r^2 \hat{x}_r \!+\! \hat{x}_r\hat{y}_r^2\right)
+ \hat{y}_c \left( \hat{x}_r^2 \hat{y}_r \!+\! \hat{y}_r\hat{x}_r^2\right)
+ 2\left(\hat{x}_r^3 \hat{x}_c  \!+\! \hat{y}_r^3 \hat{y}_c \right) \Big].
\]
The average value of $\hat{H}^2$ was calculated in Section \ref{sec-enfluc}. Average values
of other fourth-order products of operators can be calculated according to the rule (\ref{basic}), using explicit formulas
 given in Appendix \ref{ap-42}. 
However, explicit expressions in terms of all initial parameters are rather involved: 
see Equations (\ref{sigrcirc})-(\ref{sigrccirc}). For this reason, we confine ourselves here to the case of symmetric trap ($s=1$),
with $\overline{x_ry_r} = \overline{x_cy_c}=0$. Then, taking into account formula (\ref{meanmag})
and the symmetries of matrices (\ref{sigr-sigc})-(\ref{sigrc-s1}),  we find
\be
\sigma_M=\PG{{e \omega(t)}/{c}}^2\left\{2\PC{2\sigma_{11}^2 -\sigma_{14}^2} 
+ 8 \sigma_{11}\sigma_{13} + 2\sigma_{13}^2 +2\sigma_{11}\sigma_{33} +\PG{\hat{x}_r,\hat{y}_r}/2\right\}.
\ee
Nonetheless, even this formula is still rather cumbersome, as soon as each term $\sigma_{ij}$ is an inhomogeneous linear combination 
of parameters $\Upsilon$ and $\rho$ with different coefficients. Therefore, we confine ourselves here to 
the limit cases of low and high initial temperatures.

In the zero temperature case ($\rho=0$, ${\cal C}=\Upsilon=1$), 
using Equations (\ref{sigr-sigc}) and (\ref{sigrc-s1}) together with 
the identity 
\[
|\vep|^2|\dot{\vep}|^2- \mbox{Re}^2(\dot{\vep}\vep^*) = \mbox{Im}^2(\dot{\vep}\vep^*) \equiv 1,
\]
we obtain after some algebra an extremely simple formula
\be
\sigma_M^{(l)}(t)= \mu_B^2\PG{\omega(t)|\vep|^2}^2 = \PG{{\cal M}^{(l)}(t)}^2,
\ee
which shows that quantum fluctuations of the magnetic moment are always strong, even at zero temperature.

In the adiabatic case, matrix (\ref{sig-ad}) leads to the formula
\be
\sigma_M^{(ad)}(t) = (\mu_B {\cal C})^2\PG{2 + \Upsilon + \rho^2 \cos^2(2\vf) - 4\rho\cos(2\vf)}/2 - \mu_B^2/2.
\ee
In contradistinction to Equation (\ref{Madiab}) for the mean magnetic moment, the magnetic moment variance
contains the term proportional to $\Upsilon$  in the adiabatic regime. Therefore, in the high temperature case we have
$\sigma_M^{(ad)} \approx (\mu_B {\cal C})^2 \Upsilon/2 \gg {\cal M}_{ad}^2$.
Moreover, $\sigma_M^{(ad)}(t)$ is almost constant for $\Upsilon \gg 1$, since the amplitude of oscillations is much smaller than
$\Upsilon$ (remember that $\rho \le 1$).
On the other hand, in the non-adiabatic regime we obtain, taking into account only terms proportional to $\Upsilon$ in 
matrices (\ref{sigr-sigc})-(\ref{sigrc-s1}) and comparing the result with (\ref{meanmag}), the following formula 
in the high temperature case:
\be
\sigma_M^{(h)} =\PC{\mu_B {\cal C}\Upsilon}^2\PG{\omega(t)|\vep|^2-1}^2 = 4\left[{\cal M}^{(h)}\right]^2.
\ee
It is valid provided $\Upsilon\PG{\omega(t)|\vep|^2-1} \gg 1$.

It was shown in paper \cite{DDmag} that quantum fluctuations of the magnetic moment are very strong in the high-temperature
equilibrium state (when the {\em mean value\/} is small). The formulas of this section show that time-dependent magnetic fields
amplify these fluctuations. Hence, in each concrete measurement one can obtain the values of the magnetic moment of any sign,
with huge differences in outcomes of different experiments. The mean squared deviations can be much bigger than the mean values
obtained after averaging over many tests.

\section{Explicit solutions of the oscillator equation in terms of elementary functions}
\label{sec-part}

Exact solutions to Equation (\ref{eqvep}) are known for about a dozen of families of functions $\omega(t)$: see, e.g, 
a list in \cite{book3}
(since we consider the circular gauge in this section, $\omega$ means the {\em Larmor frequency}).
In the majority of cases, these solutions are expressed in terms of various special functions.
Nonetheless, there exist at least three specific examples, when solutions can be expressed in terms of elementary functions.
Two of them describe the inverse power law decrease of the magnetic field to zero value, while the third one corresponds
to the exponential-like decrease to an arbitrary final value.

\subsection{Inverse linear decrease of magnetic field}
\label{sec-om1t}

One can easily verify that Equation (\ref{eqvep}) with the function
\be
\omega(t) = \left\{
\begin{array}{ll}
\omega_{0}, & t\le 0\\
\omega_{0}t_{0}/(t+t_0) = \omega_0/\tau, & t\ge 0
\end{array}
\right.,
\qquad \tau = 1 +t/t_0
\label{omoff}
\ee
has solutions $\tau^{1/2 \pm r}$, where $r=\sqrt{1/4-u^2}$ and $u=\omega_{0}t_{0}$
(see, e.g., papers \cite{Seym65,Lewis68}). Hence,
the function $\vep(t)$
satisfying the initial conditions (\ref{invep}) has the following form at $t \ge 0$ (or $\tau \ge 1$):
\be
\vep(t) = \frac{\sqrt{\tau}}{4r\sqrt{\omega_0}}\left[( 2r + 2 i u -1)\tau^{r} + (2r -2iu +1)\tau^{-r}\right], 
\label{vep-tau}
\ee
\[
\dot\vep(t) = \frac{i\sqrt{\omega_0}}{4r\sqrt{\tau}}\left[( 2r + 2 i u +1)\tau^{r} + (2r -2iu -1)\tau^{-r}\right].
\]
Note that the adiabaticity parameters, introduced in Equation (\ref{adcond}), have very simple and time independent forms in the case involved:
$|\dot\omega/\omega^2| = (\omega_0 t_0)^{-1} =u^{-1}$, $|\ddot\omega/\omega^3| = 2u^{-1}$. 
Consequently, the adiabatic regime corresponds to values $u \gg 1$, whereas the case
of $u \ll 1$ can be considered as a smooth analog of sudden jump. 
Note that function (\ref{vep-tau}) is close to (\ref{vep-om0}) for $\tau \gg 1$ and $u\ll 1$.
However, these functions do not coincide exactly. Important consequences of this difference are shown below.

\subsubsection{Fast field variation}
\label{fastlincirc}

If $u<1/2$,  Equation (\ref{Eqfin}) results in the formula
\be
{\cal E}(t)/{\cal E}_i = \left[\left(\tau^r - \tau^{-r}\right)^2 + 16 r^2 + s_0\Upsilon \left(\tau^r - \tau^{-r}\right)^2
+4\rho r\left(\tau^{2r} - \tau^{-2r}\right) \right]
/(16\tau r^2).
\label{E-exact}
\ee
If $u \ll 1$, then $t_0 \ll \omega_0^{-1}$. Consequently, practically for all values of time variable $t$, which are not extremely small,
we have $\tau \gg 1$ (for example, if $t=\omega_0^{-1}$, then $\tau \approx u^{-1}$). 
Moreover, $2r$ is very close to unity in this case. 
Neglecting the terms $\tau^{-2r}$ and putting $r=1/2$ in coefficients of Equation (\ref{E-exact}) 
(except for the exponent $r=(1-\delta)/2$), we arrive at a simplified expression
\be
{\cal E}(t)/{\cal E}_i \approx (1 +s_0\Upsilon + 2\rho)/(4\tau^{\delta}), \quad
\delta = 1- 2r \approx 2 u^2 \ll 1.
\label{E-almostjump}
\ee
Hence, the mean energy rapidly  drops to the sudden jump value (\ref{EfEi-sudden0}),
and remains at this level for a long time interval, when $\tau^{\delta} \approx 1$.
Note that the relative accuracy of approximation (\ref{E-almostjump}) is better than $0.01$ already for $t>10 t_0$.
Finally, the energy will drop to zero anyway, but this will happen for extremely big values of $\tau$.
For example, if   $s_0\Upsilon \gg 1$, then the inequality $\tau \gg \tau_* = (s_0\Upsilon/4)^{1/\delta}$ must be fulfilled
(in order to have ${\cal E}(\tau)/{\cal E}_i <1$).
If, for instance, $u=0.1$, $s_0=1$ and $\Upsilon=40$, then $\tau_*  \approx 10^{50}$.

Equation (\ref{meanmag}) yields the following expression for the mean magnetic moment:
\[
{\cal M}(\tau) = -\mu_B{\cal C}\left\{1 + \frac{1 + s_0\Upsilon}{16 r^2}\left[(1-2r)\tau^{2r} + (1+2r)\tau^{-2r} -2\right]
+ \frac{\rho}{4r}\left[(1-2r)\tau^{2r} - (1+2r)\tau^{-2r} \right] \right\}.
\]
For $u \ll 1$ and $\tau \gg 1$, this expression can be simplified as
\be
{\cal M}(\tau) = -\mu_B{\cal C}\left\{ \frac{1 - s_0\Upsilon}{2} 
+\frac{\delta}{4} \tau^{1-\delta} \left(1 + s_0\Upsilon + 2\rho\right)
 \right\}.
 \label{M-smallutau}
\ee
Neglecting the term proportional to $\delta$ in (\ref{M-smallutau}), one arrives at the sudden jump approximation formula
(\ref{Mjump-om0}). However, this can be done provided $\tau \ll 1/\delta$ only. When $\tau \to \infty$, the magnetic moment
grows unlimitedly (maintaining the initial sign).

\begin{figure}[htb]
\includegraphics[height=2.32truein,width=3.0truein,angle=0]{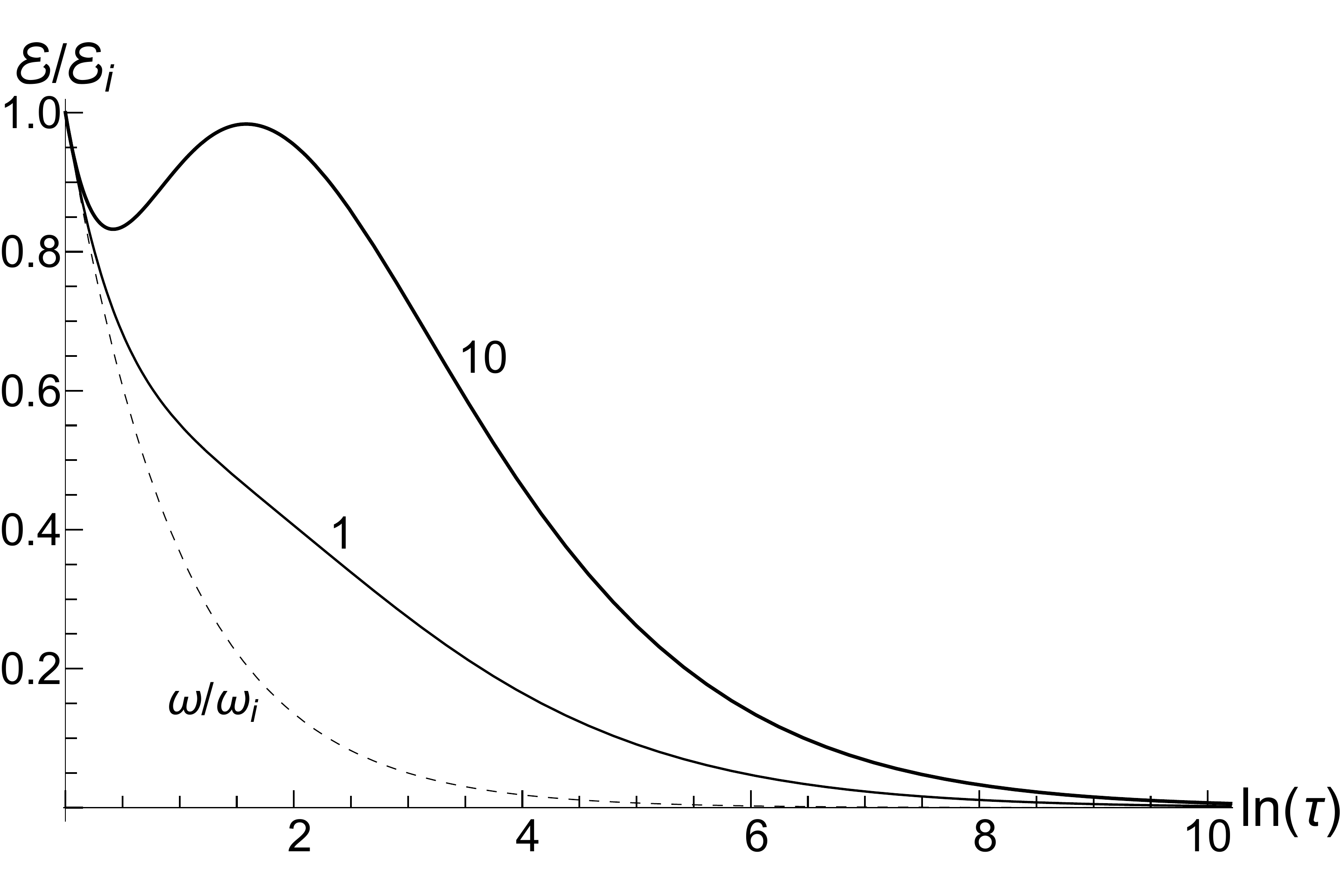}
\includegraphics[height=2.32truein,width=3.0truein,angle=0]{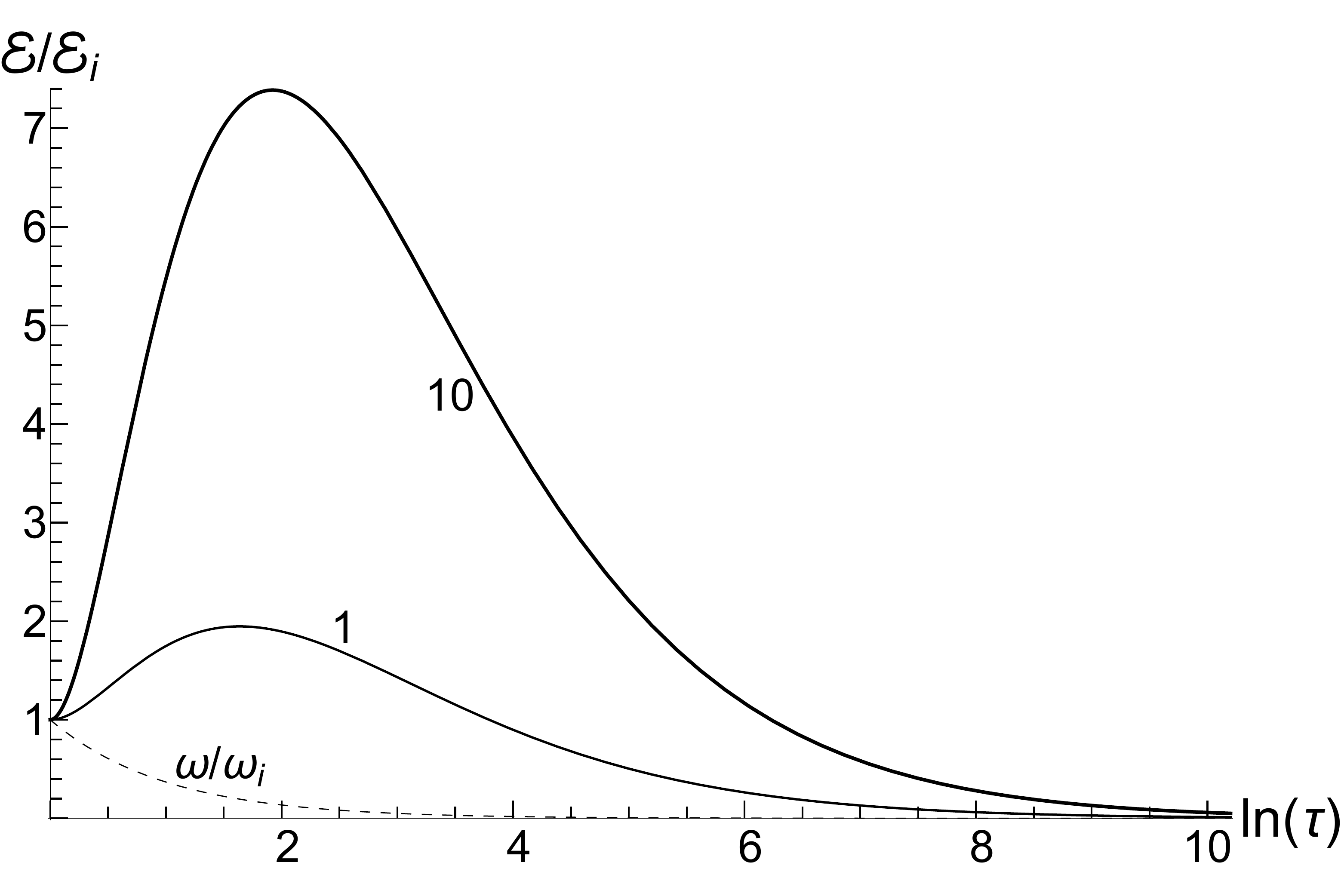}
\caption{\small The ratio ${\cal E}(\tau)/{\cal E}_i$ for different values of the asymmetry parameter $s$ (given nearby the curves)
for the inverse-linear decay of magnetic field (\ref{omoff}) with $u=\omega_0 t_0 =1/2$.
Left: the low temperature case, $\rho=0$, $\Upsilon=1$. Right: the high temperature case, $\rho=1$, $\Upsilon=10$.
The trace lines show the ratio $\omega(\tau)/\omega_i$.
 }
\label{fig-E-1tau}
\end{figure}

In the intermediate case of $u=1/2$ we have 
\be
{\cal E}(t)/{\cal E}_i = \frac{1}{\tau}  \left[ 1 + \ln^2(\tau)(1 + s_0 \Upsilon)/4
+\rho \ln(\tau) \right],
\ee
\be
{\cal M} = \mu_B{\cal C}\left\{\rho -1 -(1+s_0\Upsilon)[\ln^2(\tau) -2\ln(\tau)]/4 
-\rho\ln(\tau)\right\}.
\ee
The mean energy goes to zero value as $t\to\infty$, while the magnetic moment increases unlimitedly. 
The role of the asymmetry parameter $s_0$ is shown in Figure \ref{fig-E-1tau}.

\subsubsection{Slow field variation}
If $u>1/2$, then,
\[
\vep(t) = [\omega(t)]^{-1/2}\left[ e^{i\nu} +\sin(\nu)(2i\delta_u -1)/(2\gamma)\right], \quad
\dot\vep(t) = i[\omega(t)]^{1/2}\left[ e^{i\nu} +\sin(\nu)(2i\delta_u +1)/(2\gamma)\right],
\]
where $\gamma = \sqrt{u^2 -1/4} =|r|$, $\nu = \gamma \ln(\tau)$ and $\delta_u = u-\gamma$. 
Note that $\nu$ is close to the adiabatic phase $\int_0^t \omega(x)dx = u\ln(\tau)$ for $u \gg 1$, although
these quantities do not coincide exactly.
Now, Equation (\ref{Eqfin}) assumes the form
\be
{\cal E}(t)/{\cal E}_i = \frac{1}{\tau}  \left[ 1 + \frac{\sin^2(\nu)(2 + s_0 \Upsilon)}{2(4u^2-1)}
+\frac{\rho \sin(2\nu)}{\sqrt{4u^2 -1}} \right] .  
\label{E-exact-ubig}
\ee
This formula gives us the accuracy of the adiabatic invariant ${\cal E}(t)/\omega(t)= {\cal E}_i/\omega_i$ for $u\gg 1$.
 The peculiarity of the frequency dependence (\ref{omoff})
is that the adiabatic regime is maintained even when $\omega(\tau) \to 0$, whereas  the condition
(\ref{adcond}) fails for a generic function $\omega(t)$, if $\omega$ is close to zero: see examples in the following sections.

\begin{figure}[htb]
\includegraphics[height=2.02truein,width=3.0truein,angle=0]{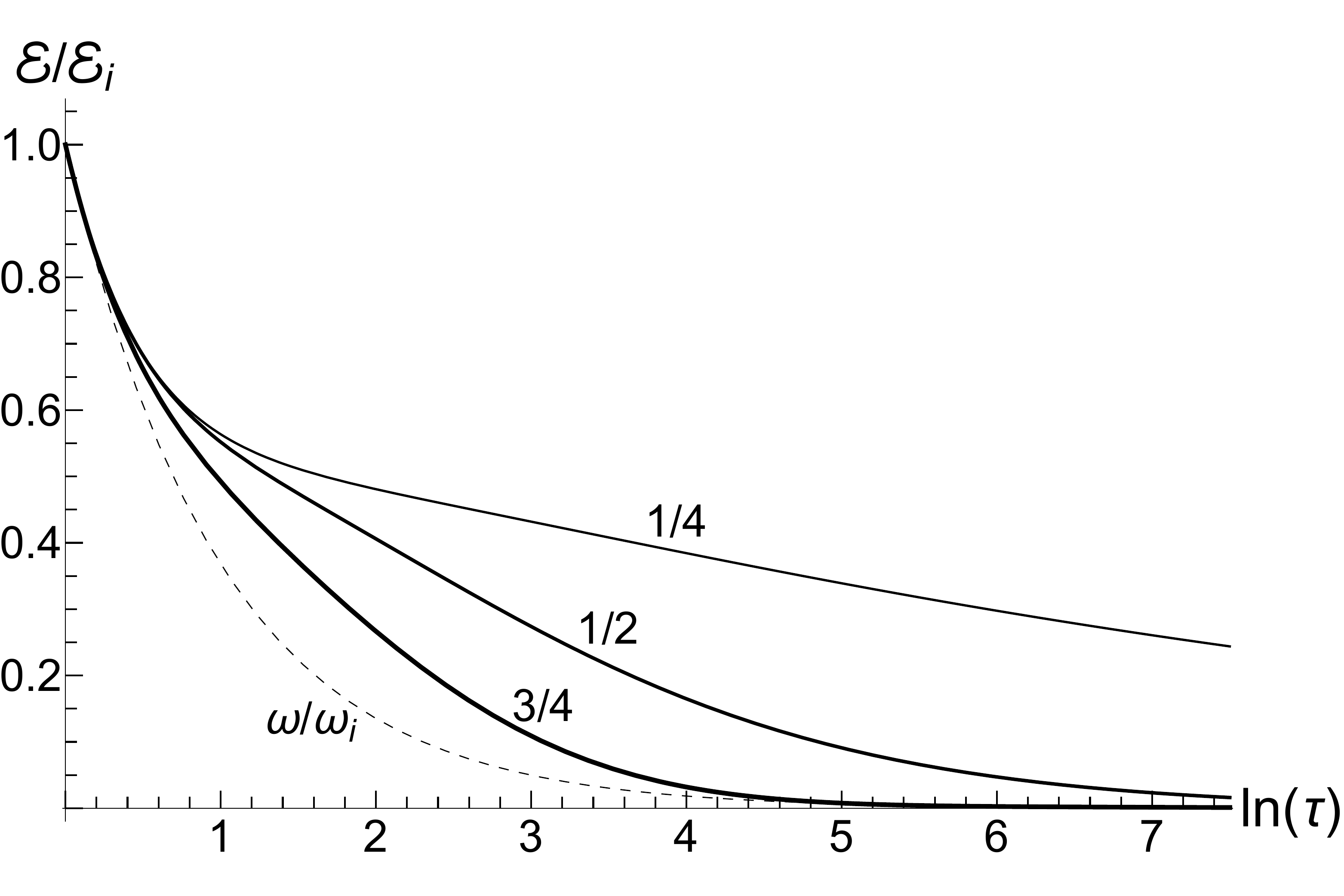}
\includegraphics[height=2.02truein,width=3.0truein,angle=0]{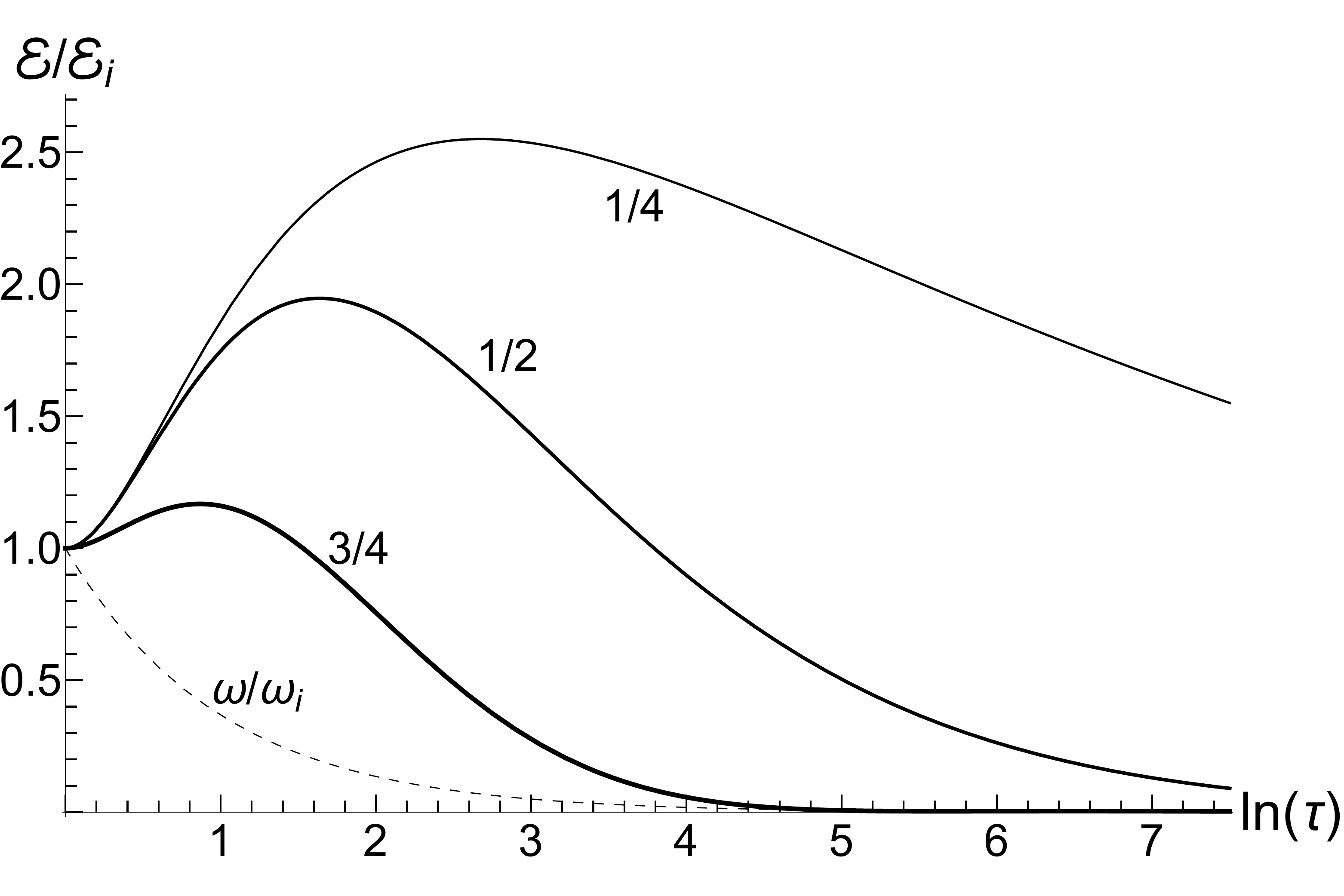}
\caption{\small The ratio ${\cal E}(\tau)/{\cal E}_i$ for different values of the evolution speed parameter $u=\omega_0 t_0$
(given nearby the curves)
for the inverse-linear decay of magnetic field (\ref{omoff}) with the asymmetry parameter $s=1$ (an isotropic trap).
Left: the low temperature case, $\rho=0$, $\Upsilon=1$. Right: the high temperature case, $\rho=1$, $\Upsilon=10$.
The trace lines show the ratio $\omega(\tau)/\omega_i$.
 }
\label{fig-E-logtau-var-u}
\end{figure}

In all the cases, the mean energy tends, finally, to the zero value, although the necessary effective time depends on the parameter $u$.
Paradoxically, this final effective time is much bigger in the ``initial fast evolution'' case ({\em almost\/} sudden jump, $u \ll 1$)
than in the ``slow evolution'' case (almost adiabatic, $u \gg 1$).
Examples of the evolution are shown in Figures \ref{fig-E-1tau} and \ref{fig-E-logtau-var-u}.
It is impressive that the mean energy is still very far from the asymptotic zero value even when the frequency
is 100 times smaller than the initial value (when $\ln\tau \approx 4.6$), if $u \le 1/2$.
Also, no proportionality between ${\cal E}(t)$ and $\omega(t)$ is observed if $s_0\Upsilon \gg 1$, even if $u >1/2$.
\begin{figure}[htb]
\includegraphics[height=2.02truein,width=3.0truein,angle=0]{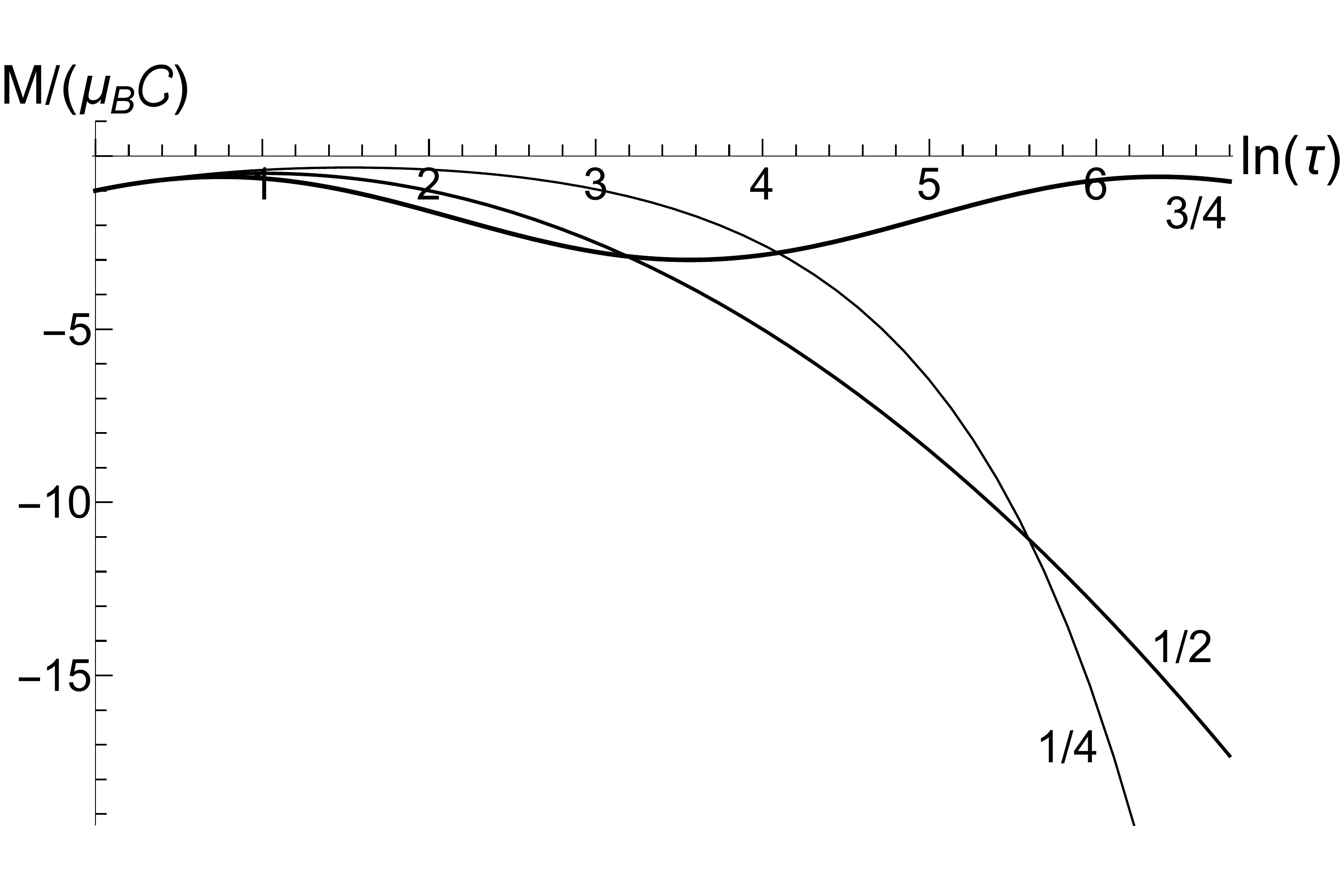}
\includegraphics[height=2.02truein,width=3.0truein,angle=0]{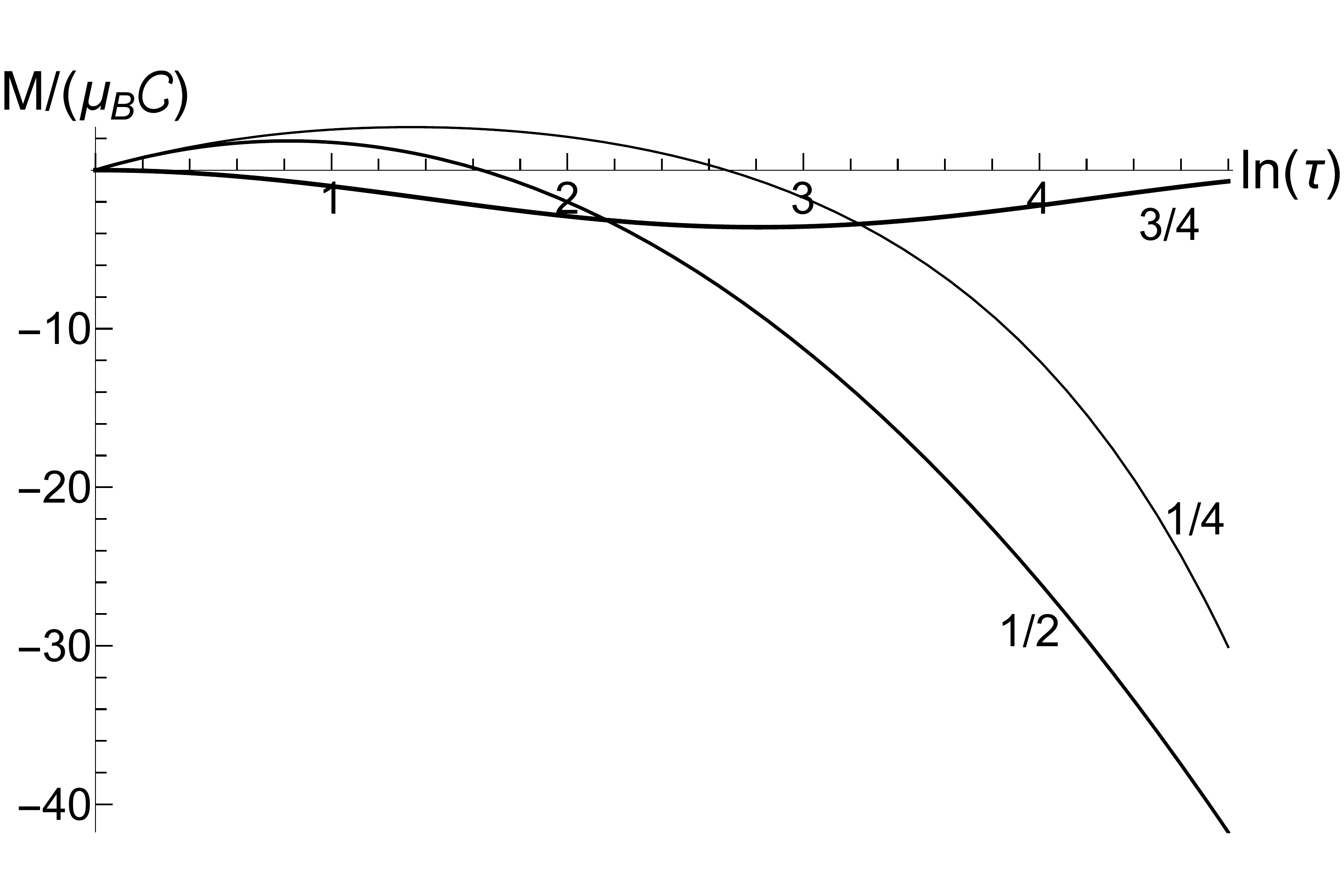}
\caption{\small The mean magnetic moment ${\cal M}(\tau)$ for different values of the evolution speed parameter $u=\omega_0 t_0$
(given nearby the curves)
for the inverse-linear decay of magnetic field (\ref{omoff}) with the asymmetry parameter $s=1$ (an isotropic trap).
Left: the low temperature case, $\rho=0$, $\Upsilon=1$. Right: the high temperature case, $\rho=1$, $\Upsilon=10$.
 }
\label{fig-M-logtau-var-u}
\end{figure}

The mean magnetic moment equals
\be
{\cal M} = \mu_B{\cal C}[\rho\cos(2\nu) -1] -\,\frac{\mu_B{\cal C}}{4u^2-1}\left\{(1+s_0\Upsilon)\left[
\sin^2(\nu) -|r|\sin(2\nu)\right] +2\rho|r|\sin(2\nu)\right\}.
\ee
This formula gives corrections to the adiabatic equation (\ref{Madiab})
(which corresponds to $u \gg 1$), demonstrating again the absence of the adiabatic invariance
for the magnetic moment.

The mean magnetic moment oscillates with a logarithmically increasing frequency in the ``adiabatic'' case $u>1/2$, while
it increases unlimitedly if $u \le 1/2$. This behavior is shown in Figure \ref{fig-M-logtau-var-u}.
We see that neither adiabatic nor sudden jump approximations work in the whole time axis, although both approximations 
can have sense inside some limited time intervals.

\subsection{Inverse quadratic decrease of magnetic field}
\label{sec-om2t}

It is interesting that there exists the function $\omega(t)$ for which the adiabatic form of solution (\ref{vep-ad})
is {\em exact}. To find it, one has to solve the equation following from formula (\ref{adterms}):
$2\omega \ddot\omega = 3\dot\omega^2$. Using the standard technique, one can transform it to the linear equation 
$dy/d\omega = 3y/\omega$ with respect to function $y = \dot\omega^2$. 
Finally, we arrive at the following function (assuming that $\omega=\omega_0 =const$ for $t\le 0$):
\be
\omega(t) = \left\{
\begin{array}{ll}
\omega_{0}, & \tau\le 1\\
 \omega_0/\tau^2, & \tau\ge 1
\end{array}
\right.,
\qquad \tau = 1 +t/t_0, \quad u= \omega_0 t_0.
\label{omoff2}
\ee
Analytic solutions to Equation (\ref{eqvep}) with this function have the form $\tau\exp(\pm iu/\tau)$ (see also, e,g., Ref. \cite{Eliezer76}).

Now, the adiabatic parameters depend on time: $|\dot\omega|/\omega^2 = 2\tau/u$, $|\ddot\omega|/\omega^3 = 6(\tau/u)^2$.
Note that $\tau/u = (t+t_0)/(\omega_0 t_0^2)$. The necessary condition for the adiabatic approximation is $u \gg 1$. However,
even under this condition, the adiabatic approximation is expected to fail asymptotically, when $t \gg \omega_0^{-1} u^2$.
 On the other hand, one can expect that the sudden jump
approximation can be quite good for $u\ll 1$ and any value of $\tau$.
But what happens in reality?

One can verify that function $\vep(t)$
satisfying the initial conditions (\ref{invep}) is the following superposition of functions 
$\tau\exp(\pm iu/\tau)$ at $t \ge 0$:
\be
\vep(t) = \frac{\tau}{u\sqrt{\omega_0}}\left[u \exp(i\vf) -\sin(\vf)\right] =
[\omega(t)]^{-1/2}\left[ \exp(i\vf) -\sin(\vf)/u\right], 
\label{vept-2}
\ee
\be
 \vf = u(1 -1/\tau)= \frac{u\omega_0 t}{u + \omega_0 t} \equiv \int_0^t \omega(x)dx.
\ee
The time derivative equals
\beqnn
\dot\vep(t) &=& \frac{\sqrt{\omega_0}}{u^2\tau}\left[u (\tau + iu)\exp(i\vf) -\tau\sin(\vf) - u\cos(\vf)\right]
\\
&=& [\omega(t)]^{1/2}\left[i \exp(i\vf)(1-i\tau/u) -(\tau\sin\vf +u\cos\vf)/u^2\right].
\eeqnn
This formula clearly shows that the condition $u \gg 1$ is not sufficient for the validity of the adiabatic approximation: an additional
condition $\tau \ll u$ must be fulfilled.
Other useful relations are
\be
F_{+}(t) \equiv \omega(t)\vep(t) + i\dot\vep(t) 
 = [\omega(t)]^{1/2} \left\{ (i\tau/u)\exp(i\vf) -\left[ (u + i\tau)\sin\vf +iu\cos\vf\right]/u^2 \right\},
\label{F+t-2} 
\ee
\be
F_{-}(t) \equiv \omega(t)\vep(t) - i\dot\vep(t) 
= [\omega(t)]^{1/2} \left\{ 2\exp(i\vf)[1-i\tau/(2u)] +\left[ (i\tau -u)\sin\vf +iu\cos\vf\right]/u^2 \right\}.
\label{F-t-2} 
\ee
The limit values at $\tau =\infty$,
\be
F_{\pm}(\infty) = \pm i\sqrt{\omega_0}\left[  u\exp(iu) - \sin(u) \right]/u^2, 
\ee
yield the following nonzero asymptotic value of the mean energy, according to Equation (\ref{Eqfin}):
\be
{\cal E}(\infty) = \frac{{\cal E}_i}{4u^4} \left\{ \left[ u^2 + \sin^2(u) - u\sin(2u)\right](1 + s_0^{-1}\Upsilon)
-2\rho \left[ u^2 \cos(2u) + \sin^2(u) - u\sin(2u)\right]\right\}.
\label{Einf-t-2}
\ee
In the limit $u \to 0$, Equation (\ref{Einf-t-2}) goes to the sudden jump approximation formula (\ref{EfEi-sudden0}),
up to terms  of the order of $u^2$.
On the other hand, the Taylor expansions of functions (\ref{F+t-2}) and (\ref{F-t-2}) for $u\ll 1$,
\[
F_{+} = -\sqrt{\omega_0}\left[ 1 -\tau^{-2}  + i(u/3)\left( 1 - \tau^{-1} \right)^3 +{\cal O}(u^2)\right], 
\]
\[
F_{-} =\sqrt{\omega_0}\left[ 1 +\tau^{-2}  + i(u/3)\left( 1 + 3\tau^{-1} -3\tau^{-2} -\tau^{-3}\right) +{\cal O}(u^2)\right],
\]
show that the accuracy of the sudden jump approximation is about 10\% already for $t=2t_0$ (or $\tau =3$). For 
$t=9t_0$ (or $\tau =10$), the accuracy is about 1\%.

\begin{figure}[htb]
\includegraphics[height=2.02truein,width=3.0truein,angle=0]{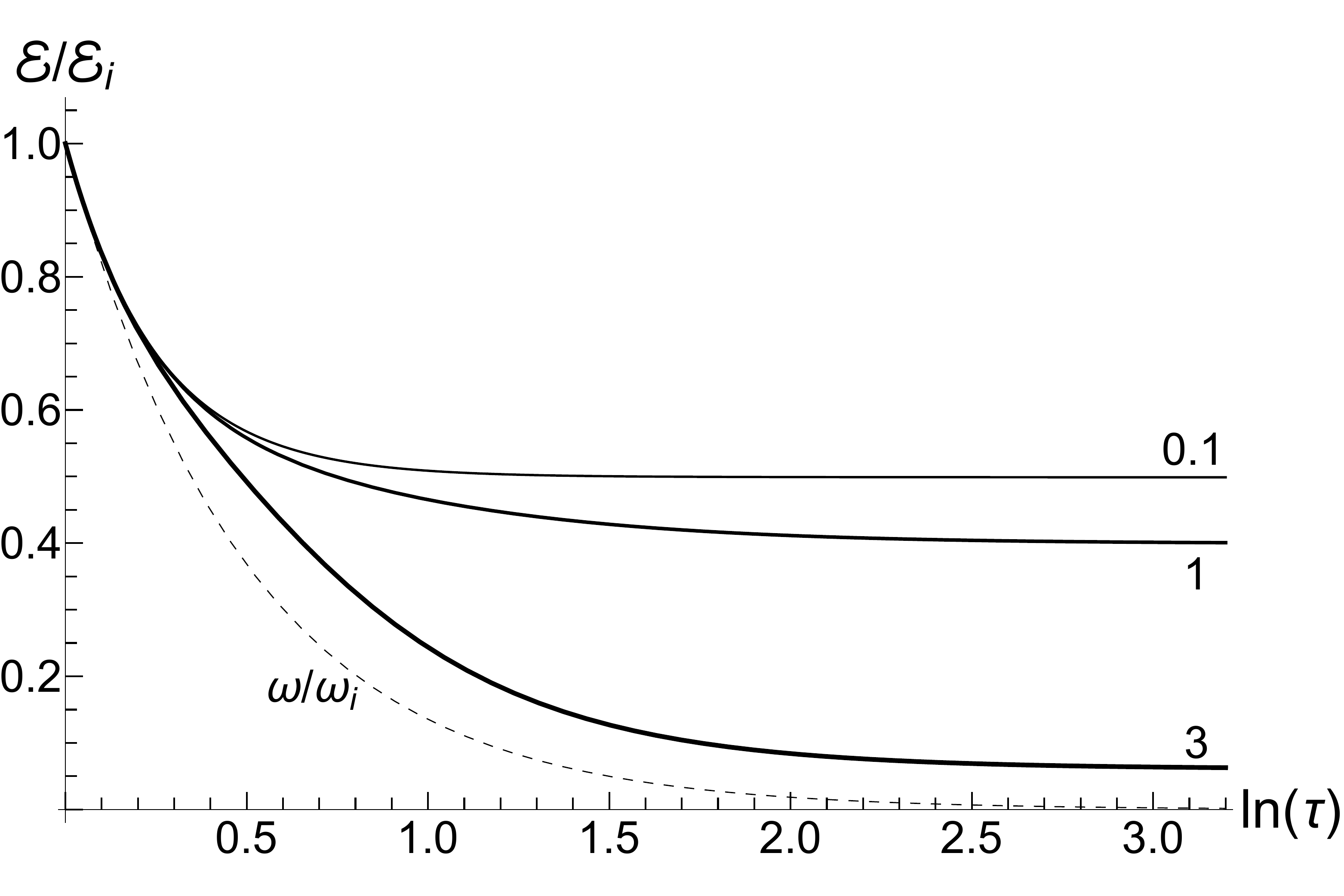}
\includegraphics[height=2.02truein,width=3.0truein,angle=0]{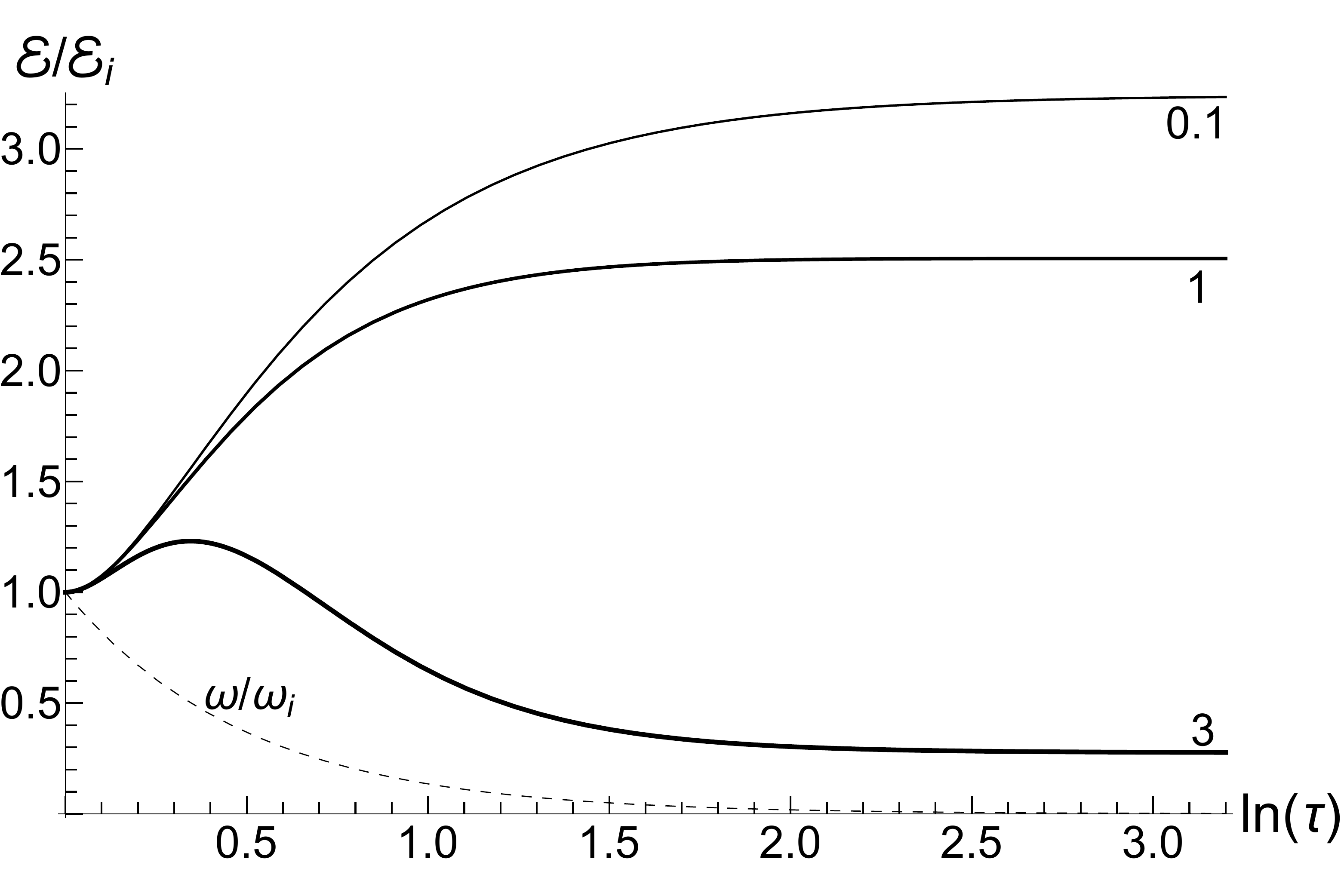}
\caption{\small The ratio ${\cal E}(\tau)/{\cal E}_i$ for different values of the evolution speed parameter $u=\omega_0 t_0$
(given nearby the curves)
for the inverse-quadratic decay of magnetic field (\ref{omoff2}) with the asymmetry parameter $s=1$ (an isotropic trap).
Left: the low temperature case, $\rho=0$, $\Upsilon=1$. Right: the high temperature case, $\rho=1$, $\Upsilon=10$.
The trace lines show the ratio $\omega(\tau)/\omega_i$.
 }
\label{fig-E-t2-var-u}
\end{figure}

\begin{figure}[htb]
\includegraphics[height=2.02truein,width=3.0truein,angle=0]{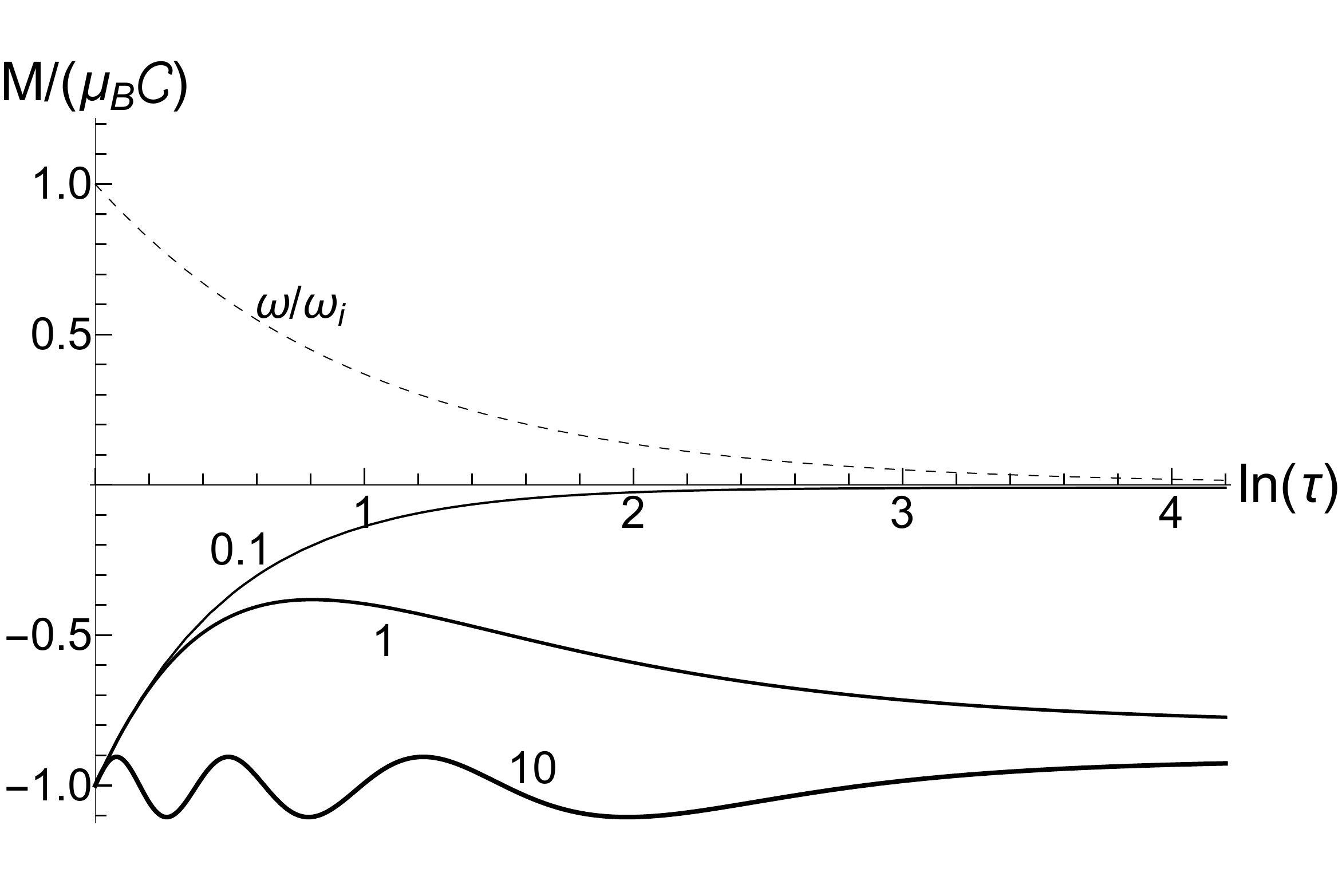}
\includegraphics[height=2.02truein,width=3.0truein,angle=0]{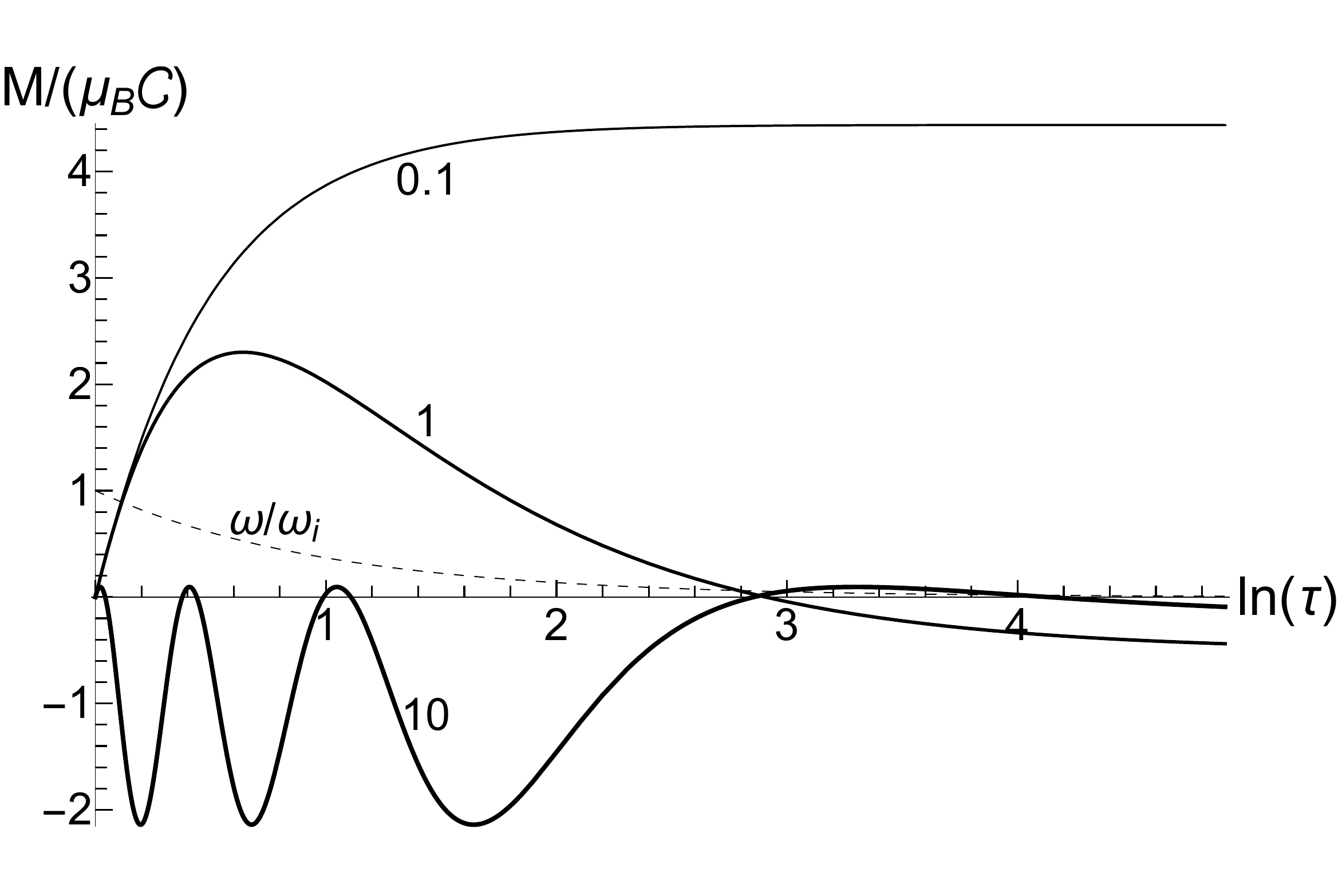}
\caption{\small The mean magnetic moment ${\cal M}(\tau)$ for different values of the evolution speed parameter $u=\omega_0 t_0$
(given nearby the curves)
for the inverse-quadratic decay of magnetic field (\ref{omoff2}) with the asymmetry parameter $s=1$ (an isotropic trap).
Left: the low temperature case, $\rho=0$, $\Upsilon=1$. Right: the high temperature case, $\rho=1$, $\Upsilon=10$.
The trace lines show the ratio $\omega(\tau)/\omega_i$.
 }
\label{fig-M-t2-var-u}
\end{figure}

Formula (\ref{meanmag}) for the mean magnetic moment assumes the  form
\be
{\cal M} = -\mu_B {\cal C}\left\{ 1 - \rho\cos(2\vf) + \frac12(1+s_0\Upsilon -2\rho)
\left[ \frac{\sin^2(\vf)}{u^2} - \frac{\sin(2\vf)}{u}\right]\right\}.
\ee
If $u\ll 1$ (the sudden jump regime), then,
\be
{\cal M}(\tau) = \frac12\mu_B {\cal C}\left[ s_0\Upsilon -1 - (1+s_0\Upsilon -2\rho)/\tau^2 + {\cal O}(u^2) \right],
\ee
in accordance with Equation (\ref{Mjump-om0}).
On the other hand, if $u \gg 1$ (the adiabatic regime), the asymptotic value 
\be
{\cal M}(\infty) = -\mu_B {\cal C}\left[ 1 - \rho\cos(2u) + {\cal O}(u^{-1}) \right],
\ee
appears to be very sensitive to the concrete value of parameter $u$.
In this case, the mean magnetic moment is preserved for the zero-temperature initial state ($\rho=0$), 
while it can be much higher than the initial
one for  high-temperature initial states ($\rho \approx 1$), for almost all values of $u$.
Figures \ref{fig-E-t2-var-u} and \ref{fig-M-t2-var-u} show functions ${\cal E}(\tau)/{\cal E}_i$ and ${\cal M}(\tau)$
for different values of parameter $u$ in the isotropic traps with $s=1$.

\subsection{Exponential-like decrease of frequency to a final value}

\begin{figure}[htb]
\includegraphics[height=2.32truein,width=3.0truein,angle=0]{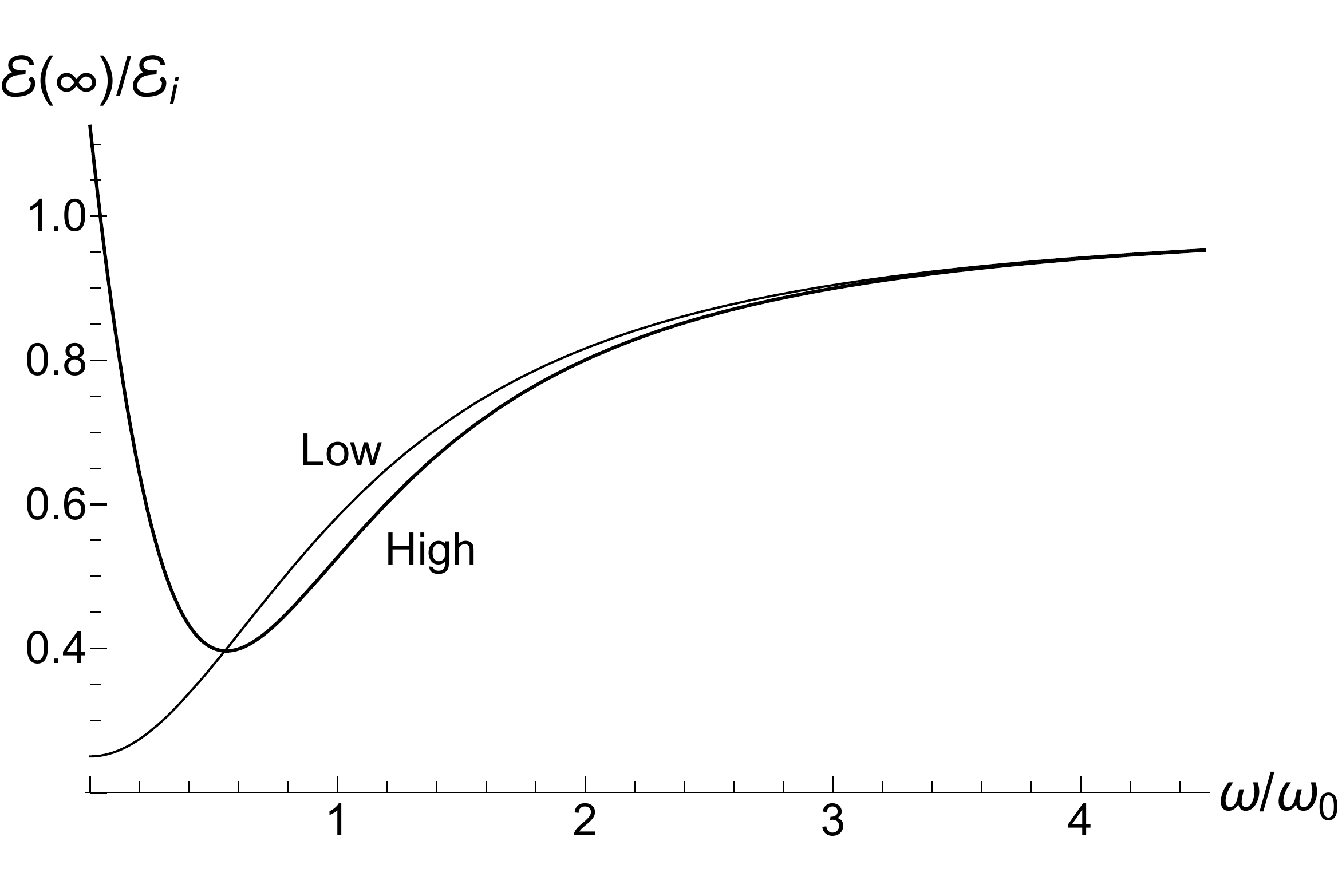}
\includegraphics[height=2.32truein,width=3.0truein,angle=0]{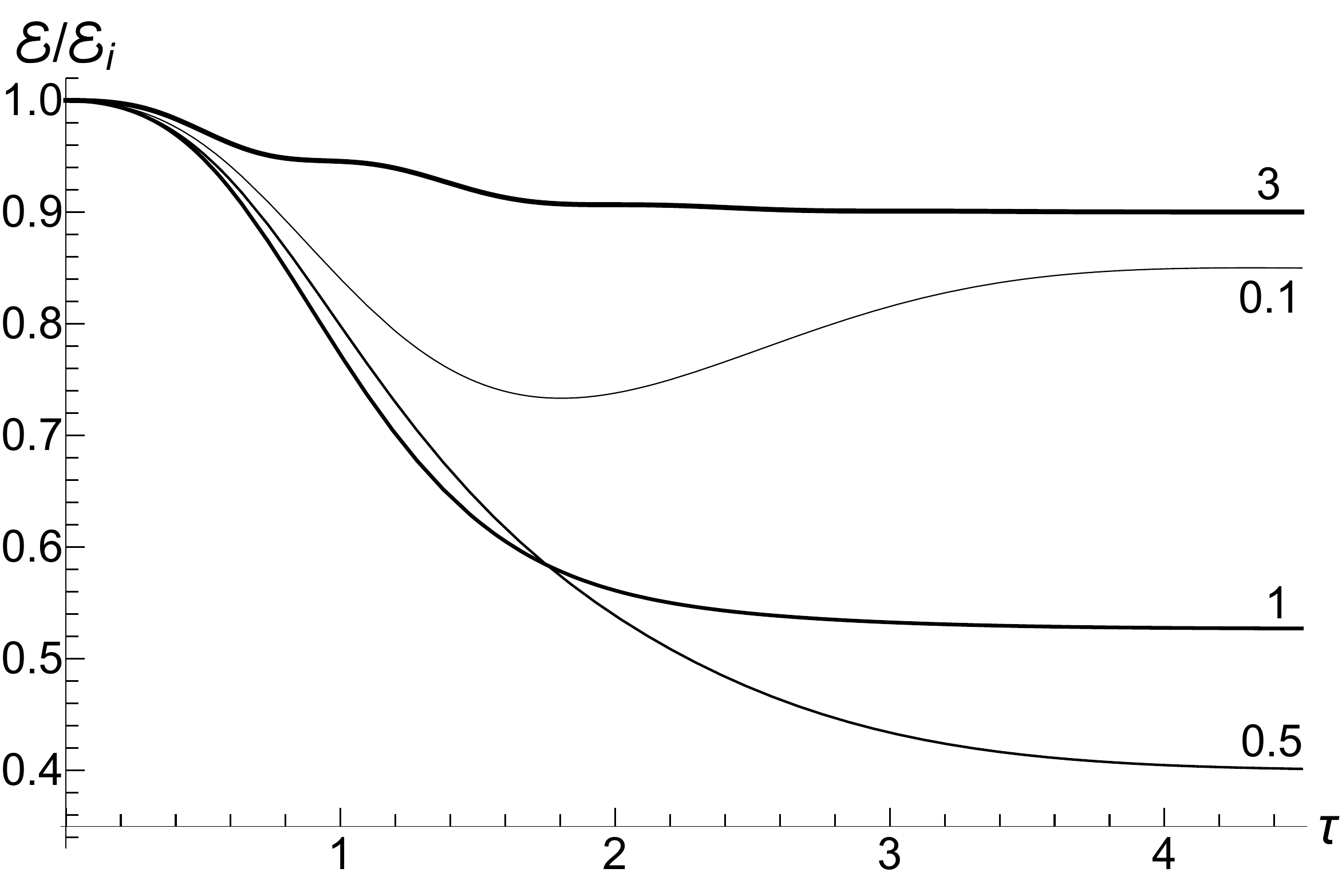}
\caption{\small Left: The asymptotic ratio  ${\cal E}(\infty)/{\cal E}_i$ as function of the ratio $\omega/\omega_0$
for the exponential-like variation of magnetic field (\ref{om2Bag}) 
in the low-temperature ($\rho=0$, $\Upsilon=1$) and high temperature ($\rho=1$, $\Upsilon=10$) cases.
Right: The time-dependent ratio  ${\cal E}(\tau)/{\cal E}_i$ in the high-temperature case ($\rho=1$, $\Upsilon=10$),
for different values of the ratio $\omega/\omega_0$ (shown nearby the related curves). The asymmetry parameter $s=1$
(the isotropic trap).
 }
\label{fig-Einf-cosh}
\end{figure}

Equation (\ref{eqvep}) can be solved in terms of trigonometric and hyperbolic functions for \cite{Bagrov}
\be
\omega^2(t) =\omega^2 + \frac{2\omega_0^2}{\cosh^2(\omega_0 t)}. 
\label{om2Bag}
\ee
This example is interesting, because it describes the evolution which is neither adiabatic nor fast.
In this case, we have $\omega_i^2 = \omega^2 + 2\omega_0^2$ and $\omega_f =\omega$.
It is convenient to introduce the ``intermediate'' frequency $\omega_1^2 = \omega^2 + \omega_0^2$.
Then, the solution satisfying the initial conditions (\ref{invep}) at $t=0$ has the form
\be
\vep(t) = D_{+} e^{i\omega t}\left[1 +i\frac{\omega_0}{\omega}\tanh(\tau)\right]
+ D_{-} e^{-i\omega t}\left[1 -i\frac{\omega_0}{\omega}\tanh(\tau)\right],
\quad D_{\pm} = \frac{\omega_1^2  \pm\omega \omega_{i}}{2\omega_1^2 \sqrt{\omega_i}}, \quad
\tau= \omega_0 t.
\label{vep-tanh}
\ee
This function becomes very close to the asymptotic form (\ref{uvsol}) already for $\tau >4$ (since $\tanh(4) \approx 0.9993$),
unless the ratio $\omega/\omega_0$ is extremely small. The coefficients $u_{\pm}$ in this case are given by the formula
$u_{\pm} = \sqrt{\omega}D_{\pm}\left(1 \pm i\omega_0/\omega\right)$.
Using Equation (\ref{EfEi-u-}), we obtain the asymptotic mean energy
\be
{\cal E}(\infty)= \frac{{\cal E}_i}{4\omega_1^2 \omega_i^2} \left[\left(\omega_1^2  +\omega \omega_{i}\right)^2 
+ s_0\Upsilon \left(\omega_1^2  -\omega \omega_{i}\right)^2 -2\rho\omega_0^4\right].
\label{Einf-cosh}
\ee
If $\omega \gg \omega_0$, then the final frequency is very close to the initial one, so ${\cal E}(\infty) \approx {\cal E}_i$
for any values of parameters $\Upsilon$ and $\rho$. 
On the other hand, if $\omega=0$, then,
\be
{\cal E}(\infty)/{\cal E}_i=(1+s_0\Upsilon-2\rho)/8.
\label{Einf-circ-exp}
\ee
The minimum $1/4$ of this ratio is achieved for the initial zero temperature and isotropic trap, while it can be quite high in the high temperature case. The ratio ${\cal E}(\infty)/{\cal E}_i$ is monotonously increasing function of the final frequency $\omega$
in the low-temperature case ($\Upsilon=1$). However, it shows a more interesting behavior as function of the ratio $\omega/\omega_0$
in the high-temperature case ($\Upsilon \gg 1$): see Figure \ref{fig-Einf-cosh}.
We do not bring here explicit formulas for the time-dependent function ${\cal E}(\tau)$, since they are rather cumbersome.

\subsubsection{A decrease to zero final frequency}

Taking the limit $\omega \to 0$ in Equation (\ref{vep-tanh}), we obtain the solution
\be
\vep(t)= \omega_i^{-1/2}\left[ 1 -\tau \tanh(\tau) + i \sqrt{2} \tanh(\tau)\right], \qquad \dot{\vep}(t)=\frac{\omega_0\left[i\sqrt{2}-\tau-\sinh(\tau)\cosh(\tau)\right]}{\omega_i^{1/2}\cosh^2(\tau)},
\label{vep-tanh-om0}
\ee
with $\omega_i=\sqrt{2}\omega_0$ and $\tau = \omega_0 t$.
Hence,
\[
|F_\pm|^2=\frac{\omega_i\PG{C_0(\tau)+C_{\mp}(\tau)-2}}{2\cosh^4(\tau)}, \qquad \mbox{Re}(F_-F_+)=\frac{\omega_i\PG{C_0(\tau)-3\cosh^2(\tau)+2}}{2\cosh^4(\tau)},
\]
\[
C_0(\tau)=\cosh^4(\tau)-2\tau\cosh(\tau)\sinh(\tau)+\tau^2\PG{2\cosh^2(\tau)-1}, \quad C_{\pm}(\tau)=\cosh^2(\tau)\PG{5\pm4\cosh(\tau)}.
\]

The evolution of mean energy is given by the formula
\be
{\cal E}(\tau)/{\cal E}_i=\frac{C_{+} + C_0 -2 + s_0\Upsilon \PC{C_{-} + C_0 -2}
 - 2\rho\PG{2+C_0-3\cosh^2(\tau)}}{8\cosh^4(\tau)}.
\label{Etau-circ} 
\ee
The asymptotic value at $\tau\to\infty$ is given by Equation (\ref{Einf-circ-exp}).

Using Equation (\ref{meanmag}), we obtain the following expression for the mean magnetic moment: 
\be
{\cal M}(\tau)=\frac{-\mu_B{\cal C}}{2\cosh(\tau)}\PG{{\cal S}+(1-s_0\Upsilon)\cosh(\tau)-2{\cal S}\tau\tanh(\tau)+\PC{{\cal S}\tau^2+2{\cal S}+8\rho}\tanh^2(\tau)}, 
\label{Mtau-circ}
\ee
where ${\cal S}=1+s_0\Upsilon-2\rho$.
The asymptotic value at $\tau\to\infty$ is always non-negative:
\be
{\cal M}(\infty)={\mu_B{\cal C}\PC{s_0\Upsilon-1}}/{2}.
\label{M-inf-circ-exp}
\ee
It equals zero only for the zero temperature initial state in the isotropic trap.

\section{Exact solutions in terms of the confluent hypergeometric and cylindrical functions}
\label{sec-hyper}

In three examples of the preceding section, the sign of frequency (or magnetic field) could not change. It appears
that the most interesting behavior can be observed in the situations when the magnetic field changes its sign. 
In this section we consider an example of exponentially varying frequency on the time semi-axis in the following form:
\be
\omega(t) = \left\{
\begin{array}{ll}
\omega_{i}, & t\le 0\\
\omega_{f} +\left(\omega_{i} - \omega_{f}\right) \exp(-\kappa t), & t\ge 0
\end{array}
\right. .
\label{omexp}
\ee
Solutions to Equation (\ref{eqvep}) with function (\ref{omexp})
were considered in \cite{Aga80}. They can be expressed in terms of the confluent hypergeometric function.
This can be achieved by means of the transformation
\[
\vep = x^{(c-1)/2}\exp(-x/2)y(x), \quad x=x_0 \exp(-\kappa t).
\]
Then, Equation (\ref{eqvep}) assumes the
canonical form of the equation for the confluent hypergeometric function,
\be
x d^2 y/dx^2 +(c-x)dy/dx -ay =0,
\label{confeq}
\ee
with the following set of parameters:
\be
x_0 = 2i\mu, \quad a=1/2, \quad c= 1-2i\gamma,
\quad
\mu = \left(\omega_{i} -\omega_{f} \right)/\kappa, \quad \gamma =\omega_{f}/\kappa.
\label{acmugam}
\ee
Choosing the solution to Equation (\ref{confeq}) which is regular at $x=0$ \cite{BE},  
\be
\Phi(a;c;x) = \sum_{n=0}^{\infty} \frac{a(a+1)\ldots(a+n-1) x^n}{c(c+1)\ldots(c+n-1) n!}, 
\label{def-Phi}
\ee
we obtain the time dependent solution to Equation (\ref{eqvep}) which is regular at $t=\infty$:
\be
\vep_1(t) = \omega_i^{-1/2}\exp[i\phi(t)] \frac{\Phi[1/2; 1-2i\gamma; 2i\mu\xi(t)]}{\Phi(1/2; 1-2i\gamma; 2i\mu)},
\label{vep1}
\ee
\[
\xi(t) = \exp(-\kappa t), \quad \phi(t) =\omega_{f} t +\mu[1-\xi(t)].
\]
However, although function (\ref{vep1}) satisfies the first initial condition (\ref{invep}), $\vep_1(0) = \omega_{i}^{-1/2}$,
it does not satisfy the second condition, due to the nonzero time derivative of function $\Phi[1/2; 1-2i\gamma; 2i\mu\xi(t)]$.
Therefore, the correct complex solution to Equation (\ref{eqvep}), satisfying (\ref{invep}), 
should be constructed as a linear combination of functions $\vep_1(t)$ and $\vep_1^*(t)$:
\be
\vep(t) = D_{+}\vep_1(t) +D_{-} \vep_1^*(t),
\label{vepD1D2}
\ee
\be
D_{+} = \frac{1-\lambda^*/2}{1 - \mbox{Re}\lambda}, \quad D_{-} = -\,\frac{\lambda/2}{1 - \mbox{Re}\lambda} = 1 - D_{+},
\qquad
\lambda = \frac{2\left(\omega_{i} -\omega_{f} \right)\Phi^{\prime}(1/2; 1-2i\gamma; 2i\mu)}
{\omega_{i}\, \Phi(1/2; 1-2i\gamma; 2i\mu)}.
\label{D+D-}
\ee
Here $\Phi^{\prime}$ is the derivative of function $\Phi(a;c;x)$ with respect to its argument $x$.
It can be written  as \cite{BE}
\[
\Phi^{\prime}(a; c; x) = (a/c)\Phi(a+1; c+1; x).
\]
Hence, parameter $\lambda$ can be also written as
\be
\lambda = \frac{\left(\omega_{i} -\omega_{f} \right)\Phi(3/2; 2-2i\gamma; 2i\mu)}
{\omega_{i}(1-2i\gamma) \Phi(1/2; 1-2i\gamma; 2i\mu)}.
\label{lambda}
\ee

When $t \to \infty$, then $\xi \to 0$ and $\Phi(a;c;2i\mu\xi) \to 1$. Therefore, we have asymptotically
\[ 
\vep_1(t) = \frac{\exp\left[i\left(\omega_{f} t +\mu\right)\right]}{\omega_{i}^{1/2} \Phi(1/2; 1-2i\gamma; 2i\mu)}, \quad
\dot\vep_1(t) = i\omega_{f} \vep_1(t).
\]
This means that 
\be
u_{+} =  \sqrt{\frac{|\omega_f|}{\omega_i}} \times \left\{
\begin{array}{cc}
\displaystyle{
\frac{ D_{+}\exp(i\mu)}{\Phi(1/2; 1-2i\gamma; 2i\mu)} }, & \omega_f >0
\\[3mm] 
\displaystyle{
\frac{D_{-}\exp(-i\mu)}{[\Phi(1/2; 1-2i\gamma; 2i\mu)]^*}}, & \omega_f <0
\end{array} \right. ,
\label{u+-D+-a}
\ee
\be
u_{-} =  \sqrt{\frac{|\omega_f|}{\omega_i}} \times \left\{
\begin{array}{cc}
\displaystyle{
\frac{D_{-}\exp(-i\mu)}{[\Phi(1/2; 1-2i\gamma; 2i\mu)]^*} }, & \omega_f >0
\\[3mm]
\displaystyle{
\frac{ D_{+}\exp(i\mu)}{\Phi(1/2; 1-2i\gamma; 2i\mu)} }, & \omega_f <0
\end{array} \right. .
\label{u+-D+-}
\ee
Then, the identity (\ref{uvcond}) takes the form (for positive as well as for negative values of $\omega_f$)
\be
(\omega_{f}/\omega_{i})\left[(1 - \mbox{Re}\lambda)|\Phi(1/2; 1-2i\gamma; 2i\mu)|^2\right]^{-1} =1.
\label{ident}
\ee
Hence, the signs of $\omega_f$ and $1 - \mbox{Re}\lambda$ coincide.
Other consequences of (\ref{ident}) are the formulas
\be
|u_{-}|^2 = \frac{|\lambda|^2 }{4[1-\mbox{Re}(\lambda)]}, \;
\quad \omega_f >0;
\qquad
|u_{+}|^2 = \frac{|\lambda|^2 }{4[\mbox{Re}(\lambda) -1]}, \;
\quad \omega_f <0.
\label{upm2}
\ee
\be
u_{+} u_{-} = \frac{\left(|\lambda|^2 -2 \lambda\right) \omega_f}{4[1-\mbox{Re}(\lambda)]|\omega_f|}. 
\label{u+u-}
\ee

\subsection{Mean energy}
\label{E-semiaxis}

Equations (\ref{EfEi-u-}), (\ref{upm2}) and (\ref{u+u-}) yield the following ratio between the final and initial mean energies:
\be
\frac{{\cal E}_f}{{\cal E}_i} = 
\frac{\omega_f\left[(1 + s_0 \Upsilon -2\rho)|\lambda|^2 +4\rho \mbox{Re}(\lambda)\right]}
{4\omega_i[1-\mbox{Re}(\lambda)]}
+ \frac{\omega_f}{\omega_i} \times
\left\{
\begin{array}{cc}
1 , & \omega_f >0
\\
s_0\Upsilon, & \omega_f <0
\end{array} \right. .
\label{EfEi-gen}
\ee
In the case of initial zero temperature and isotropic trap ($\rho=0$ and $s_0\Upsilon=1$), we have
\be
\frac{{\cal E}_f}{{\cal E}_i} = \frac{\omega_{f}}{\omega_i} \left( 1 +\frac{|\lambda|^2/2 }{ 1 - \mbox{Re}\lambda } \right).
\label{Omlam1}
\ee
\begin{figure}[htb]
\begin{center}
\includegraphics[height=2.32truein,width=3.0truein,angle=0]{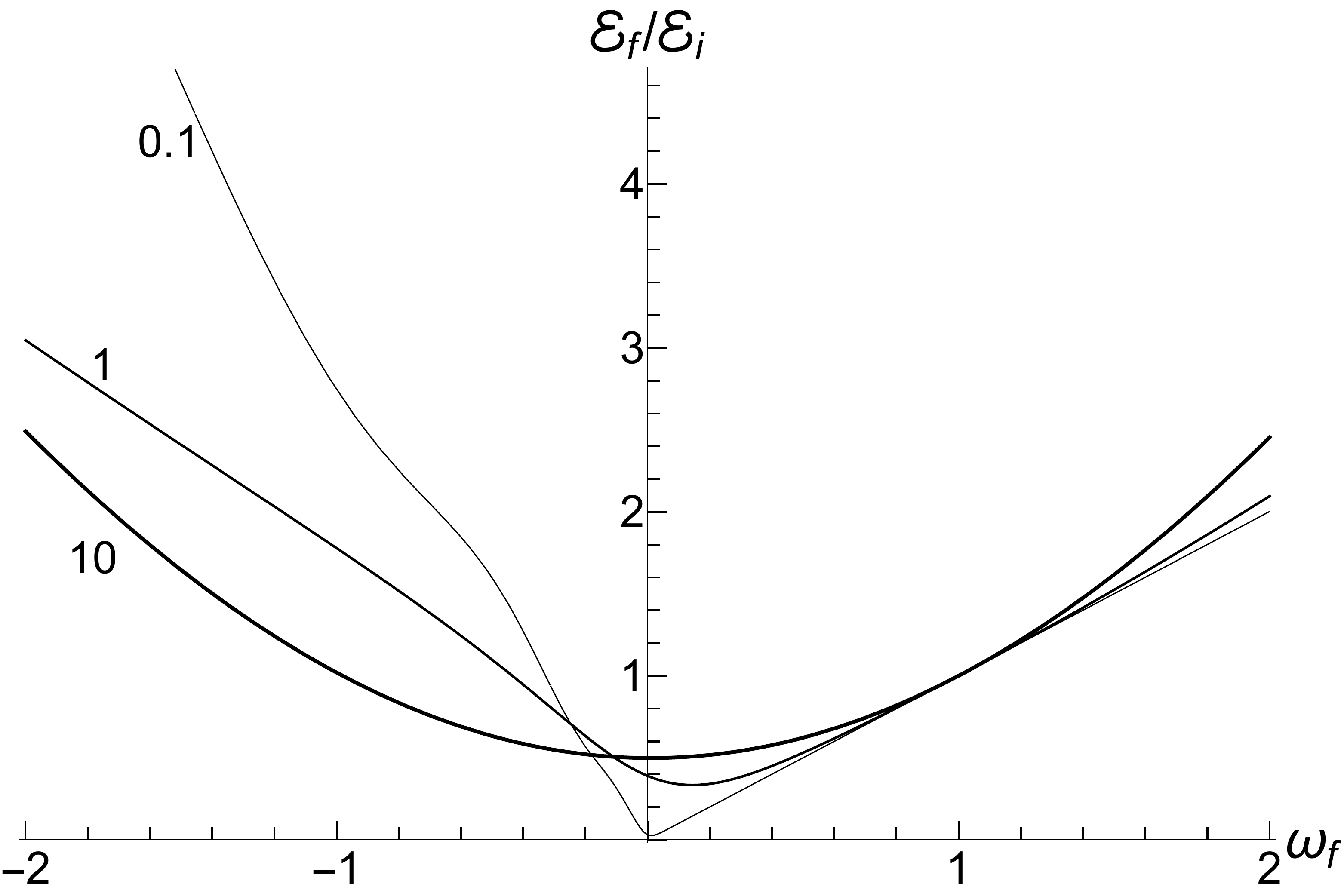}
\includegraphics[height=2.32truein,width=3.0truein,angle=0]{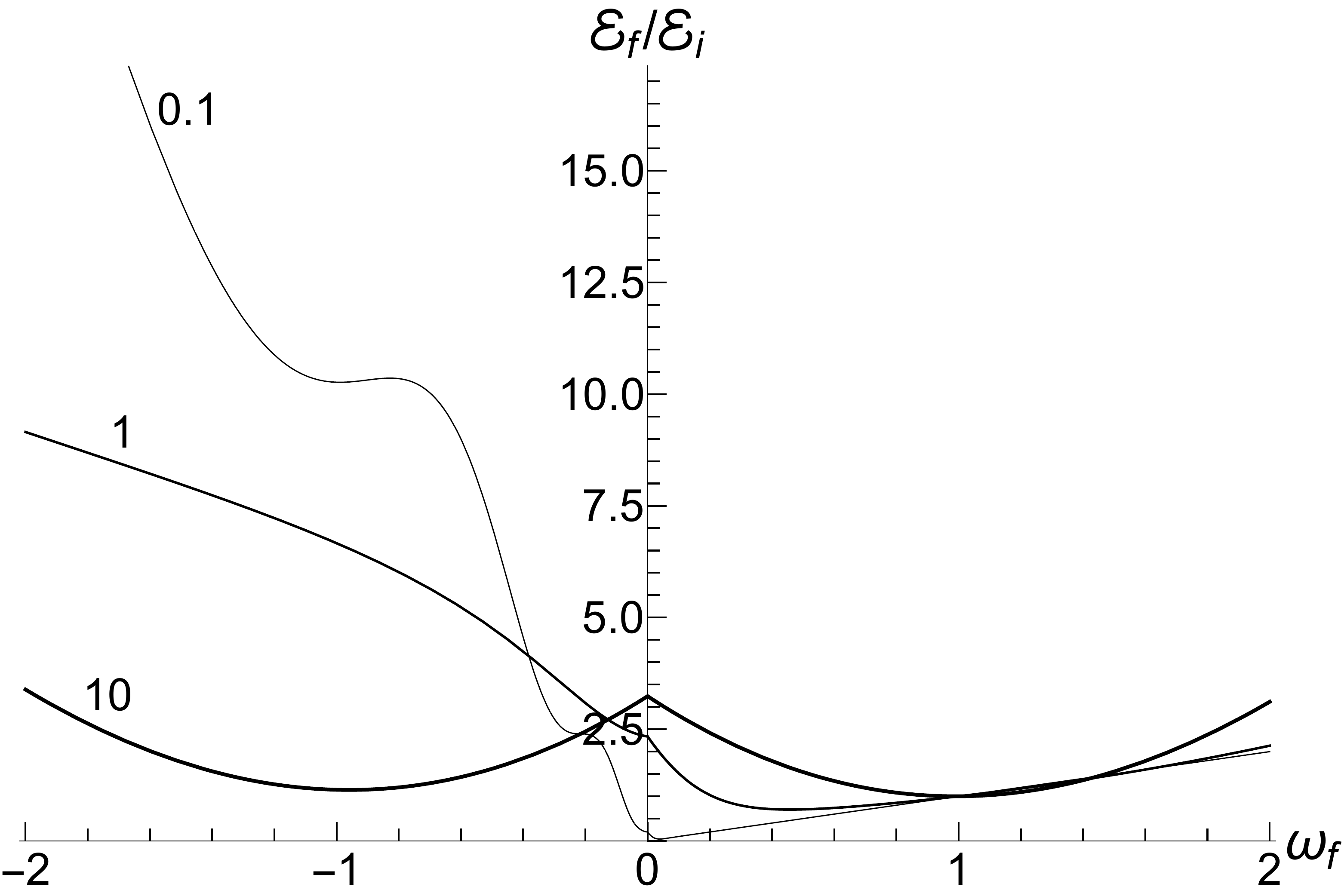}
\end{center}
\caption{\small The ratio ${\cal E}_f/{\cal E}_i$ versus the final frequency $\omega_{f}$ for different values of parameter $\kappa$
(shown nearby the respective lines) in the case of exponentially varying frequency on the time semi-axis (\ref{omexp}).
 The initial frequency is $\omega_{i}=1$.  Left: $\rho=0$, $s_0\Upsilon=1$.
Right: $\rho=1$, $s_0\Upsilon=10$.
 }
\label{fig-EfEi-wf}
\end{figure}  
Figure \ref{fig-EfEi-wf} shows the ratio ${\cal E}_f/{\cal E}_i$ 
as function of ratio $\omega_{f}/\omega_{i}$ with several fixed values of parameter $\kappa$,  for the initial zero-temperature and high-temperatures states.
The accuracy of numerical calculations (performed with the aid of Mathematica and Mapple) was checked by the
fulfillment of identity (\ref{ident}). The case of $\kappa=10\omega_i$ corresponds to the sudden jump approximation discussed 
in Section \ref{sec-jump-gen}. One can see the symmetry with respect to the change of sign of the final frequency $\omega_f$, 
as well as the cusp at $\omega_f=0$ in the high-temperature regime.
However, the symmetry is broken for moderate values of $\kappa$,
and the striken asymmetry is observed for $\kappa \ll \omega_{i}$.
For example, the curve ${\cal E}_f(\omega_{f})$ is practically the straight line ${\cal E}_f = {\cal E}_i\omega_{f}/\omega_{i}$ 
for $\kappa / \omega_{i} =0.1$ and $\omega_{f} >0$ in the low-temperature regime. But if $\omega_{f} < 0$, we see the straight line 
${\cal E}_f = 3{\cal E}_i|\omega_{f}|/\omega_{i}$ for $|\omega_{f}| \ll \omega_{i}$. 
This asymmetry (including the ``strange'' coefficient $3$) is explained in Appendix \ref{sec-asymp}.

\begin{figure}[htb]
\includegraphics[height=2.32truein,width=3.0truein,angle=0]{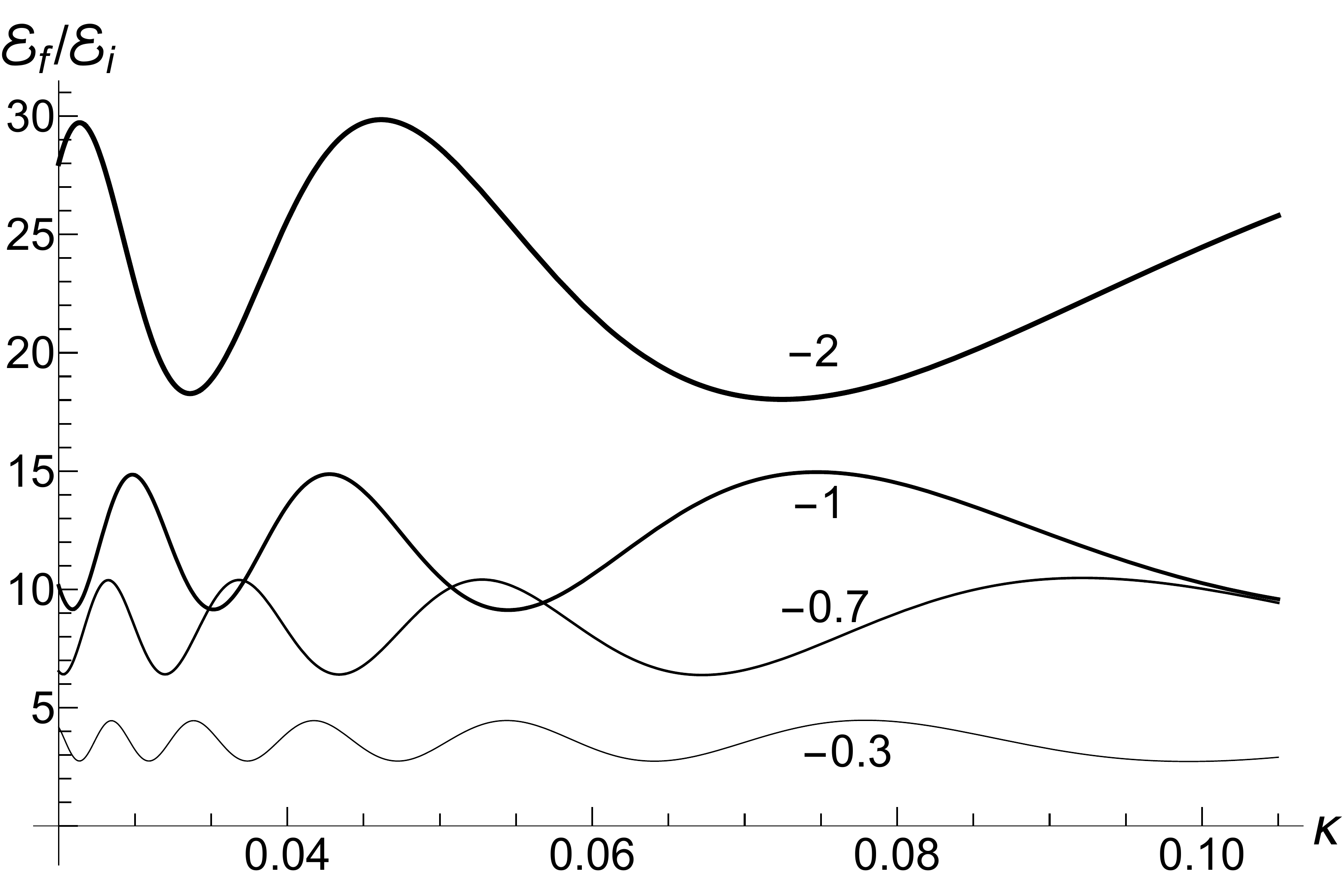}
\includegraphics[height=2.32truein,width=3.0truein,angle=0]{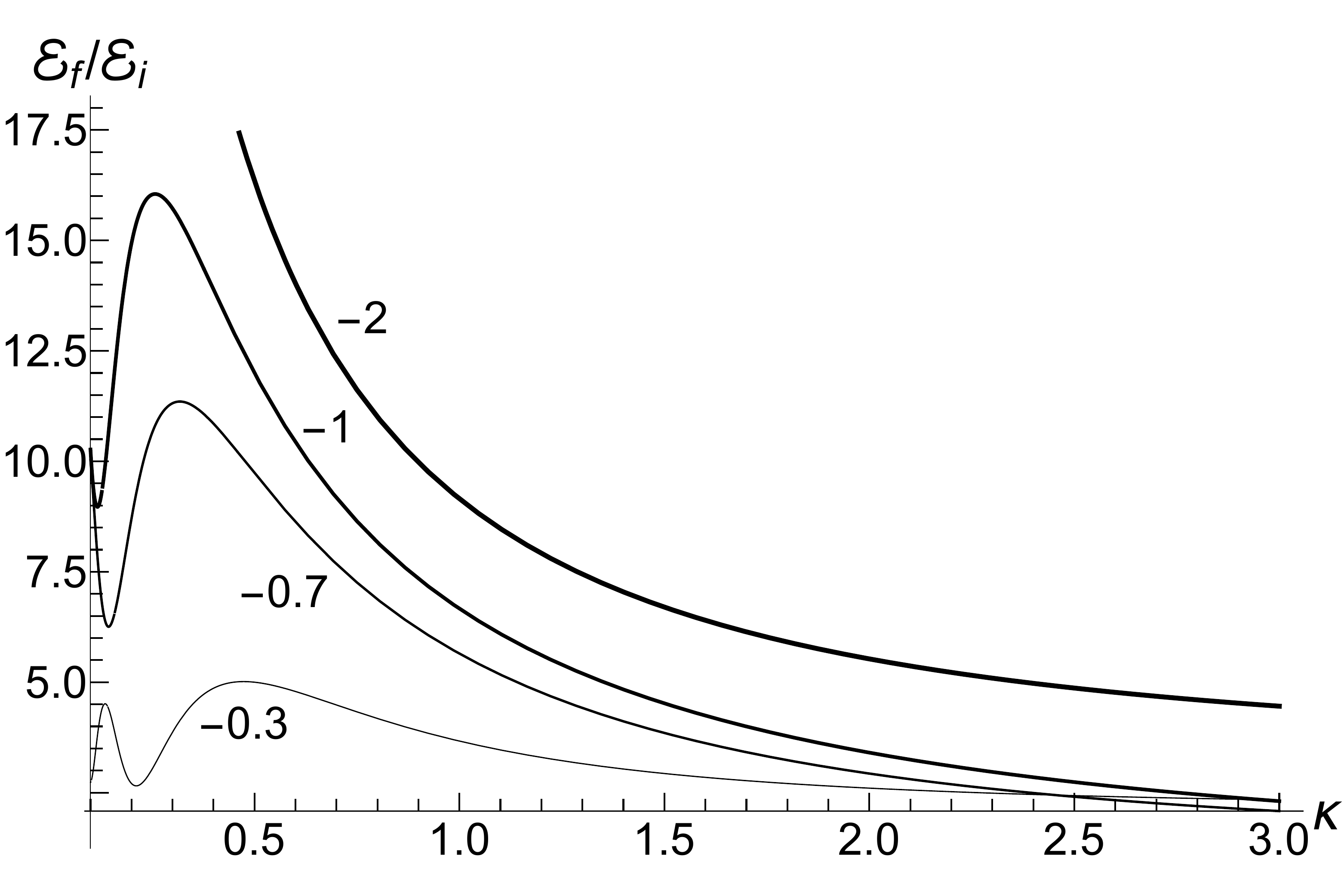}
\caption{\small The ratio ${\cal E}_f/{\cal E}_i$ versus parameter $\kappa$ for different negative values of the final frequency $\omega_{f}$
(shown nearby the respective lines) in the case of exponentially varying frequency on the time semi-axis (\ref{omexp}). 
The initial frequency is taken as $\omega_{i}=1$.
Other parameters are: $\rho=1$, $s_0\Upsilon=10$.
 }
\label{fig-EfEi-k1-10}
\end{figure}

Figure \ref{fig-EfEi-k1} shows the ratio ${\cal E}_f/{\cal E}_i$ 
as function of ratio $\kappa/\omega_{i}$ for positive values of the final frequency $\omega_f$.
The dependence is rather weak, except for the case of small values of $\omega_f$, when the final energy turns out to be much higher
than the initial one in the almost sudden-jump regime with $\kappa/\omega_{i} \gg 1$, especially in the high-temperature case.
For negative values of $\omega_f$, the ratio ${\cal E}_f/{\cal E}_i$ is shown in Figure \ref{fig-EfEi-k1-10} for the high-temperature case 
($\rho=1$, $s_0\Upsilon=10$). Plots in the low-temperature case look similar, only the vertical scale is diminished.

According to Figure \ref{fig-EfEi-k1}, the sudden jump approximation seems to be quite reasonable already for $\kappa> 5\omega_{i}$.
In principle, one can expect this approximation to be valid under the condition $\kappa \gg \omega_{i}$. 
Indeed, if $\kappa \gg \omega_{i,f}$, then coefficients $\mu$ and $\gamma$ are very small. 
Putting $\gamma=\mu=0$ in the arguments of hypergeometric functions in Equation (\ref{lambda}), one 
obtains $\lambda = (\omega_i -\omega_f)/\omega_i$. Then, it is easy to verify that formulas (\ref{u+-D+-a}) and (\ref{u+-D+-}) 
coincide with
 the instantaneous jump expressions (\ref{upmjump}) for the coefficients $u_{\pm}$. 
More precise estimations of the accuracy of this approximation are given in Appendix \ref{sec-ap-jump}. 
\begin{figure}[htb]
\includegraphics[height=2.32truein,width=3.0truein,angle=0]{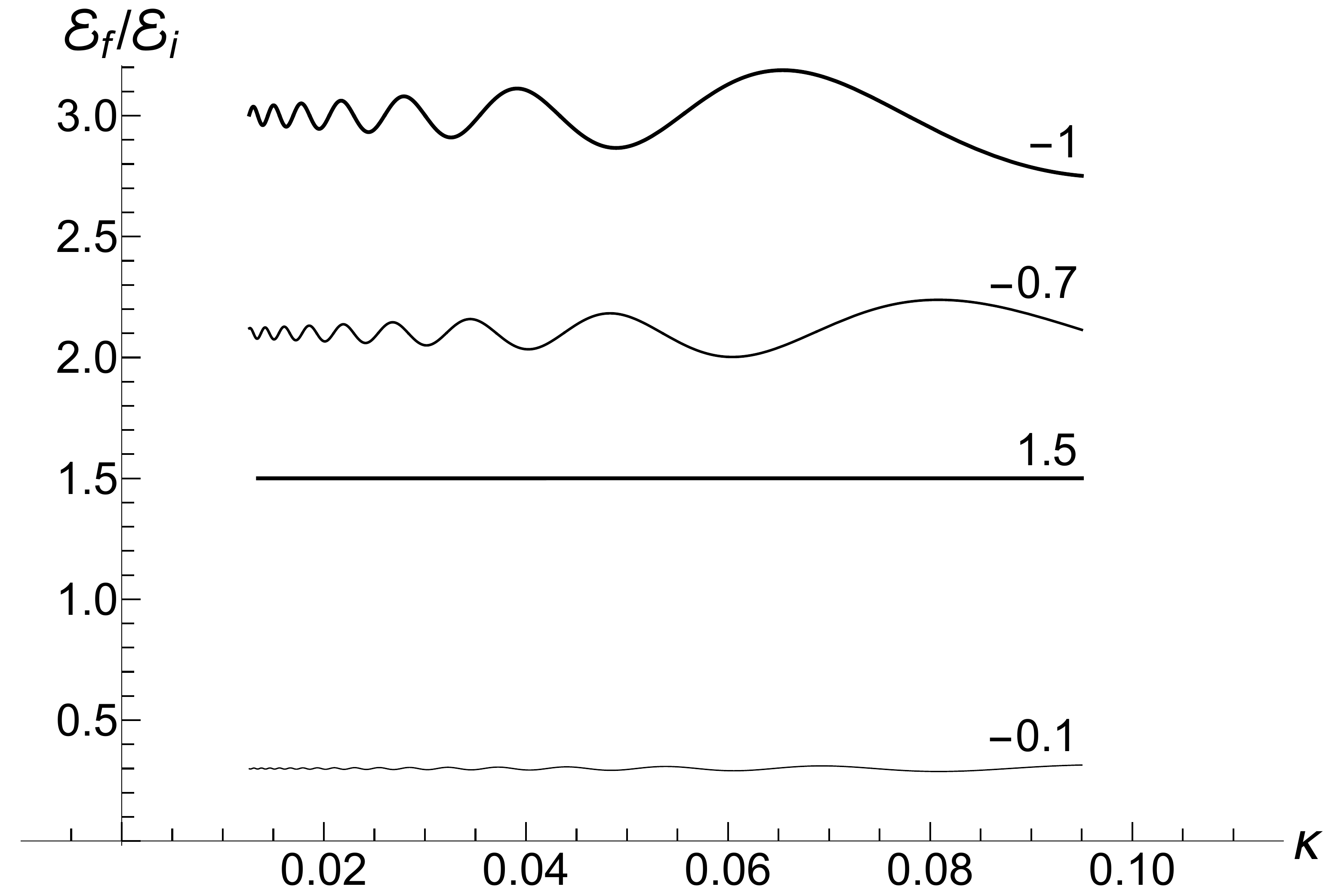}
\includegraphics[height=2.32truein,width=3.0truein,angle=0]{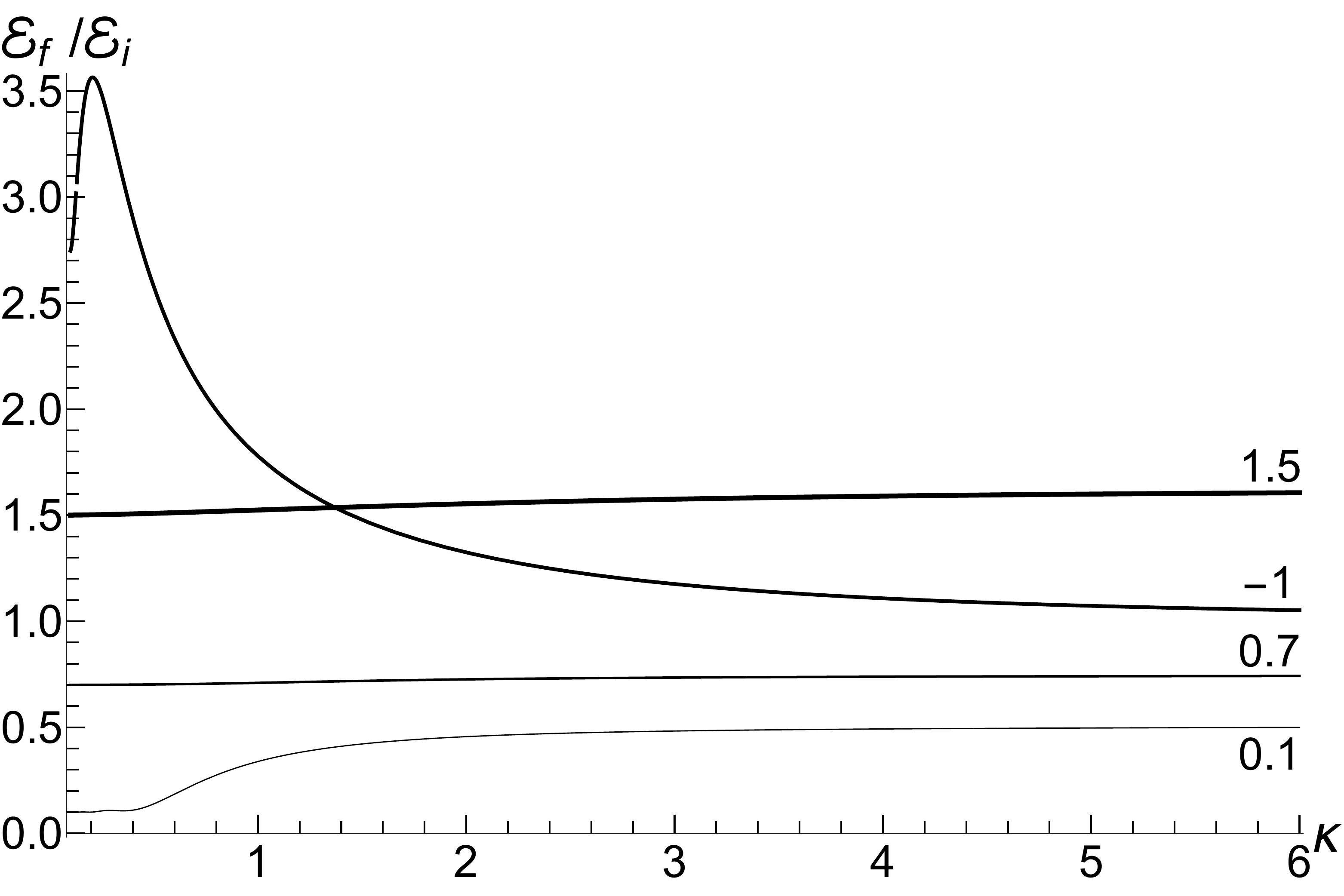}
\caption{\small The ratio ${\cal E}_f/{\cal E}_i$ versus parameter $\kappa$ for different positive values of the final frequency $\omega_{f}$
(shown nearby the respective lines) in the case of exponentially varying frequency on the time semi-axis (\ref{omexp}). 
The initial frequency is $\omega_{i}=1$.
Left: $\rho=0$, $s_0\Upsilon=1$. Right: $\rho=1$, $s_0\Upsilon=10$.
 }
\label{fig-EfEi-k1}
\end{figure}  

\subsection{Mean magnetic moment in the asymptotic regime}

\begin{figure}[htb]
\includegraphics[height=2.32truein,width=3.0truein,angle=0]{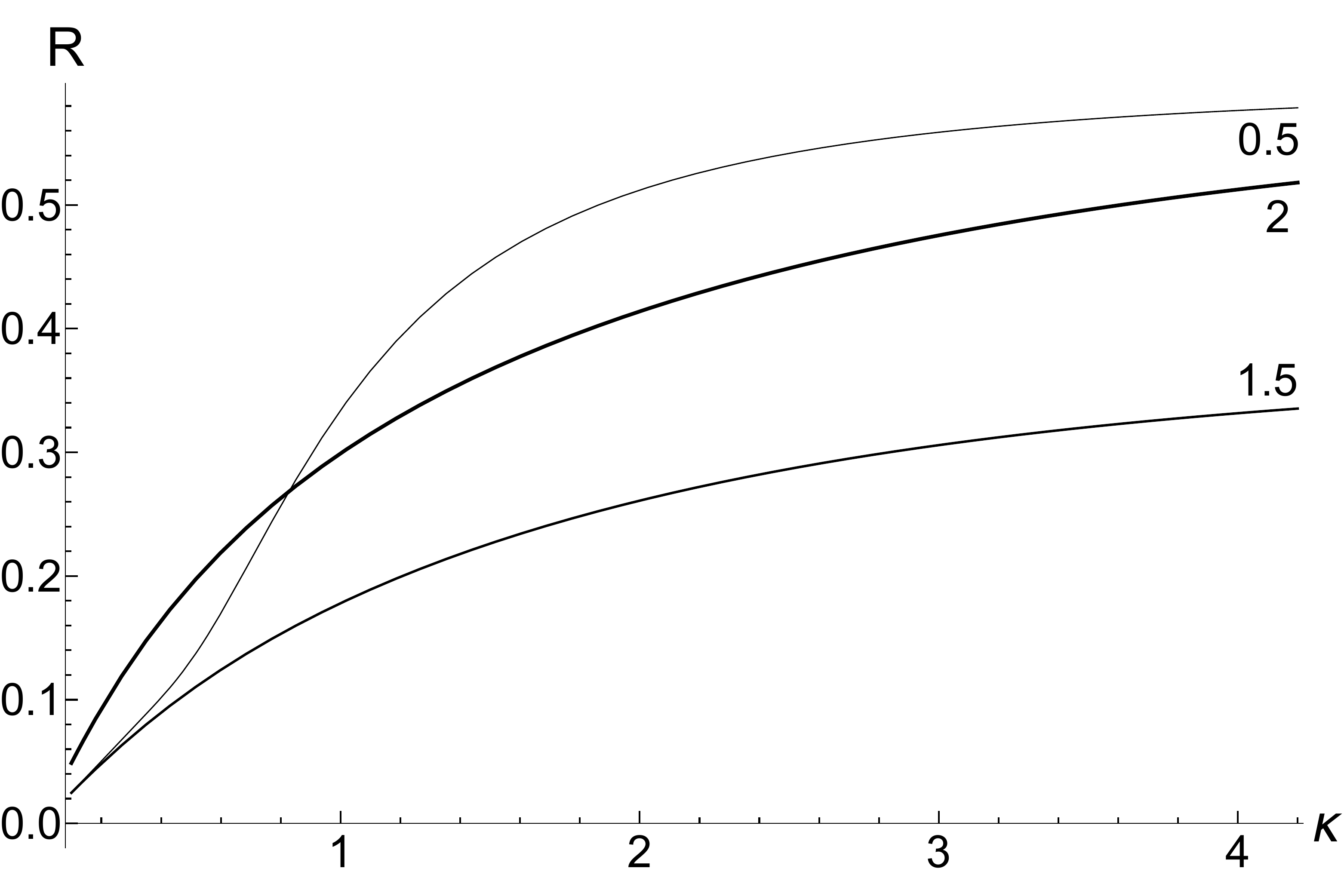}
\includegraphics[height=2.32truein,width=3.0truein,angle=0]{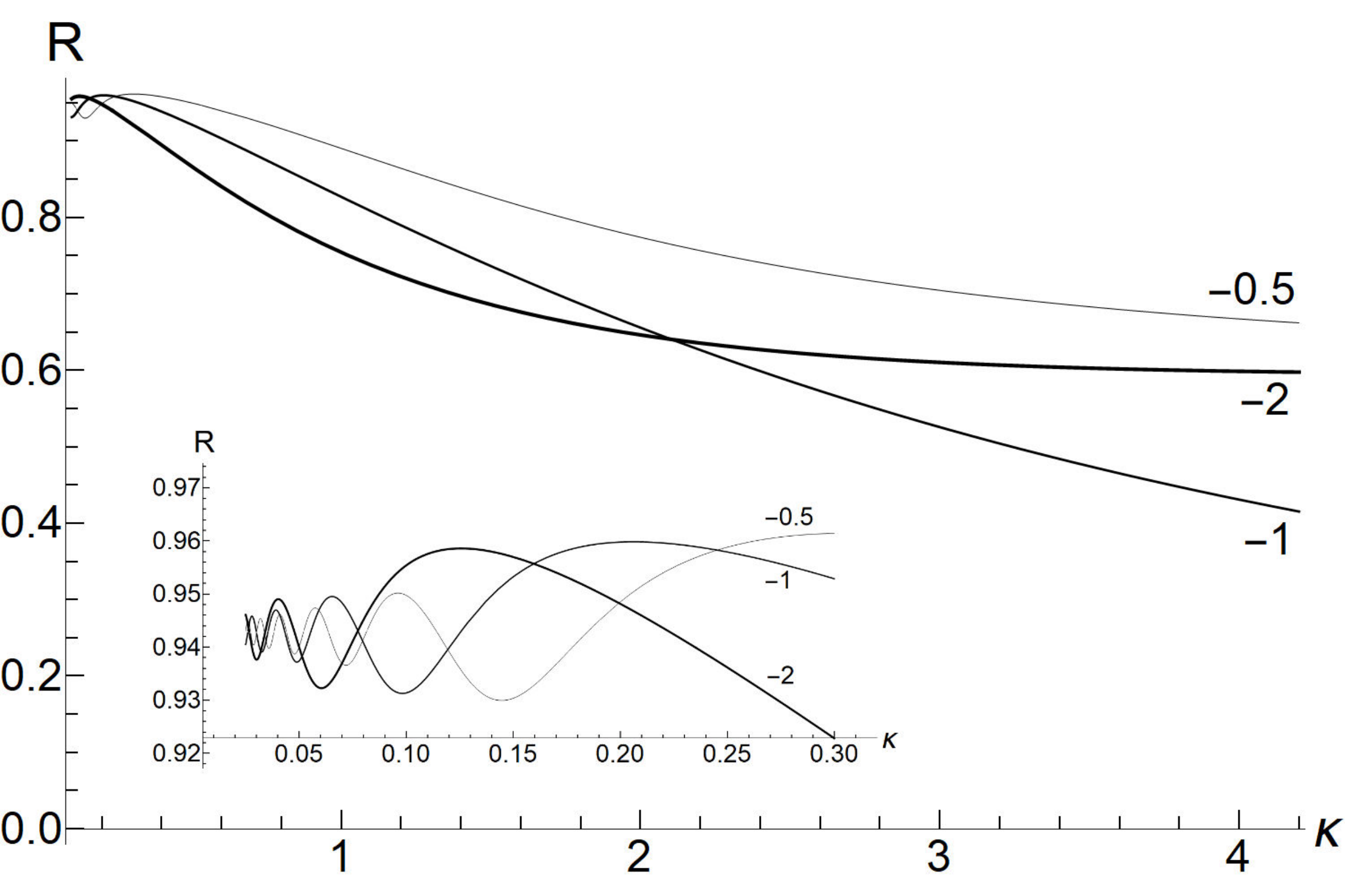}
\caption{\small The ratio $R = |\widetilde{\Delta {\cal M}}|/|\langle\langle {\cal M}\rangle\rangle|$
in the case of zero initial temperature ($\rho=0$ and $s_0\Upsilon=1$)
 versus parameter $\kappa$ for different  values of the final frequency $\omega_{f}$ (shown nearby the respective lines)
 in the case of exponentially varying frequency on the time semi-axis (\ref{omexp}). 
 The initial frequency is taken as $\omega_{i}=1$.
{ Left:} $\omega_{f} >0$. { Right:} $\omega_{f} <0$.
 }
\label{fig-R(k)}
\end{figure}

\begin{figure}[htb]
\includegraphics[height=2.32truein,width=3.0truein,angle=0]{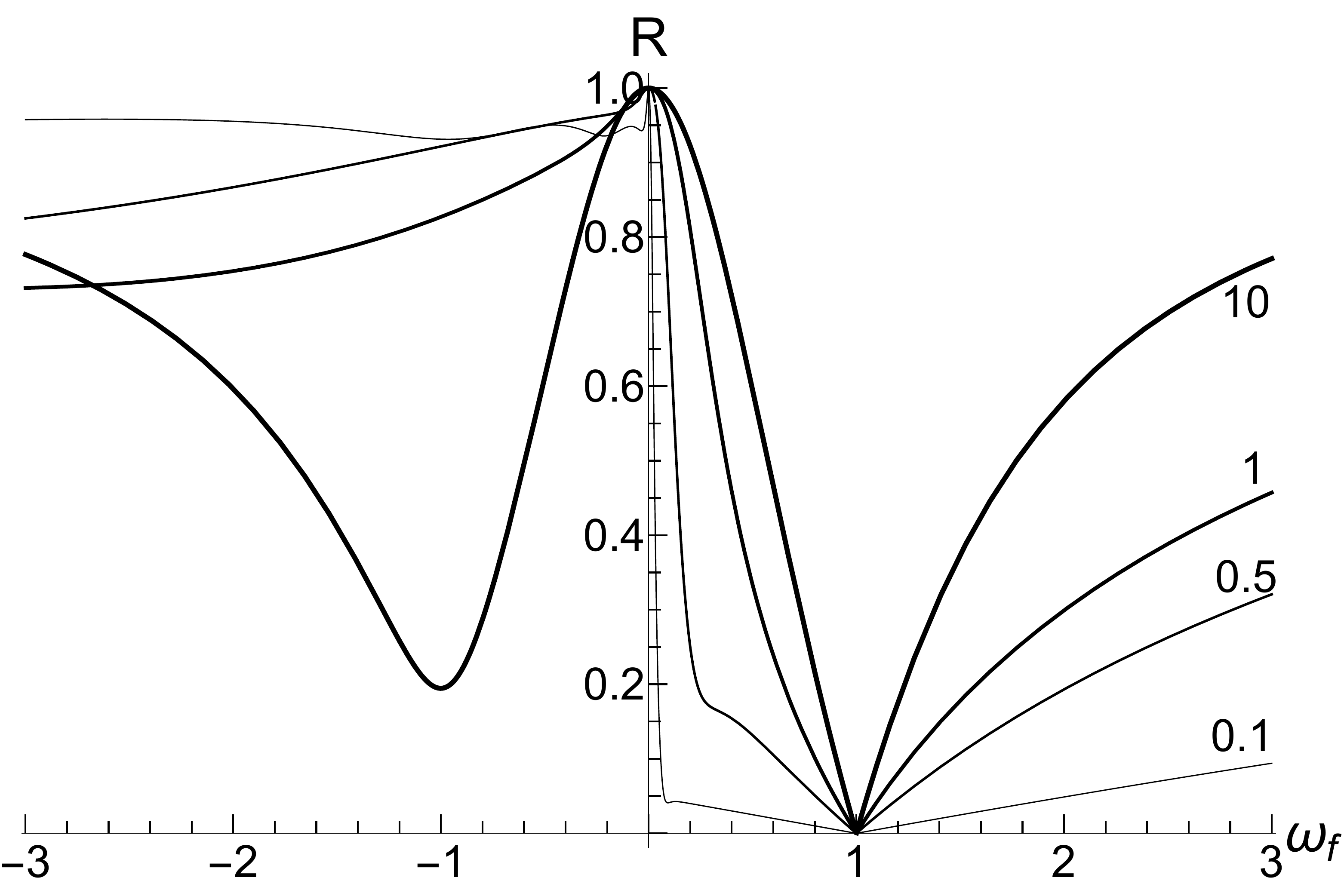}
\includegraphics[height=2.32truein,width=3.0truein,angle=0]{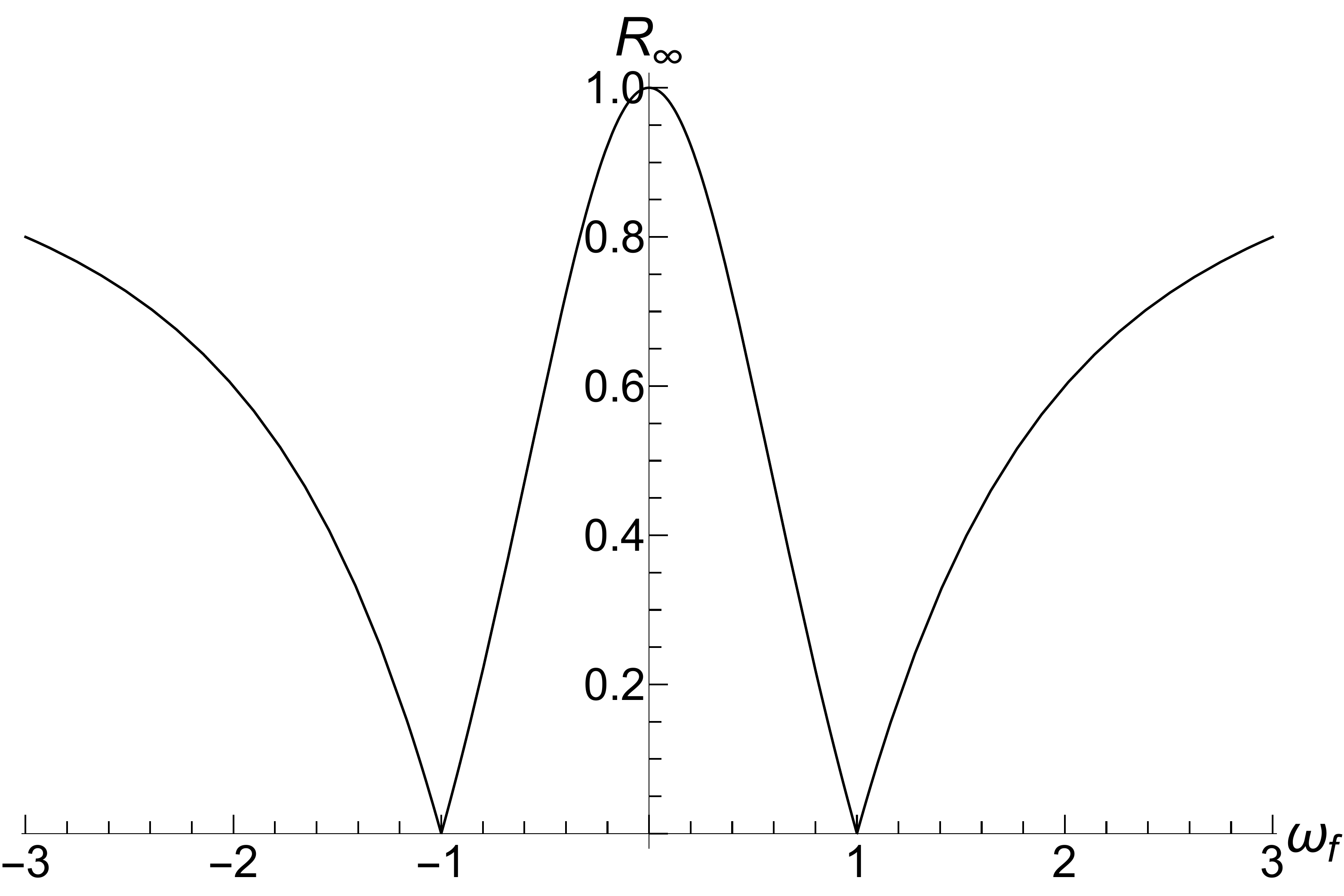}
\caption{\small The ratio $R = |\widetilde{\Delta {\cal M}}|/|\langle\langle {\cal M}\rangle\rangle|$
in the case of zero initial temperature ($\rho=0$ and $s_0\Upsilon=1$)
and exponentially varying frequency (\ref{omexp}) 
 versus the final frequency  $\omega_{f}$.
Left:  for different finite values of the parameter $\kappa$ (shown nearby the respective lines). 
Right: for the limit case of $\kappa=\infty$ (the sudden jump).
 The initial frequency is taken as $\omega_{i}=1$.
 }
\label{fig-R(wf)}
\end{figure}

In view of Equations (\ref{M-av-fluc})-(\ref{flucM}),
one needs two coefficients, $|u_{-}|^2$ (or $|u_{+}|^2$) and $u_{+} u_{-}$, to calculate the mean magnetic moment in the asymptotic regime.
They are given by formulas (\ref{upm2}) and (\ref{u+u-}).
Explicit expressions are rather cumbersome. We bring here only the simple result for the ratio 
$R = |\widetilde{\Delta {\cal M}}|/|\langle\langle {\cal M}\rangle\rangle|$ in the case of zero initial temperature,
when Equations (\ref{Msimple-low}) and (\ref{u+u-}) yield
\be
R = \frac{|\lambda|\sqrt{|\lambda|^2 + 4[1-\mbox{Re}(\lambda)]}}{|\lambda|^2 + 2[1-\mbox{Re}(\lambda)]}.
\label{R-magmom}
\ee
Figures \ref{fig-R(k)} and \ref{fig-R(wf)} show the ratio (\ref{R-magmom}) as function of $\kappa$ for different fixed values of the final frequency $\omega_f$ (assuming $\omega_i=1$)
and as function of $\omega_f$ for different values of $\kappa$.
We see that the dependence $R(\kappa)$ is quite different for finite positive and negative values of the final frequency $\omega_f$,
especially if $\kappa \ll \omega_i$ (a slow evolution).
Function $R(\omega_f;\kappa)$ also shows a strong asymmetry for small and moderate values of the fixed parameter $\kappa$. 
A symmetry with respect to the sign of frequency $\omega_f$ is restored for
$\kappa \gg 1$, when $R(\omega_f;\infty)$ coincides with the sudden jump formula (\ref{Rjumplow}).

According to Figure \ref{fig-R(wf)}, we see that $R=1$ for $\omega_f=0$ and any value of parameter $\kappa$.
This result can be derived from formula (\ref{R-magmom}) in the following way. If $\omega_f=0$, then parameter $\gamma$ defined
in Equation (\ref{acmugam}) equals zero. In this case, we can use the known formula relating the confluent hypergeometric function with
the Bessel function \cite{BE}:
\be
\Phi(1/2; 1; 2i\mu) = J_0(\mu) e^{i\mu}, \qquad \Phi(1/2; 1; x) = J_0(x/2i) e^{x/2}.
\ee
Then, $d\Phi(1/2; 1; x)/dx = (1/2)e^{x/2}\left[J_0(x/2i) -iJ_0^{\prime}(x/2i)\right]$. Using the formula 
$J_0^{\prime}(x) = -J_1(x)$ and the formula for $\lambda$ in Equation (\ref{D+D-}), we obtain the expression
$
\lambda = 1 + i J_1(\mu)/[2J_0(\mu)]$.
Since $\mbox{Re}(\lambda) =1$ in this approximation, Equation (\ref{R-magmom}) yields $R=1$.

The identity (\ref{ident}) shows that the fraction $\left[(1 - \mbox{Re}\lambda)\right]^{-1}$ behaves as 
$(\omega_{i}/\omega_{f})$ when $\omega_f \to 0$. Then, Equation (\ref{upm2}) tells us that coefficients $u^2_{\pm}$ diverge as
$|\omega_{i}/\omega_{f}|$ in this limit. In view of Equation (\ref{<<M>>}), we conclude that the average magnetic moment
$\langle\langle {\cal M}\rangle\rangle$ grows unlimitedly with time if $\omega_f=0$.

\subsection{Exponential switching off the field: solutions in terms of the Hankel functions}

To understand better the behavior of the mean energy
in the case of $\omega_f=0$, we notice that
the substitution $x=\mu \exp(-\kappa t)$ with $\mu =\omega_{i}/\kappa$ transforms equation (\ref{eqvep})
with function $\omega(t) = \omega_i \exp(-\kappa t)$  to the Bessel equation
\be
x^2 f^{\prime\prime} + x f^{\prime} + x^2 f =0.
\label{EqBes}
\ee
Complex solutions to this equation can be written as linear combinations of the Hankel functions of zero order,
$H_0(x) = J_0(x) +iY_0(x)$ and $H_0^*(x)$, where $J_0(x)$ is the Bessel function and $Y_0(x)$ the Neumann function \cite{BE}. 
Then, function $\vep(t)$ can be written in the form (\ref{vepD1D2}) with
\be
\vep_1 = \frac{H_0(\mu\xi)}{\sqrt{\omega_{i}} H_0(\mu)}, \quad 
\frac{d\vep_1}{dt} =\frac{\xi \sqrt{\omega_{i}}H_1(\mu\xi)}{ H_0(\mu)}, \quad \xi = e^{-\kappa t},
\ee
so 
\be
D_+ = \frac{1 + i\eta^*}{2\mbox{Im}(\eta)}, \quad D_- = -\,\frac{1 + i\eta}{2\mbox{Im}(\eta)}, \quad
\eta = \frac{H_1(\mu)}{ H_0(\mu)}, \quad
\mbox{Im}(\eta)= -2\left[\pi\mu|H_0(\mu)|^2\right]^{-1}.
\label{DpmR}
\ee
The following known formulas  were used here:
\be
 H_0^{\prime}(x) = - H_1(x), \quad
H_0(x)H_0^{\prime*}(x) - H_0^{\prime}(x)H_0^{*}(x) = -4i/(\pi x).
\label{ident-Hank}
\ee 
The functions $F_{\pm}(t)$, introduced in Equation (\ref{def-Fpm}), can be written as follows,
\be
F_{\pm}(\xi) = \xi \sqrt{\omega_i}\left[ D_{+} h_{\pm}(\xi) + D_{-} h_{\mp}^*(\xi)\right], \quad
h_{\pm}(\xi) = \left[H_0(\mu\xi) \pm i H_1(\mu\xi)\right]/H_{0}(\mu).
\label{def-hpm}
\ee

\subsubsection{Mean energy}

\begin{figure}[htb]
\begin{center}
\includegraphics[height=2.02truein,width=3.0truein,angle=0]{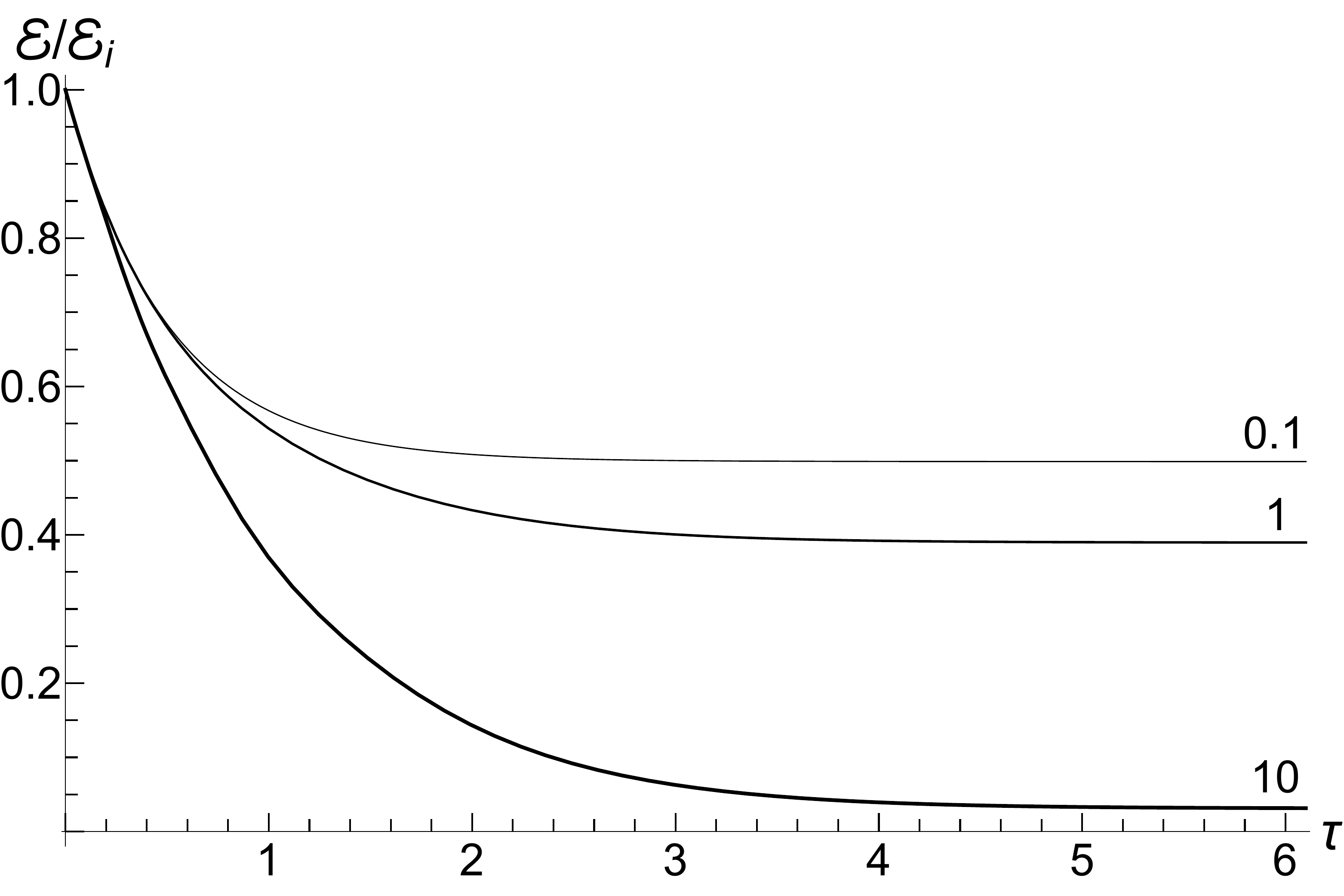}
\includegraphics[height=2.02truein,width=3.0truein,angle=0]{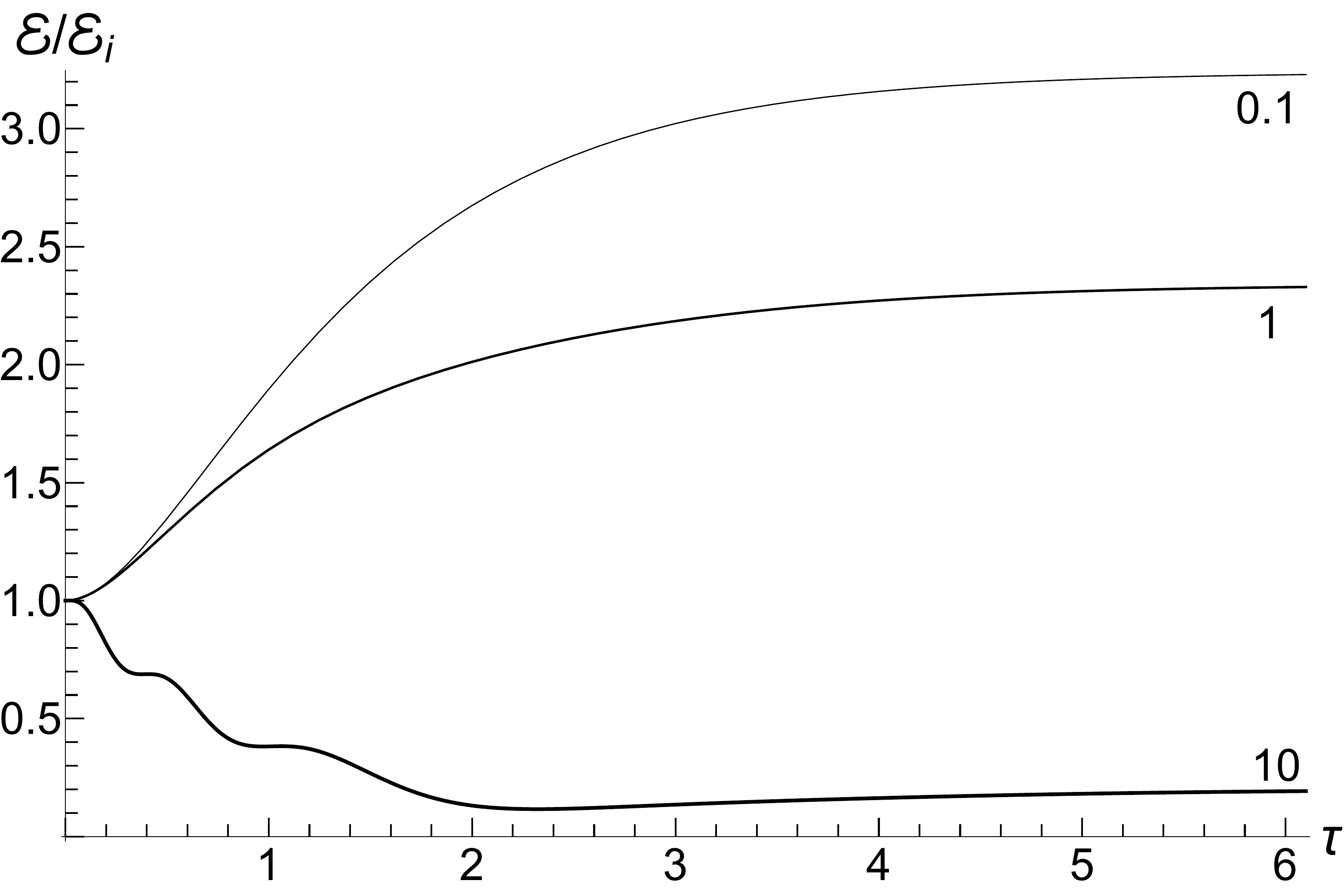}
\end{center}
\caption{\small 
The ratio ${\cal E}/{\cal E}_i$ versus the dimensionless time $\tau =\kappa t$ for different values of parameter $\mu=\omega_i/\kappa$ 
(shown nearby the respective lines) in the case of exponentially varying frequency (\ref{omexp}) with $\omega_f=0$. 
The initial frequency is $\omega_{i}=1$.
Left: $\rho=0$, $s_0\Upsilon=1$.
Right: $\rho=1$, $s_0\Upsilon=10$.
 }
\label{fig-wf=0-1}
\end{figure}

\begin{figure}[htb]
\begin{center}
\includegraphics[height=2.02truein,width=3.0truein,angle=0]{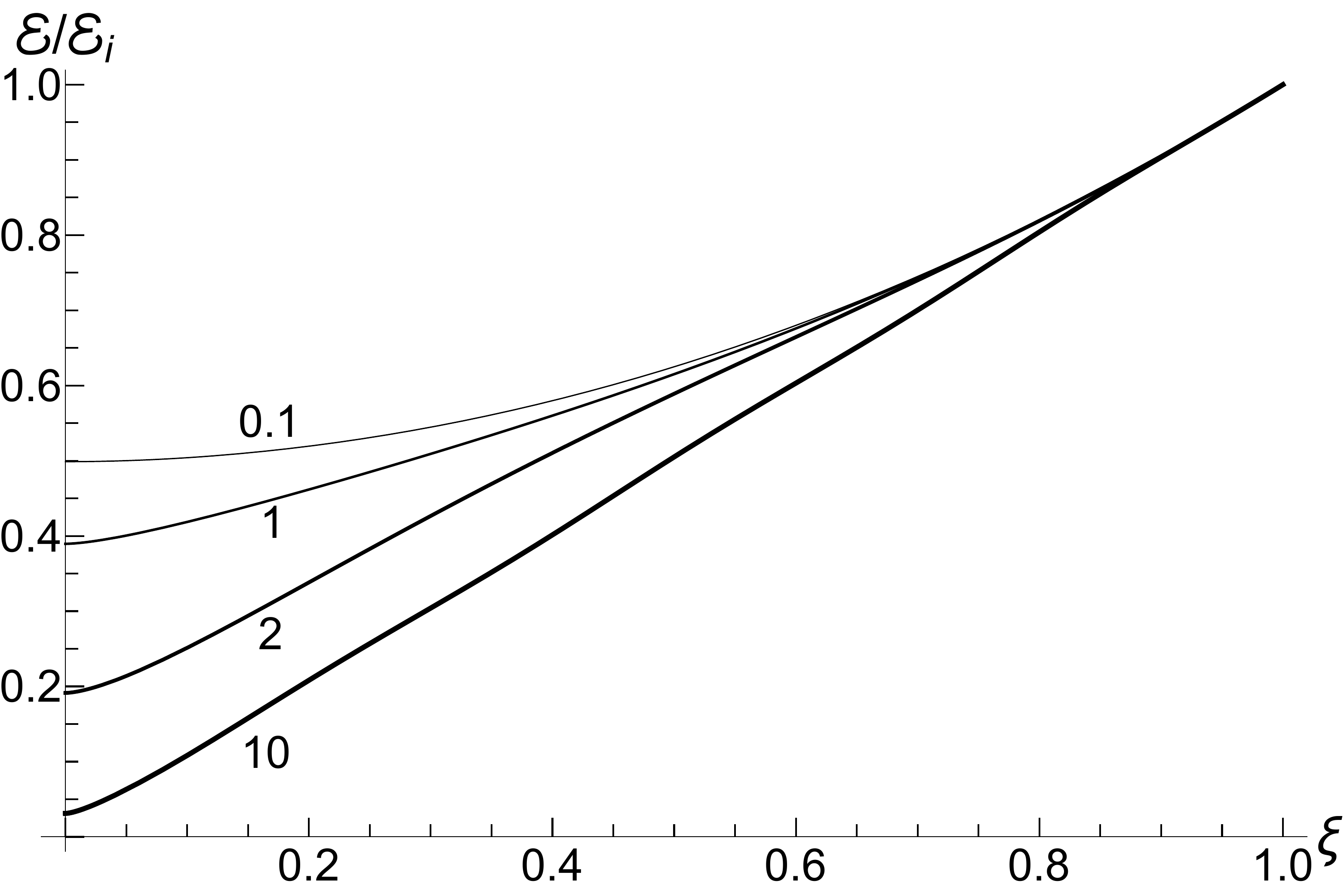}
\includegraphics[height=2.02truein,width=3.0truein,angle=0]{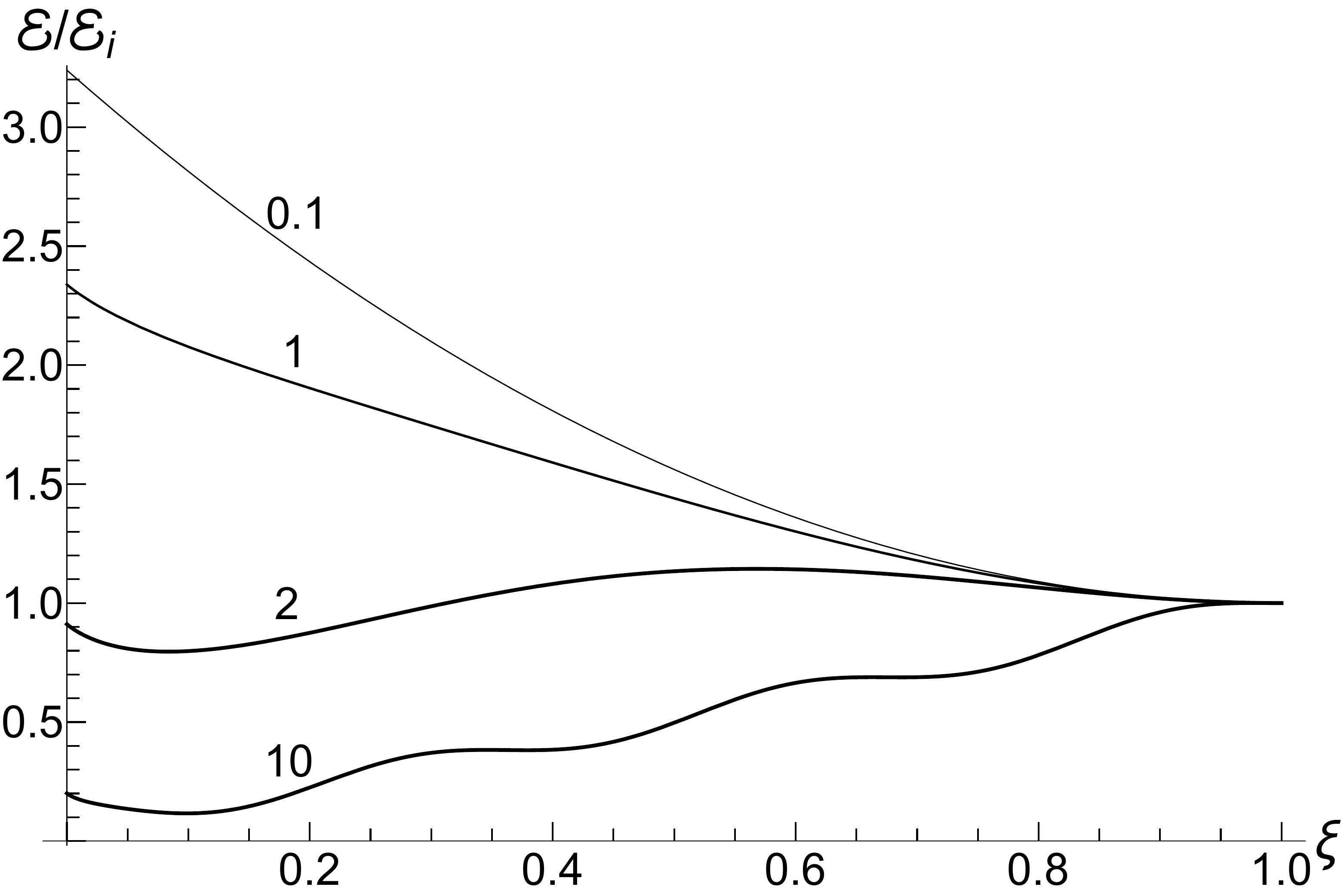}
\end{center}
\caption{\small 
The ratio ${\cal E}/{\cal E}_i$ versus the variable $\xi=\omega(t)/\omega_i$ for different values of parameter $\mu=\omega_i/\kappa$ 
(shown nearby the respective lines) in the case of exponentially varying frequency (\ref{omexp}) with $\omega_f=0$. 
The initial frequency is taken as $\omega_{i}=1$.
Left: $\rho=0$, $s_0\Upsilon=1$.
Right: $\rho=1$, $s_0\Upsilon=10$.
 }
\label{fig-Exi}
\end{figure}

The time-dependent mean energy is given by Equation (\ref{Eqfin}) with the following coefficients:
\be
|F_{\pm}(\xi)|^2  = \frac{(\pi\mu\xi)^2 \omega_i}{8}\left\{
V_{+}(\xi)V_{+}(1) - \mbox{Re}\left[U_{+}^*(1)U_{+}(\xi)\right] \mp \frac{16}{(\pi\mu)^2 \xi} \right\},
\ee
\be
\mbox{Re}[F_{-}(\xi)F_{+}(\xi)]  = \frac{(\pi\mu\xi)^2 \omega_i}{8}\left\{
  \mbox{Re}\left[U_{-}^*(1)U_{+}(\xi)\right] - V_{+}(\xi)V_{-}(1) \right\},
\ee
where
\be
V_{\pm}(\xi) = |H_0(\mu\xi)|^2 \pm |H_1(\mu\xi)|^2, \quad U_{\pm}(\xi) = H_0^2(\mu\xi) \pm H_1^2(\mu\xi).
\ee
The most simple expression can be written for the initial zero-temperature state:
\be
{\cal E}(t)/{\cal E}_i = (\pi\mu\xi)^2 \left\{V_{+}(1)V_{+}(\xi) - \mbox{Re}\left[U_{+}^*(1)U_{+}(\xi)\right]\right\}/{16}.
\label{Omeg0-0}
\ee
Typical plots of the ratio ${\cal E}/{\cal E}_i$ as function of the dimensionless parameter $\tau=\kappa t$ are given in 
Figure \ref{fig-wf=0-1}.
Similar plots for ${\cal E}/{\cal E}_i$ as function of variable $\xi$ are given in Figure \ref{fig-Exi}.
Note that small values of parameter $\mu =\omega_i/\kappa$ correspond to almost instant ``jump'' of the frequency to the final zero value,
whereas the case of $\mu \gg 1$ corresponds to a slow (quasi-adiabatic) frequency decay to zero.
The left-hand side of Figure \ref{fig-Exi} with $\mu=10$ shows practically adiabatic evolution ${\cal E}/{\cal E}_i =\xi$
up to very small values of $\xi$. However, the adiabaticity is always broken at the final stage of evolution, when the
mean energy tends to a nonzero final value, even at zero temperature. On the other hand, the adiabatic evolution becomes
very approximate in the high-temperature case, as one can see in the right-hand side of Figure \ref{fig-Exi}, where the line
with the same value $\mu=10$ clearly shows oscillations around the straight line ${\cal E}/{\cal E}_i =\xi$.
The final energy is smaller than the initial one for any value of $\mu$ in the zero-temperature regime. But it can be much higher
than ${\cal E}_i$ in the high-temperature case with $\mu <2$, as one can see in Figures \ref{fig-wf=0-1} and \ref{fig-Exi}.
\begin{figure}[htb]
\begin{center}
\includegraphics[height=2.32truein,width=3.0truein,angle=0]{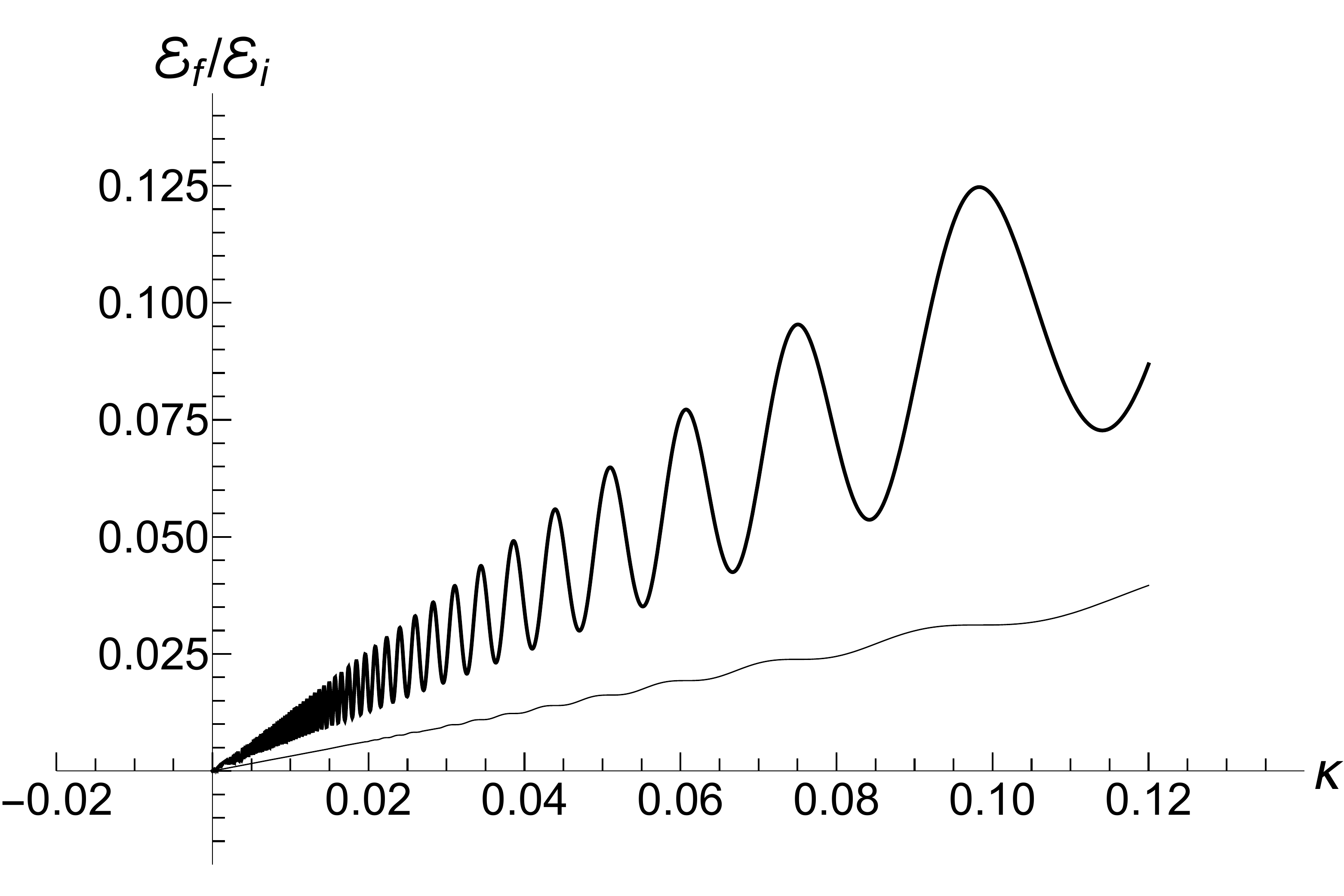}
\includegraphics[height=2.32truein,width=3.0truein,angle=0]{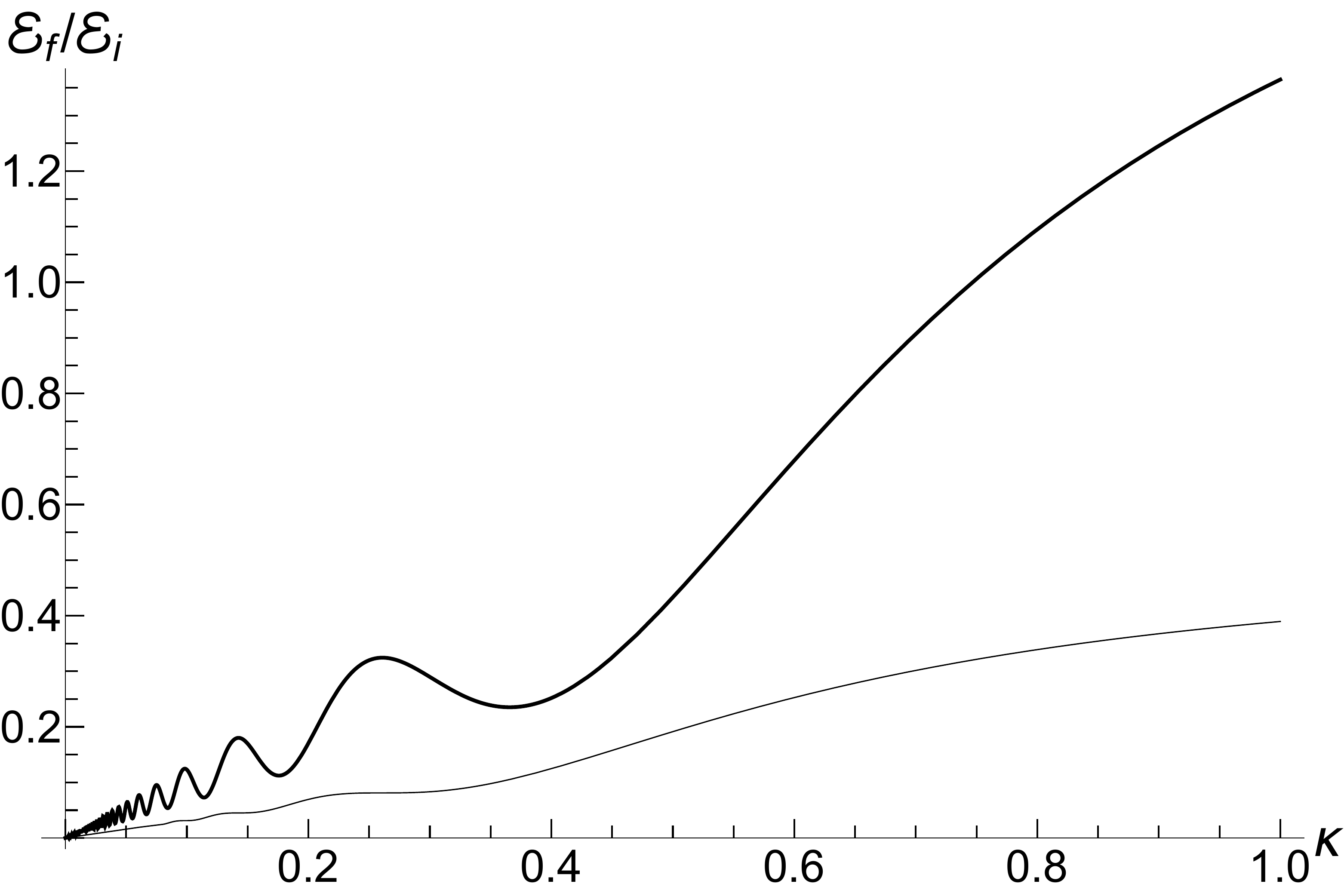}
\end{center}
\caption{\small 
The ratio ${\cal E}_f /{\cal E}_i$ versus parameter $\kappa $ 
in the case of exponentially varying frequency (\ref{omexp}) with $\omega_f=0$ and $\omega_{i}=1$.
Lower curves: $\rho=0$, $s_0\Upsilon=1$.
Upper curves: $\rho=1$, $s_0\Upsilon= 5$. 
}
\label{fig-wf=0-2}
\end{figure}

If $t \to\infty$, then $\xi\to 0$, so  \cite{BE} $\xi H_0(\mu\xi) \to 0$, but  $\xi H_1(\mu\xi) \to -2i/(\pi\mu)$.
Using these relations,  one can obtain after some algebra the following formula for the final mean energy 
${\cal E}_f = {\cal E}(\infty)$:
\be
{\cal E}_f /{\cal E}_i =  \left\{ (1 + s_0\Upsilon) \left[J_0^2(\mu) + J_1^2(\mu)\right] + 
2\rho \left[J_0^2(\mu) - J_1^2(\mu)\right]\right\}/4.
\label{Om0inf}
\ee
The right-hand side of this equation  is shown in Figure \ref{fig-wf=0-2} as function of parameter  $\kappa$.

If $\mu \ll 1$, formula (\ref{Om0inf}) assumes the form 
\[
{\cal E}_f /{\cal E}_i \approx   \left\{ (1 + s_0\Upsilon) \left(1-\mu^2/4\right) + 
2\rho \left(1-3\mu^2/4\right)\right\}/4.
\]
Putting $\mu=0$, we arrive at the instant jump approximation formula (\ref{EfEi-sudden0}).
We see that the relative accuracy of this approximation is of the order of $\mu^2/4$.

In the adiabatic limit, $\kappa \ll \omega_{i}$, the known asymptotic formulas for $\mu \gg 1$,
\[
J_0(\mu) \approx \sqrt{{2}/(\pi\mu)} \cos(\mu - \pi/4), \quad J_1(\mu) \approx \sqrt{{2}/(\pi\mu)} \sin(\mu - \pi/4),
\]
lead to the relation
\be
{\cal E}_f /{\cal E}_i \approx  \kappa \left[ 1 + s_0\Upsilon  + 
2\rho \sin(2\omega_i/\kappa)\right]/(2\pi \omega_i).
\label{EfEi-om0adiab}
\ee
The  proportionality of the ratio ${\cal E}_f /{\cal E}_i$ to $\kappa$ when $\kappa \ll \omega_i$ 
in the zero-temperature case is clearly seen in Figure \ref{fig-wf=0-2}.
On the other hand, this ratio demonstrates strong oscillations as function of $\kappa$ in the high-temperature regime.

\section{The Epstein--Eckart profiles: solutions in terms of the Gauss hypergeometric functions}
\label{sec-tanh}

Exact solutions in terms of the Gauss hypergeometric function 
\be
F(a,b;c;x) = \sum_{n=1}^{\infty} \frac{(a)_n (b)_n x^n}{(c)_n n!},
\ee
satisfying the equation
\be
x(1-x)F^{\prime\prime} +(c-(a+b+1)x)F^{\prime} -abF =0,
\label{eqF}
\ee
 can be found for the family of the Epstein--Eckart profiles \cite{Eckart,Epstein},
which are combinations of some fractions containing exponential functions of time. The total family has four constant parameters.
In order to simplify the analysis, we confine ourselves here with two simple subfamilies containing two or three parameters. 
 
\subsection{Evolution on the whole time axis}
\label{sec-EEfullline}

The first example corresponds to the Larmor frequency of the form
\be
\omega(t) = \frac{\omega_f \exp(\kappa t) +\omega_i}{\exp(\kappa t) +1},
 \quad -\infty < t < \infty, \quad \kappa >0.
\label{omtanh}
\ee
One can verify (see Appendix \ref{sec-appEE-1}) that Equation (\ref{eqvep}) with $\omega (t)$ given by Equation (\ref{omtanh})  
has the solution
\be
\vep(t) = \omega_i^{-1/2} e^{i\omega_i t} (1+\zeta)^{d} F(a, b; c; -\zeta), \quad \zeta = e^{\kappa t},
\label{sol-F}
\ee
with the following parameters:
\be
d = 1/2 - \sqrt{1/4 - \left(\tilde\omega_i -\tilde\omega_f \right)^2}, 
\quad
a = d + i\left(\tilde\omega_i +|\tilde\omega_f| \right), \quad b = d + i\left(\tilde\omega_i -|\tilde\omega_f| \right),
\quad c = 1 +2i\tilde\omega_i ,
\label{d-}
\ee
where $\tilde\omega_{i,k} \equiv \omega_{i,k}/\kappa$.
There exists also the solution with $d = 1/2  + \sqrt{...}$,
but namely
the choice (\ref{d-}) leads to the desired solution $\omega_i^{-1/2}\exp(i\omega_i t)$ if $\omega_i = \omega_f$.
Since  $\zeta= d\zeta/dt =0$ for $t = -\infty$,  function (\ref{sol-F}) behaves exactly as $\omega_i^{-1/2}\exp(i\omega_i t)$ at
$t \to -\infty$.

Note, however, that function (\ref{sol-F}) is the solution to Equation (\ref{eqvep}) for $t\le 0$ only, when $\zeta \le 1$. 
For $t \ge 0$, one should use the
analytic continuation of the hypergeometric function, given by  formula 2.10(2) from \cite{BE}: 
\be
F(a, b; c; -\zeta) = B_1 \zeta^{-a} F(a, 1-c +a; 1-b +a; -\zeta^{-1})
 + B_2 \zeta^{-b} F(b, 1-c +b; 1-a +b; -\zeta^{-1}),
  \label{hyper2}
\ee
\be
B_1 =\frac{\Gamma(c)\Gamma(b-a)}{\Gamma(b)\Gamma(c-a)}, \quad B_2 = \frac{\Gamma(c)\Gamma(a-b)}{\Gamma(a)\Gamma(c-b)}.
\label{B1B2}
\ee
Therefore, at $\zeta \to \infty$ we arrive at 
 the  form (\ref{uvsol}) of $\vep(t)$ at $t \to \infty$, with the following coefficients $u_{\pm}$:
\be
u_{\pm} = \frac{\left(|\omega_f|/\omega_i\right)^{1/2}\Gamma\left(1 + 2i\tilde\omega_i\right)\Gamma\left( \pm 2i|\tilde\omega_f|\right)}
{\Gamma\left[d + i\left(\tilde\omega_i \pm |\tilde\omega_f|\right)\right]
\Gamma\left[ 1 + i\left(\tilde\omega_i \pm |\tilde\omega_f|\right) -d\right]}.
\label{upmhyper}
\ee
If $|\tilde\omega_{i,f}| \ll 1$, then $d \approx \left(\tilde\omega_i - \tilde\omega_f\right)^2$. Hence,
 formula $\Gamma(x) = \Gamma(1+x)/x \approx 1/x$ (valid for $|x| \ll 1$) leads immediately to 
 the sudden jump relations (\ref{upmjump}). 
Analyzing the maximum of ratio 
$|\dot\omega/\omega^2|$ as function of time for the Epstein--Eckart profile (\ref{omtanh})
with $\omega_f >0$, we obtain the condition of the adiabatic approximation $\kappa |\omega_f -\omega_i|/(\omega_f \omega_i) \ll 1$,
which is equivalent to $\kappa \ll \mbox{min}(\omega_f, \omega_i)$, if the initial and final frequencies 
are well different. 

In the case of $\omega_f = -\omega_i$, the formula $\Gamma(z)\Gamma(1-z) = \pi/\sin(\pi z)$ leads to
a simple expression
\be
u_{-} = \frac{i\sin(\pi d)}{\sinh\left(2\pi\tilde\omega_i\right)} = 
\frac{i \cos\left(\pi \sqrt{1/4 - 4\tilde\omega_i^2}\right)}
{\sinh\left(2\pi\tilde\omega_i\right)}.
\label{u--f}
\ee 
In the fast transition limit, $\tilde\omega_i \ll 1$, we have $u_{-} \approx 2i\tilde\omega_i$, so ${\cal E}_f$ is close
to ${\cal E}_i$, in accordance with the sudden jump approximation.
In the adiabatic limit, $\tilde\omega_i \gg 1$, we have
$
u_{-} \approx i\coth\left(2\pi\tilde\omega_i\right)$. Note that parameter $\rho$ is not very important for the mean energy:
$\rho=0$ at zero temperature and $\rho \ll \s_0\Upsilon$ in the high-temperature case. Taking into account this observation,
we obtain the following limit ratio for $\tilde\omega_i \gg 1$ (and $\omega_f = -\omega_i$):
${\cal E}_f/{\cal E}_i \approx 2 + s_0\Upsilon$ (i.e., ${\cal E}_f/{\cal E}_i \approx 3$ at zero temperature).
Using Equation (\ref{<<M>>simple}), one can obtain the following simple expressions 
for the magnetic moment in the case of $\omega_f = -\omega_i$:
\be
\frac{\langle\langle {\cal M}\rangle\rangle}{\mu_B{\cal C}} \approx
\left\{ \begin{array}{ll}
s_0\Upsilon & \tilde\omega_i \ll 1
\\
1 + 2s_0\Upsilon & \tilde\omega_i \gg 1
\end{array} \right., \qquad
\frac{|\widetilde{\Delta {\cal M}}|}{\mu_B{\cal C}} \approx
(1 +s_0\Upsilon) \times \left\{
\begin{array}{ll}
2\tilde\omega_i  & \tilde\omega_i \ll 1
\\
\sqrt{2} & \tilde\omega_i \gg 1
\end{array} \right..
\ee

Neglecting the term proportional to $\rho$ in Equation (\ref{EfEi-u-}), we need to know the only quantity $|u_-|^2$. 
Using the formula \cite{PBM}
$|\Gamma(ix)|^2 = {\pi}[x\sinh(\pi x)]^{-1}$,
we can write
\be
|u_-|^2 = {\pi^2} \left| \Gamma\left[d + i\left(\tilde\omega_i - |\tilde\omega_f|\right)\right]
\Gamma\left[ 1 + i\left(\tilde\omega_i - |\tilde\omega_f|\right) -d\right]\right|^{-2}
\left[ \sinh\left(2\pi\tilde\omega_i\right)\sinh\left(2\pi|\tilde\omega_f|\right)\right]^{-1}.
\label{modu-2}
\ee
The right-hand side of this equation diverges when $\omega_f \to 0$. Consequently, the magnetic moment grows
unlimitedly with time if $\omega_f=0$.

For  negative values of $\omega_f$ with $|\tilde\omega_f| \gg 1$, we have
$
d = 1/2 \pm i\left(\tilde\omega_i +|\tilde\omega_f|\right) +{\cal O}\left(|\tilde\omega_f|^{-1}\right)$. Then,
the product of two Gamma functions in (\ref{modu-2}) takes the form
$\Gamma\left[1/2 + 2i\tilde\omega_i \right) \Gamma\left( 1/2 - 2i |\tilde\omega_f|\right)$.
Hence, using the relation \cite{PBM} $|\Gamma(1/2 +ix)|^2 = {\pi}/\cosh(\pi x)$, we obtain the formula
$|u_{-}|^2 \approx \coth\left(2\pi\tilde\omega_i\right)\coth\left(2\pi|\tilde\omega_f|\right)$,
so that ${\cal E}_f/{\cal E}_i \approx (|\omega_f|/\omega_i)(2 + s_0\Upsilon)$ 
in the limit of $\kappa \to 0$.

\begin{figure}[htb]
\begin{center}
\includegraphics[height=1.42truein,width=3.0truein,angle=0]{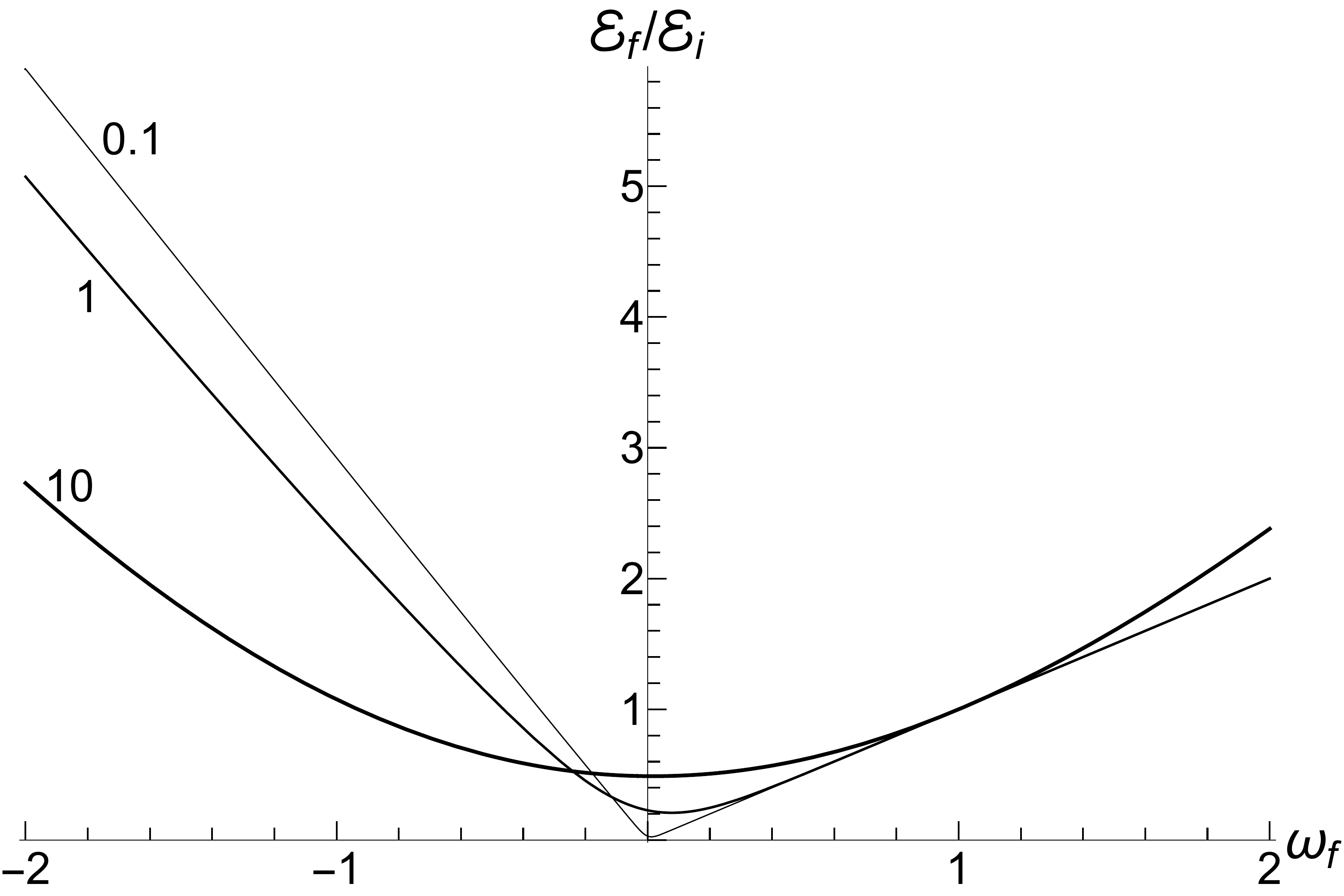}
\includegraphics[height=1.42truein,width=3.0truein,angle=0]{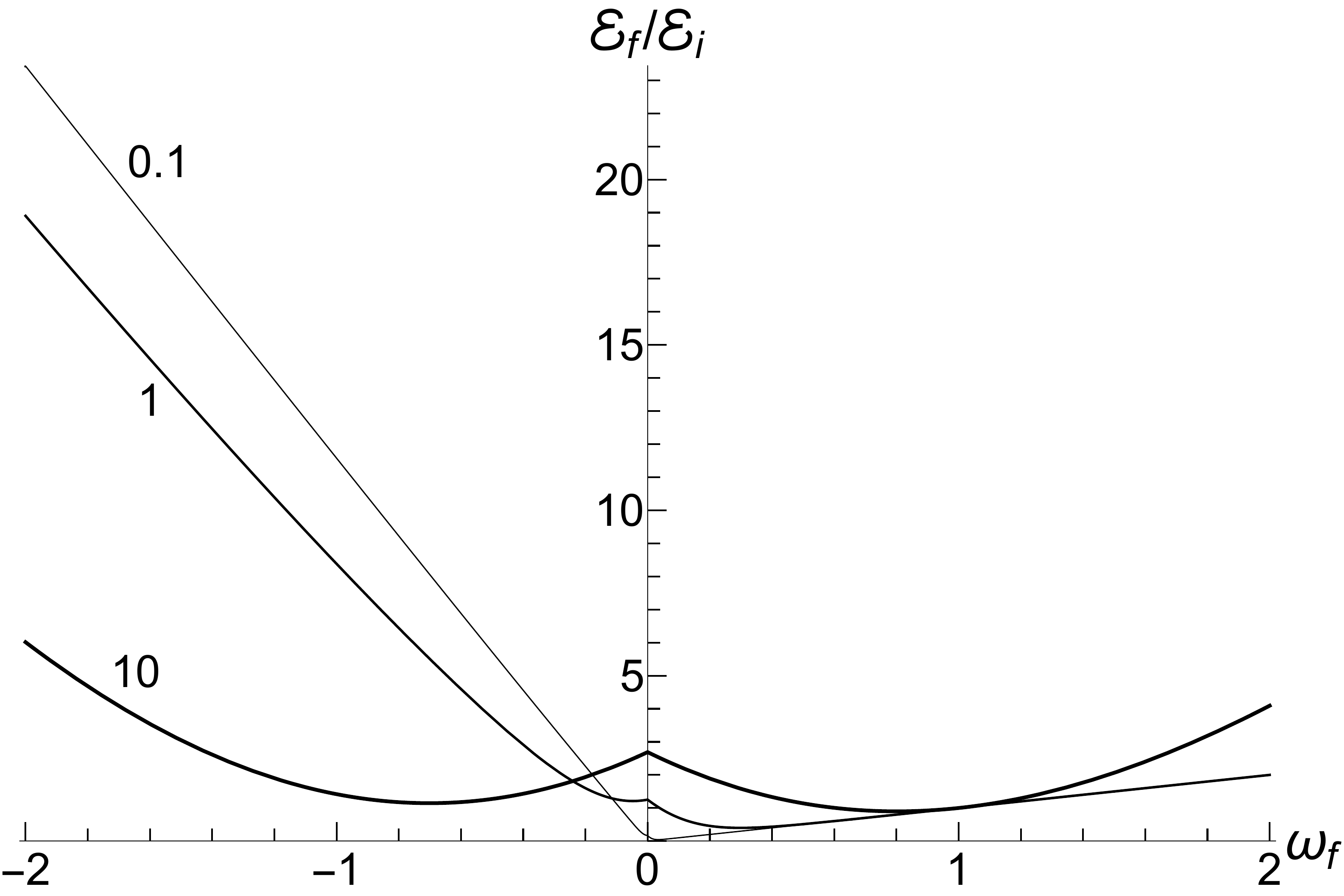}
\end{center}
\caption{\small The ratio ${\cal E}_f/{\cal E}_i$ versus the final frequency $\omega_{f}$ for different values of parameter $\kappa$
(shown nearby the respective lines) in the case of the Epstein--Eckart profile (\ref{omtanh}).
 The initial frequency is taken as $\omega_{i}=1$.  Left: $\rho=0$, $s_0\Upsilon=1$.
Right: $\rho=1$, $s_0\Upsilon=10$.
 }
\label{fig-EfEi-wfEps}
\end{figure}

\begin{figure}[htb]
\includegraphics[height=1.42truein,width=3.0truein,angle=0]{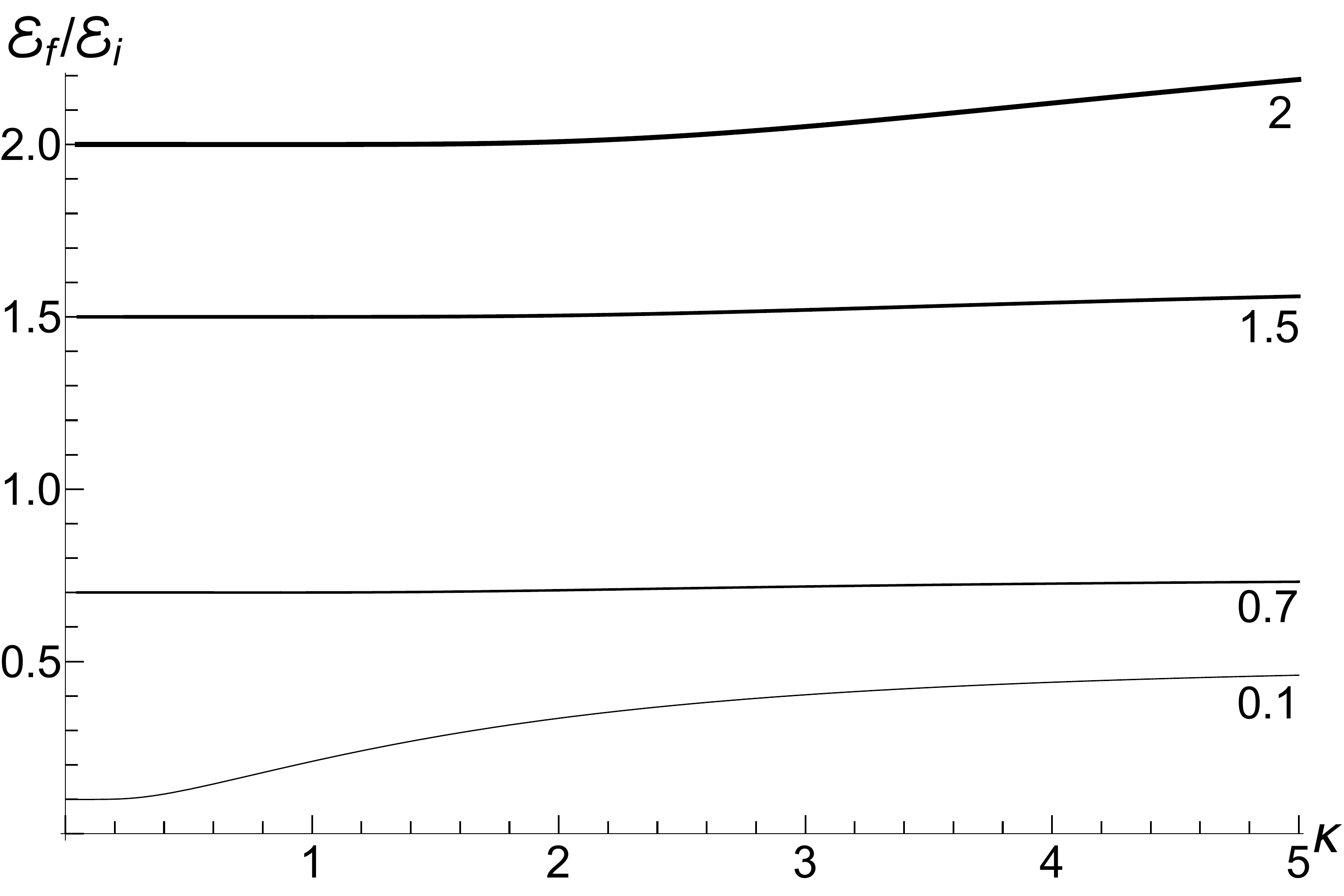}
\includegraphics[height=1.42truein,width=3.0truein,angle=0]{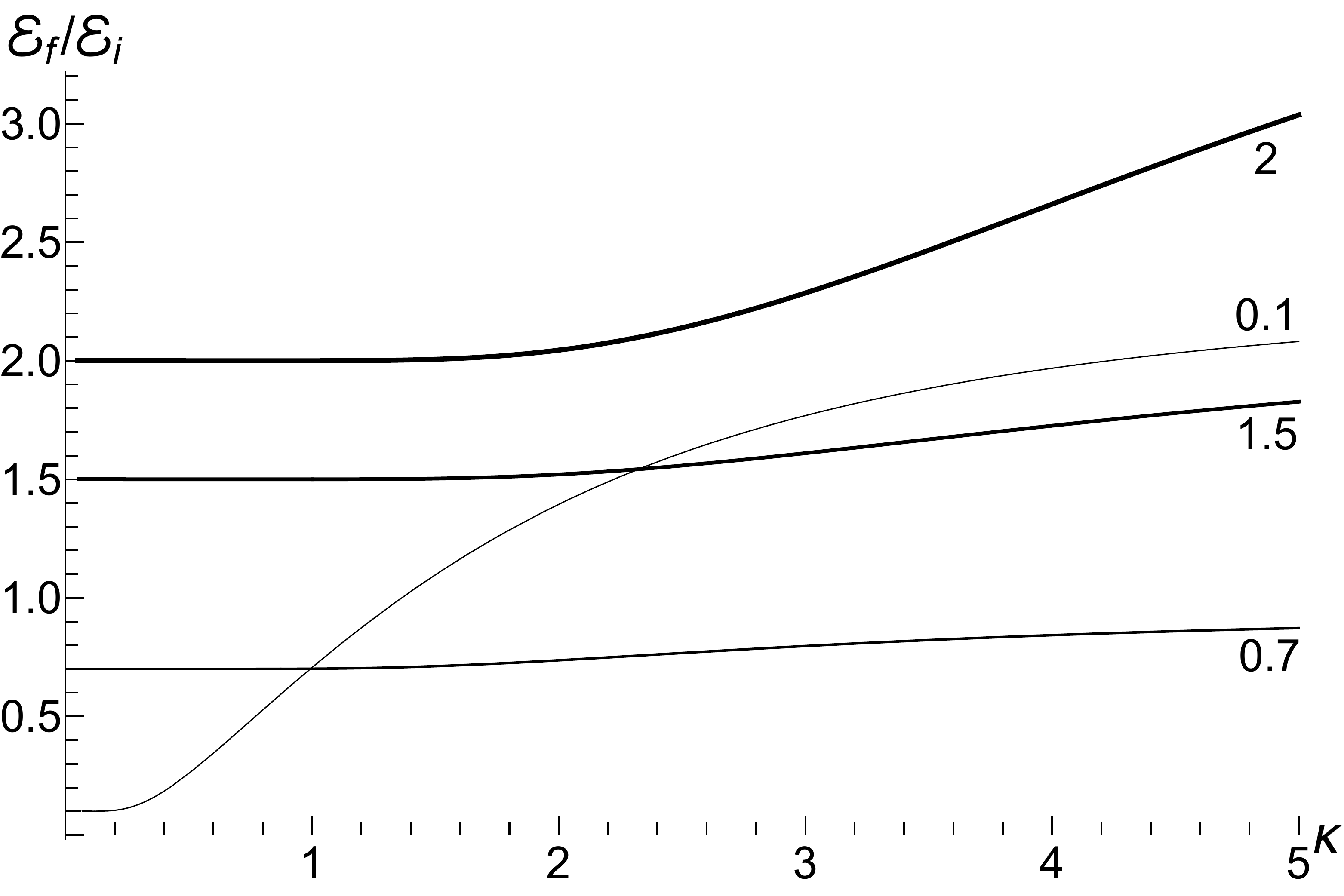}
\caption{\small The ratio ${\cal E}_f/{\cal E}_i$ versus parameter $\kappa$ for different positive values of the final frequency $\omega_{f}$
(shown nearby the respective lines) in the case of the Epstein--Eckart profile (\ref{omtanh}). 
The initial frequency is taken as $\omega_{i}=1$.
Left: $\rho=0$, $s_0\Upsilon=1$. Right: $\rho=1$, $s_0\Upsilon=10$.
 }
\label{fig-EfEi-k1Eps}
\end{figure}

\begin{figure}[hbt]
\includegraphics[height=1.42truein,width=3.0truein,angle=0]{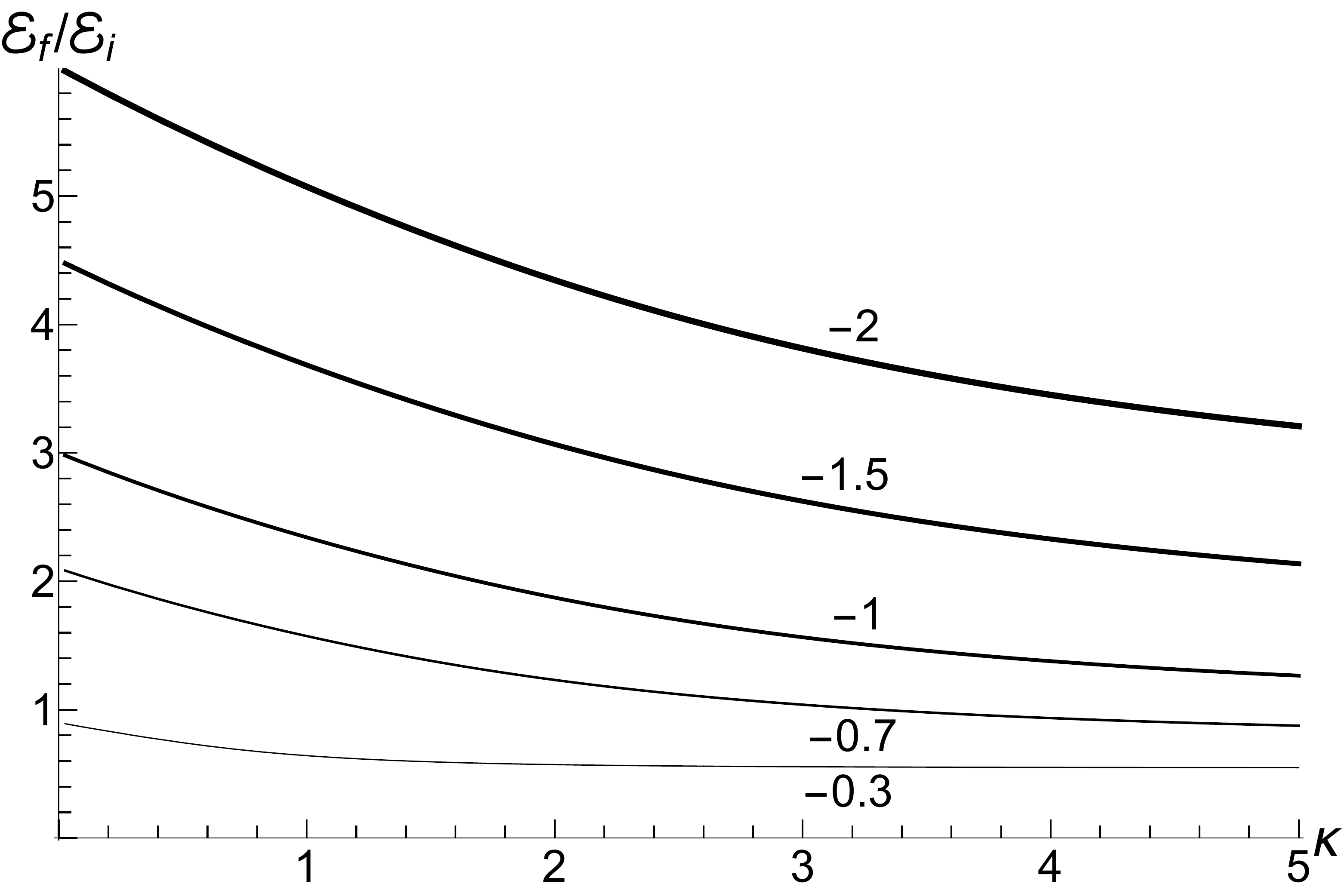}
\includegraphics[height=1.42truein,width=3.0truein,angle=0]{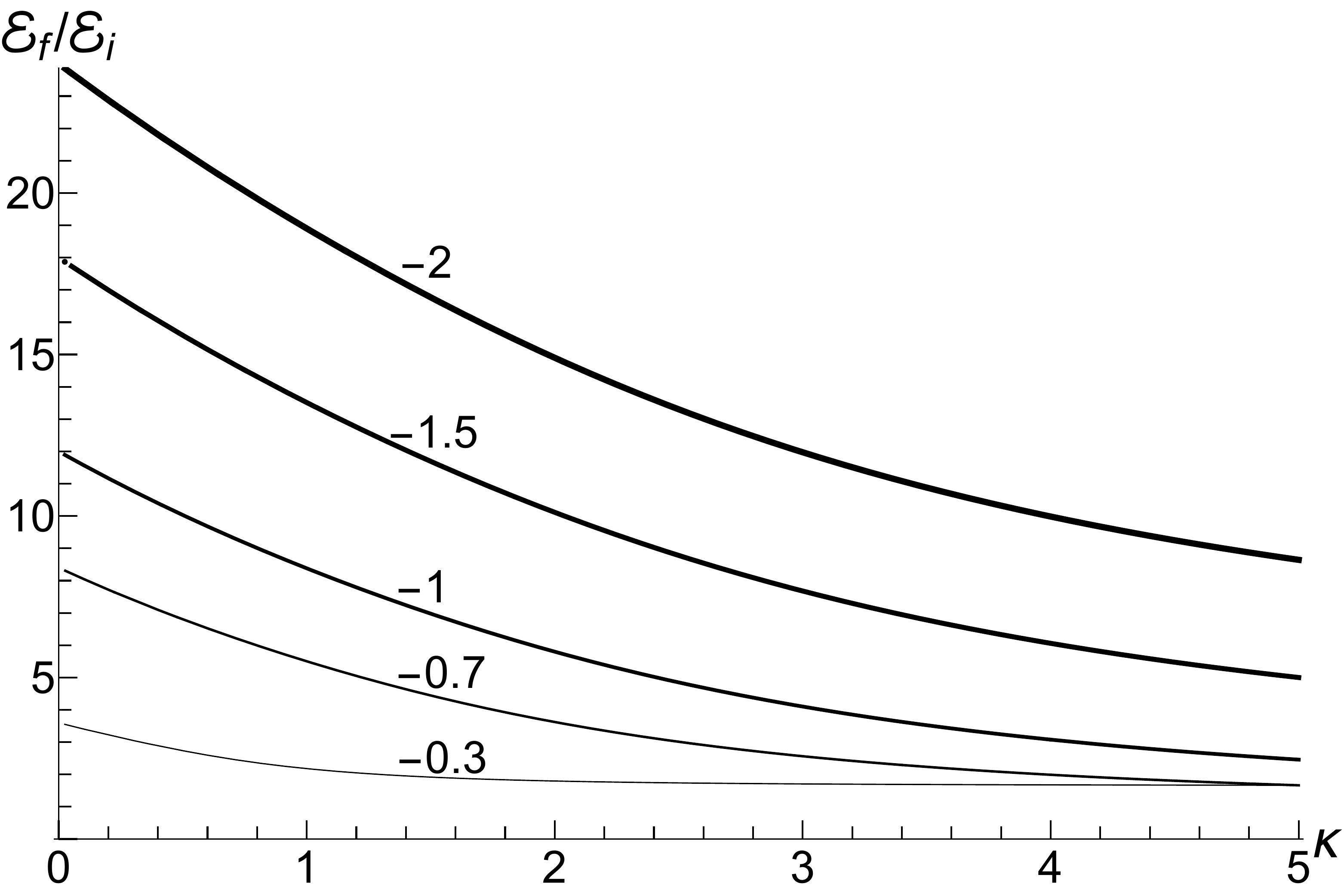}
\caption{\small The ratio ${\cal E}_f/{\cal E}_i$ versus parameter $\kappa$ for different negative values of the final frequency $\omega_{f}$
(shown nearby the respective lines) in the case of the Epstein--Eckart profile (\ref{omtanh}). 
The initial frequency is taken as $\omega_{i}=1$.
Left: $\rho=0$, $s_0\Upsilon=1$. Right: $\rho=1$, $s_0\Upsilon=10$.
 }
\label{fig-EfEi-k1Eps-}
\end{figure}

On the other hand, if $\omega_f >0$ and $\tilde\omega_f \gg 1$, then 
$d \approx 1/2 \pm i\left(\tilde\omega_i -\tilde\omega_f\right)$, and
the product of two Gamma functions in (\ref{modu-2}) takes the form
$\Gamma\left[1/2 + 2i\left(\tilde\omega_i -\tilde\omega_f\right) \right]\Gamma\left( 1/2 \right)$.
Then, using the consequence of the Stirling formula  \cite{PBM},
\[
|\Gamma(x+iy)|^2 \approx 2\pi |y|^{2x-1} e^{-\pi |y|}, \quad |y| \gg 1,
\label{Stirmod}
\]
we obtain
\[
|u_-|^2 \approx \exp\left(2\pi|\tilde\omega_i -\tilde\omega_f|\right)
\left[\sinh\left(2\pi\tilde\omega_i\right)\sinh\left(2\pi\tilde\omega_f\right)\right]^{-1}
\approx 2\exp\left[  -4\pi\,\mbox{min}\left(\tilde\omega_i, \tilde\omega_f\right)\right] \ll 1.
\]
In this case, we have the known adiabatic invariant ${\cal E}_f/{\cal E}_i \approx \omega_f/\omega_i$. 
 Figures \ref{fig-EfEi-wfEps}--\ref{fig-EfEi-k1Eps-}
 show the ratio ${\cal E}_f/{\cal E}_i$ for the same values of $\omega_f$ and $\kappa$ 
as in Figures \ref{fig-EfEi-wf}--\ref{fig-EfEi-k1-10}, for $\omega_i =1$, using Equations (\ref{EfEi-u-})  and (\ref{modu-2}). 

We see that Figures \ref{fig-EfEi-wfEps} and \ref{fig-EfEi-k1Eps} look  similar to Figures \ref{fig-EfEi-wf} and \ref{fig-EfEi-k1}.
Especially expressive is Figure \ref{fig-EfEi-wfEps} with straight lines in the adiabatic regime $\kappa =0.1$, but with different
inclinations for positive and negative values of the final frequency $\omega_f$.
On the other hand, Figures \ref{fig-EfEi-k1Eps-} and \ref{fig-EfEi-k1-10} for negative values of the final frequency $\omega_f$
are different: there are no oscillations for small values of $\kappa$ in Figure \ref{fig-EfEi-k1Eps-}, whereas such oscillations
are well pronounced in  Figure \ref{fig-EfEi-k1-10}.

\subsection{``Mild'' transition to the exponential decay on the semi-axis}
\label{sec-mild}

In all examples of the evolution starting at $t=0$, considered in the preceding sections, the frequency $\omega(t)$ had a discontinuity
of the derivative at the initial instant. This drawback can be removed for the time-dependent frequency 
\be
\omega_m(t) = \omega_i/\cosh(\kappa t), \qquad \omega_m^2(t) = \omega_i^2\left[1 -\tanh^2(\kappa t)\right].
\label{omcosh}
\ee
Note that $\omega_m(t) > \omega_i \exp(-\kappa t)$ for $t>0$ and $\omega_m(t) \approx 2\omega_i \exp(-\kappa t)$ for $\kappa t \gg 1$.

An example (\ref{vep-tanh-om0}) shows that the solution to Equation (\ref{eqvep}) with frequency $\omega_m(t)$ 
can be expressed in terms of $\tanh(\kappa t)$ in the special case when
$(\omega_i/\kappa)^2=2$. Therefore, it seems reasonable to introduce the new variable $\xi = \tanh(\kappa t)$. Using the transformation
of derivatives $d\psi/dt = \kappa\left(1-\xi^2\right)d\psi/d\xi$, one can transform Equation (\ref{eqvep}) with the
time-dependent frequency (\ref{omcosh}) to the {\em Legendre equation\/}
\be
\left(1-\xi^2\right)d^2\vep/d\xi^2 - 2\xi d\vep/d\xi + (\omega_i/\kappa)^2 \vep =0.
\label{Legeq}
\ee
Its general solution is a superposition of the {\em Legendre functions\/} of the first and second kind, $P_{\nu}(\xi)$
and $Q_{\nu}(\xi)$ \cite{Grad}
\be
\vep(t) = D_{p}P_{\nu}(\xi) + D_{q} Q_{\nu}(\xi), \qquad \nu = -1/2 + r, \qquad r = \sqrt{1/4 + (\omega_i/\kappa)^2}.
\label{vepgen-cosh}
\ee
[One can verify that the second solution of the equation $\nu(\nu+1) = (\omega_i/\kappa)^2$, $\nu = -1/2 - r$, 
results in the same expression (\ref{vepgen-cosh}) due to the properties of functions $P_{\nu}(\xi)$
and $Q_{\nu}(\xi)$].
Constant complex coefficients $D_{p}$ and $D_q$ are determined by the initial conditions (\ref{incondvep}). 
The following relations are useful for our purposes \cite{Grad} (remembering that $0\le \xi <1$):
\be
P_{\nu}(\xi) = F\left(-\nu, \nu+1; 1; \frac{1-\xi}{2}\right) = F\left( 1/2 - r, 1/2 + r;1; 
\frac{\exp(-\kappa t)}{2\cosh(\kappa t)}\right),
\label{Pnu-F}
\ee
\be
Q_{\nu}(\xi) = \frac{\pi}{2\sin(\nu\pi)}\left[\cos(\nu\pi)P_{\nu}(\xi) - P_{\nu}(-\xi)\right], \qquad \nu \neq 0, \pm 1, \pm 2, \ldots,
\label{Qnu-P}
\ee
\be
P_0(\xi) =1, \quad P_1(\xi) = \xi, \qquad Q_0(\xi) =  \frac{1}{2}\ln\left(\frac{1+\xi}{1-\xi}\right), \quad
Q_1(\xi) = \frac{\xi}{2}\ln\left(\frac{1+\xi}{1-\xi}\right) -1,
\label{P1Q1xi}
\ee
\be
P_{\nu}(0) = -\,\frac{\sin(\nu\pi)}{2\pi^{3/2}}\Gamma\left(\frac{\nu+1}{2}\right)\Gamma\left(-\,\frac{\nu}{2}\right), \qquad
Q_{\nu}(0) = \frac{1-\cos(\nu\pi)}{4\pi^{1/2}}\Gamma\left(\frac{\nu+1}{2}\right)\Gamma\left(-\,\frac{\nu}{2}\right), 
\label{P0Q0}
\ee
\be
\left(1-\xi^2\right)dP_{\nu}(\xi)/d\xi = (\nu+1)\left[\xi P_{\nu}(\xi) - P_{\nu+1}(\xi)\right], 
\label{Leg-derivP}
\ee
\be
\left(1-\xi^2\right)dQ_{\nu}(\xi)/d\xi = (\nu+1)\left[\xi Q_{\nu}(\xi) - Q_{\nu+1}(\xi)\right],
\label{Leg-deriv}
\ee
Using Equations (\ref{incondvep}), (\ref{vepgen-cosh}), (\ref{Leg-derivP}) and (\ref{Leg-deriv}), we find the coefficients
\be
D_p = \frac{(\nu+1)Q_{\nu+1}(0) + i\mu Q_{\nu}(0)}{\omega_i^{1/2}(\nu+1)\left[P_{\nu}(0)Q_{\nu+1}(0) -
Q_{\nu}(0)P_{\nu+1}(0) \right]},
\label{DpDq1a}
\ee
\be
D_q = -\,\frac{(\nu+1)P_{\nu+1}(0) + i\mu P_{\nu}(0)}{\omega_i^{1/2}(\nu+1)\left[P_{\nu}(0)Q_{\nu+1}(0) -
Q_{\nu}(0)P_{\nu+1}(0) \right]},
\label{DpDq1}
\ee
where $\mu= \omega_i/\kappa$.
Expressions in Equations (\ref{DpDq1a}) and (\ref{DpDq1}) can be simplified with the aid of Equation (\ref{P0Q0}) 
and the known formulas for the products
of Gamma-functions, such as $\Gamma(x)\Gamma(1-x) = \pi/\sin(\pi x)$ and  $\Gamma(x)\Gamma(-x) = -\pi/[x\sin(\pi x)]$.
Then, the following relation can be verified:
\[
(\nu+1)\left[P_{\nu}(0)Q_{\nu+1}(0) - Q_{\nu}(0)P_{\nu+1}(0) \right] = -1.
\]
Consequently,
\be
D_p = \sqrt{\pi/\omega_i}\left\{\cos(\nu\pi/2)
\frac{\Gamma\left[(\nu+2)/{2}\right]}{\Gamma\left[(\nu+1)/{2}\right]}
+i(\mu/\nu) \sin(\nu\pi/2)
\frac{\Gamma\left[(\nu+1)/{2}\right]}{\Gamma\left[{\nu}/{2}\right]}\right\},
\label{DpGam}
\ee
\be
D_q = \frac{2}{\sqrt{\pi\omega_i}}\left\{-\sin(\nu\pi/2)
\frac{\Gamma\left[(\nu+2)/{2}\right]}{\Gamma\left[(\nu+1)/{2}\right]}
+i(\mu/\nu) \cos(\nu\pi/2)
\frac{\Gamma\left[(\nu+1)/{2}\right]}{\Gamma\left[{\nu}/{2}\right]}\right\}.
\label{DqGam}
\ee
In the special case of $\mu =\sqrt{2}$, when $\nu=1$, Equations (\ref{vepgen-cosh}), (\ref{P1Q1xi}), (\ref{DpGam}) and (\ref{DqGam})
yield the solution (\ref{vep-tanh-om0}).

\subsubsection{Mean energy}
\label{mild-Energy}

\begin{figure}[htb]
\includegraphics[height=1.42truein,width=3.0truein,angle=0]{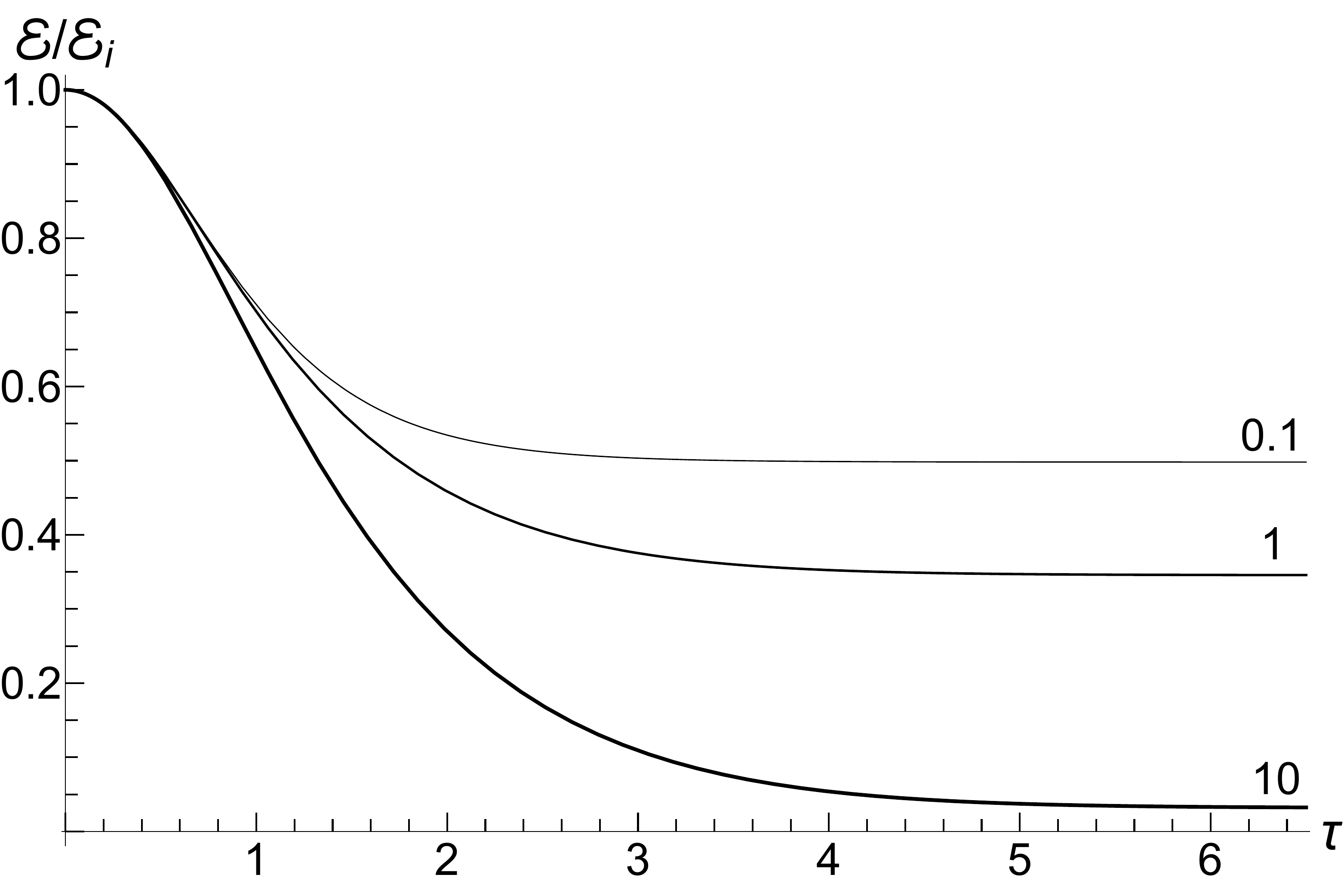}
\includegraphics[height=1.42truein,width=3.0truein,angle=0]{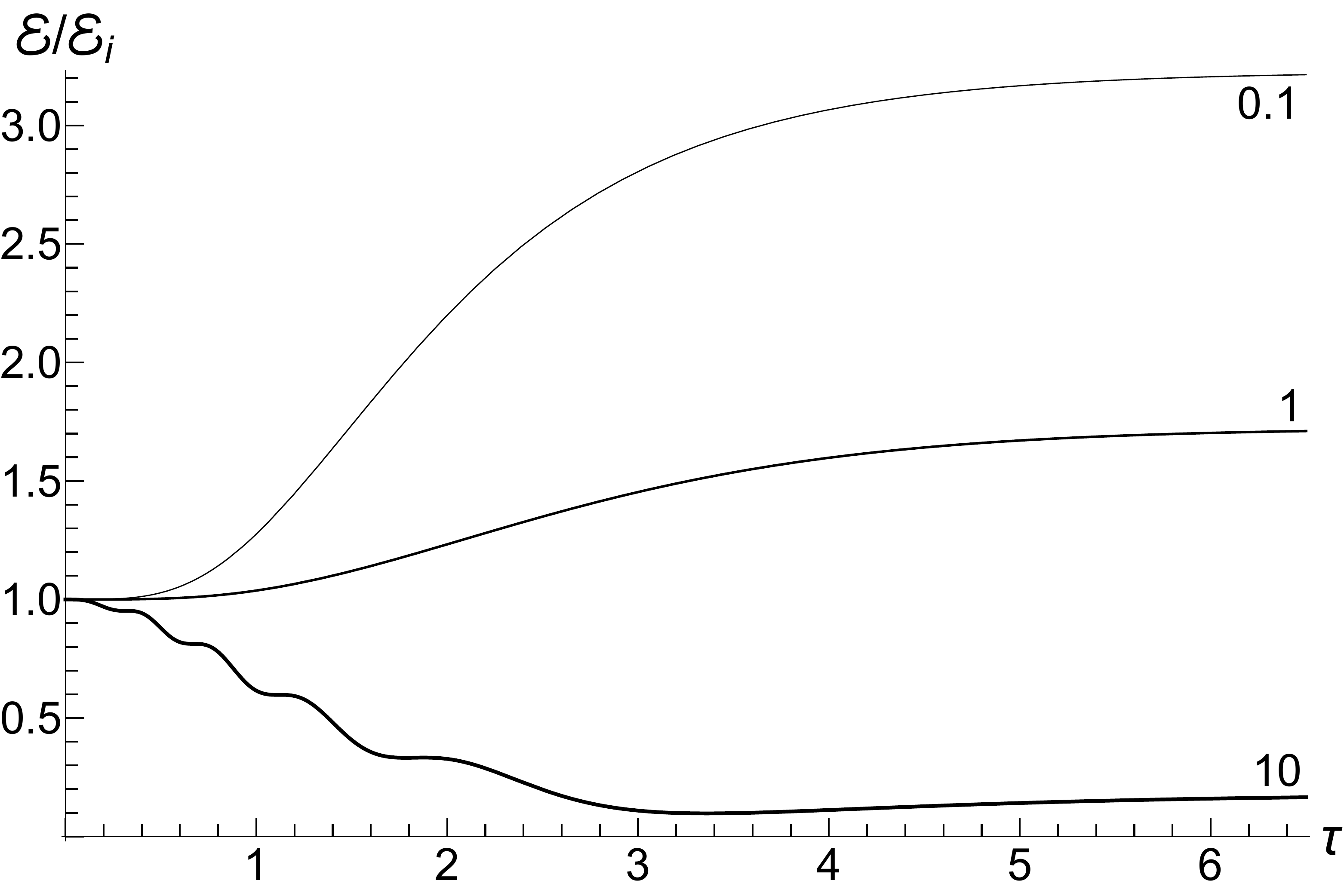}
\caption{\small The ratio ${\cal E}(\tau)/{\cal E}_i$ for the ``mild'' exponential frequency decay (\ref{omcosh})
with $\mu = 0.1, 1.0, 10,0$. Left: $\rho=0$, $s_0\Upsilon=1$. Right: $\rho=1$, $s_0\Upsilon=10$.
 }
\label{fig-Emild}
\end{figure}

The functions $F_{\pm}(\xi)$ determining the mean energy in accordance with Equations (\ref{def-Fpm}) and (\ref{Eqfin}), can be written
as follows,
\beqn
F_{\pm}(\xi) &=& \omega_i \sqrt{1-\xi^2} \left[ D_{p}P_{\nu}(\xi) + D_{q} Q_{\nu}(\xi)\right]
\nonumber \\ &&
 \pm i \kappa(\nu+1)\left\{ D_p \left[\xi P_{\nu}(\xi) -P_{\nu+1}(\xi)\right] 
+ D_q \left[\xi Q_{\nu}(\xi) -Q_{\nu+1}(\xi)\right]
\right\}.
\label{F+-xi}
\eeqn
Figure \ref{fig-Emild} shows the evolution of the ratio ${\cal E}(\tau)/{\cal E}_i$ 
in the low- and high-temperature regimes.
Pay attention to small oscillations for $\mu=10$ in the right plot. They arise due to oscillatory nature of functions $P_{\nu}(\xi)$
and $Q_{\nu}(\xi)$ with big values of index $\nu$ (remember that $P_{\nu}(\xi)$ is the Legendre polynomial if $\nu$ is an integer).
These oscillations are suppressed in the low-temperature regime, but the high value of parameter $\Upsilon$ amplifies the 
oscillations during the initial stage of the evolution. 

\begin{figure}[htb]
\includegraphics[height=2.32truein,width=3.0truein,angle=0]{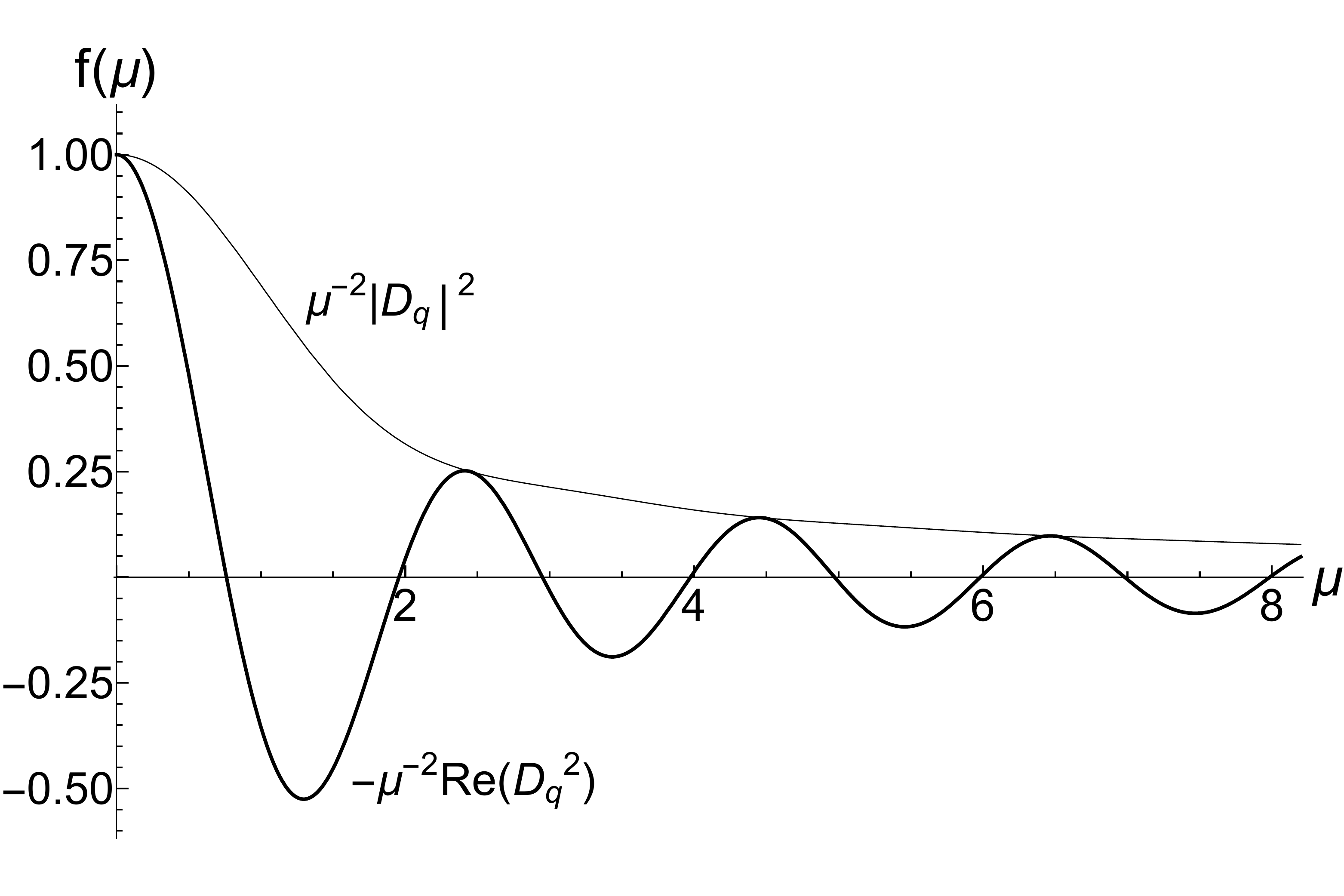}
\includegraphics[height=2.32truein,width=3.0truein,angle=0]{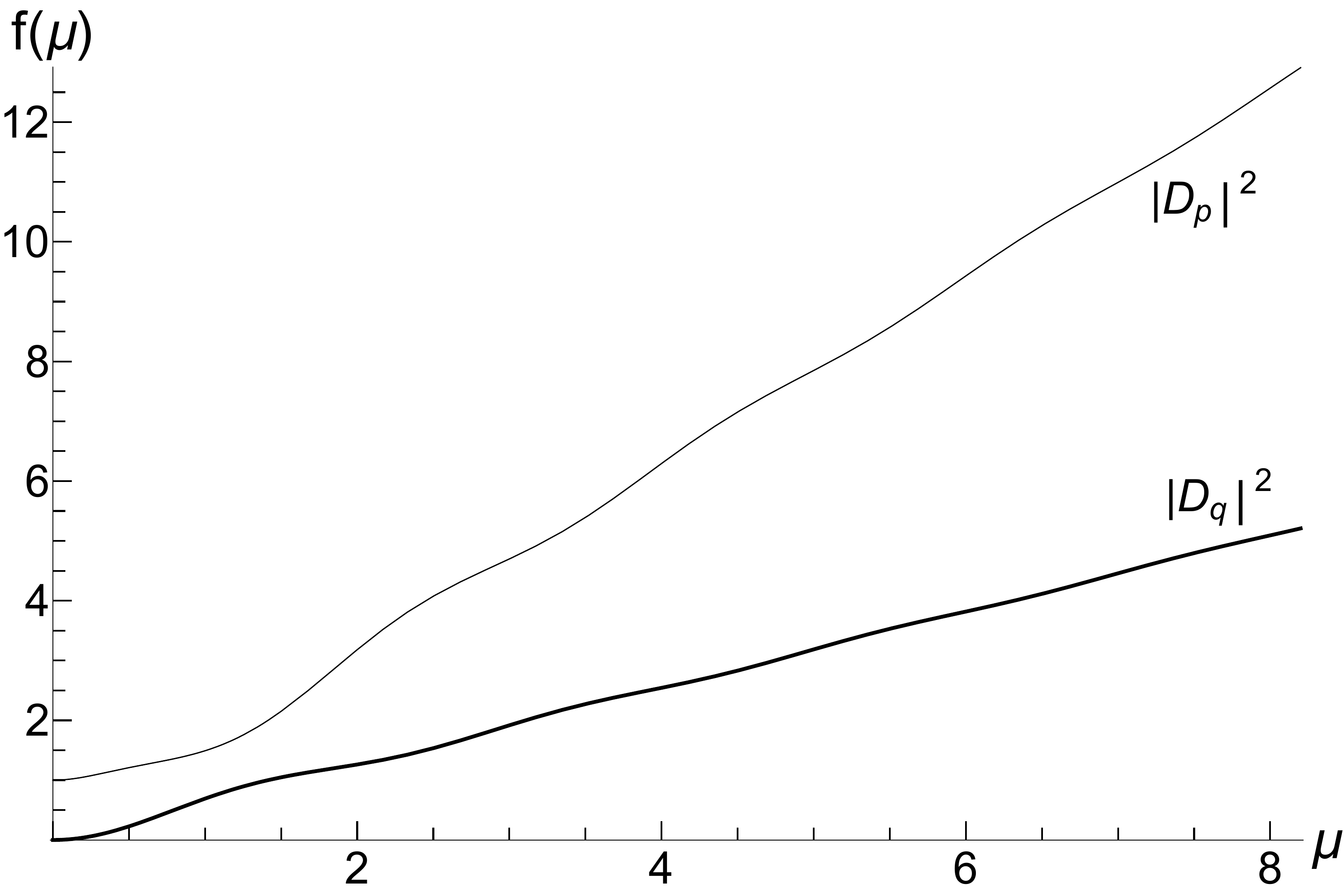}
\caption{\small Left: functions $\mu^{-2}|D_q(\mu)|^2$ and $- \mu^{-2} \mbox{Re}\left[D_q^2(\mu)\right]$ with $\omega_i=1$.
Right: functions $|D_q(\mu)|^2$ and $|D_p(\mu)|^2$.
 }
\label{fig-Dq-mod}
\end{figure}

 The asymptotic value of the mean energy at $t\to\infty$ is determined by the limit values $F_{\pm}(1)$.
 Since $P_{\nu}(1) = 1$ for any value $\nu$, the coefficient $D_p$ does not contribute to these limit values:
\[
F_{\pm}(1) = \omega_i D_{q} \lim_{\xi \to 1}\left\{\sqrt{1-\xi^2} Q_{\nu}(\xi) \pm i \mu^{-1}(\nu+1)
\left[\xi Q_{\nu}(\xi) -Q_{\nu+1}(\xi)\right]\right\}.
\]
The following representation of function $Q_{\nu}(\xi)$ is useful here (see, e.g., Section 3.6.1 in Ref. \cite{BE}):
\[
Q_{\nu}(\xi) =  P_{\nu}(\xi)\left[ \frac12\ln\left(\frac{1+\xi}{1-\xi}\right) -\gamma - \psi(\nu+1) \right]
+\sum_{l=1}^{\infty} c_l (1-\xi)^l,
\]
where $\gamma$ is the Euler constant and $\psi(z) = d\ln[\Gamma(z)]/dz$ is the logarithmic derivative of the Gamma-function.
The explicit form of coefficients $c_l$ is not important for our purpose, as soon as the last series goes to zero for $\xi=1$.
Since the divergence of function $Q_{\nu}(\xi)$ at $\xi=1$ is only logarithmic, 
$\lim_{\xi \to 1}\left[\sqrt{1-\xi^2} Q_{\nu}(\xi)\right] =0$.
Then, using the relation $\psi(1+z) - \psi(z) = 1/z$ (see, e.g., Equation 1.7(8) from \cite{BE}), we arrive at the simple formula
$ F_{\pm}(1) = \pm i \kappa D_q$. Hence, the final mean energy equals [see Equation (\ref{Eqfin})]
\be
{\cal E}_f = \frac{\omega_i {\cal E}_i}{4\mu^2}\left[ |D_q|^2 \left(1 + s_0\Upsilon \right) - 2 \rho \mbox{Re}\left(D_q^2\right)\right]
\label{Ff-mild}
\ee
 Plots of functions $\mu^{-2}|D_q(\mu)|^2$ and $- \mu^{-2} \mbox{Re}\left[D_q^2(\mu)\right]$ are shown in Figure \ref{fig-Dq-mod}
(assuming $\omega_i=1$).

For $\mu \ll 1$ we have $\nu \approx \mu^2$ and $D_q \approx i \mu/\sqrt{\omega_i}$. Then, Equation (\ref{Ff-mild}) goes to the sudden jump
formula (\ref{EfEi-sudden0}). To see the dynamics of the ``fast jump'', we can approximate Equation (\ref{F+-xi}), 
taking $D_p\approx 1/\sqrt{\omega_i}$ and replacing
functions $P_{\nu}(\xi)$ and  $Q_{\nu}(\xi)$ with $P_0(\xi)$ and  $Q_0(\xi)$ from Equation (\ref{P1Q1xi}). Then, we obtain
$F_{\pm}(\xi) \approx \omega_i^{1/2}\left[1/\cosh(\tau) \mp 1\right]$ and
\be
{\cal E}(\tau)/{\cal E}_i = \frac14\left\{ \left[1 + 1/\cosh(\tau) \right]^2 + s_0\Upsilon \left[1 -1/\cosh(\tau) \right]^2
+ 2\rho \tanh^2(\tau)\right\}, \qquad \tau \equiv \kappa t.
\label{Efratio-tau-0}
\ee

In the ``adiabatic'' limit $\mu \gg 1$ we have $\nu \approx \mu -1/2$. Then, using the Stirling formula for the Gamma-functions,
we find $D_q \approx i \exp(i\nu\pi/2)\sqrt{2\nu/(\pi\omega_i)}$ and $D_p \approx \exp(i\nu\pi/2)\sqrt{\pi\nu/(2\omega_i)}$. 
The final energy is very close to that given by 
Equation (\ref{EfEi-om0adiab}), but the frequency of oscillations is different:
\be
{\cal E}_f /{\cal E}_i \approx  \kappa \left[ 1 + s_0\Upsilon  + 
2\rho \sin(\pi\omega_i/\kappa)\right]/(2\pi \omega_i).
\label{EfEi-om0adiab-mild}
\ee

\subsubsection{Mean magnetic moment}
\label{sec-MMM-mu0}

\begin{figure}[htb]
\includegraphics[height=2.32truein,width=3.0truein,angle=0]{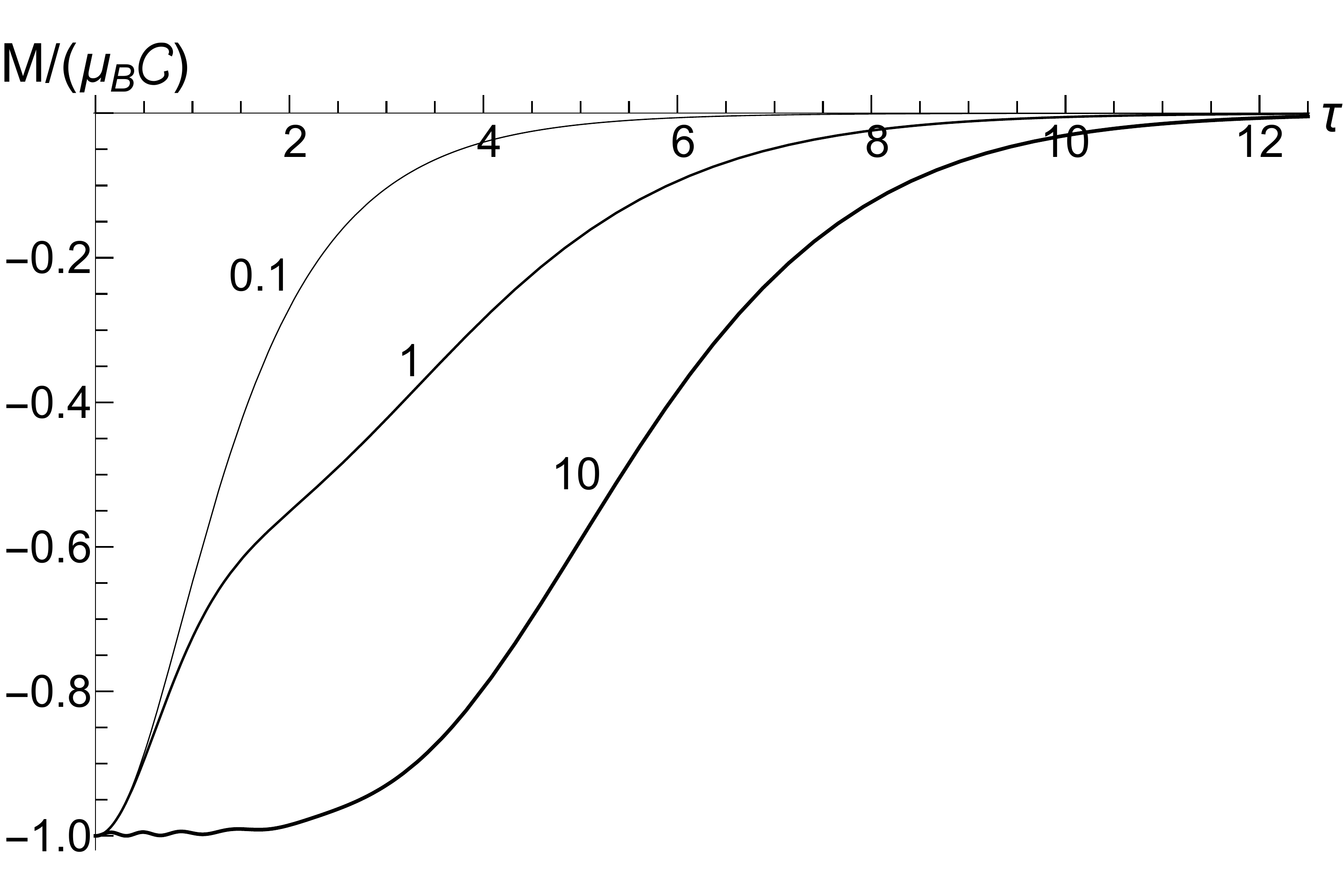}
\includegraphics[height=2.32truein,width=3.0truein,angle=0]{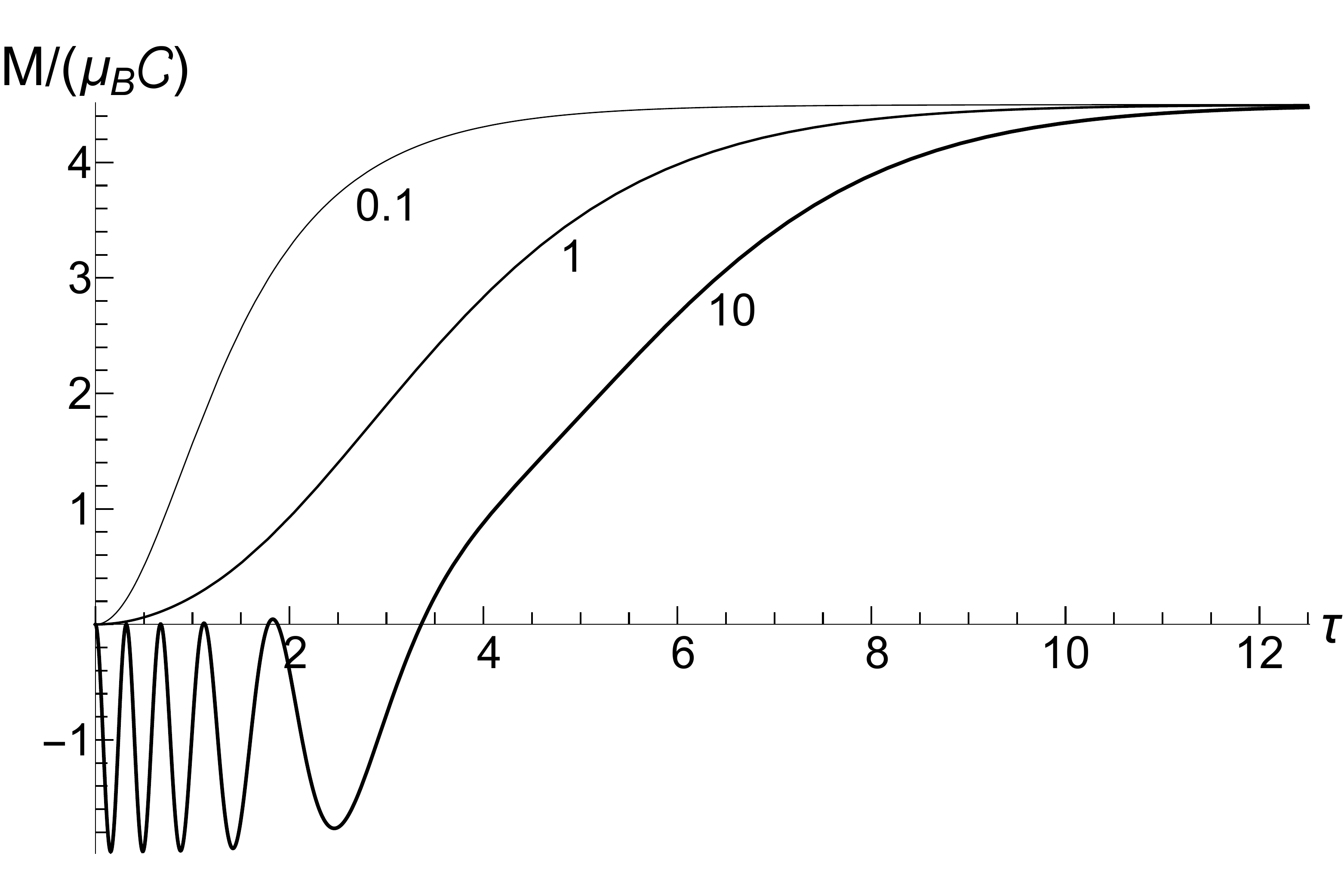}
\caption{\small The normalized mean magnetic moment as function of dimensionless time $\tau= \kappa t$ 
for the ``mild'' exponential frequency decay (\ref{omcosh})
with $\mu = 0.1, 1.0, 10,0$. Left: $\rho=0$, $s_0\Upsilon=1$. Right: $\rho=1$, $s_0\Upsilon=10$.
 }
\label{fig-Mmild}
\end{figure}

Figure \ref{fig-Mmild} shows the evolution of the mean magnetic moment as function of dimensionless time $\tau= \kappa t$,
calculated in accordance with Equation (\ref{meanmag}).
The oscillations at the initial stage of the evolution are distinctly pronounced here.
%
One can verify that the product $\sqrt{\omega(t)}\vep(t) = \sqrt{\omega_i}\left(1-\xi^2\right)^{1/4}\vep(\xi)$ goes to zero as
$t \to \infty$ (or $\xi \to 1$) for any value of parameter $\nu$ (because the divergence of function $Q_{\nu}(\xi)$ at $\xi \to 1$
is only logarithmic). Consequently, Equation (\ref{meanmag}) results in the asymptotic value of the magnetic moment
(\ref{M-inf-circ-exp}) for all values of the ratio $\kappa/\omega_i$.
Function $\vep(t)$ (\ref{vepgen-cosh}) in the case of $\mu \ll 1$ has the form 
$\vep = \omega_i^{-1/2}\left( 1 + i\mu\tau \right) = \omega_i^{-1/2}\left( 1 + i\omega_i t \right)$.
However, the term $(\mu\tau)^2$ can be neglected in the formula for $\omega(\tau)|\vep(\tau)|^2$ when $\mu \ll 1$
(due to the exponential decrease of the frequency). Hence, formula (\ref{meanmag}) for the time-dependent mean magnetic moment
assumes the form
\be
{\cal M}(t)  = 
 -(\mu_B {\cal C}/2)\PH{1/\cosh(\tau) +1 + \Upsilon s_0 \PG{1/\cosh(\tau) -1}
-2\rho /\cosh(\tau)}.
\label{meanmag-mild}
\ee

\section{Landau gauge: general formulas}
\label{sec-Land}

For $\alpha=1$, the set of equations (\ref{dotx})-(\ref{dotpy}) takes the form
\[
 \dot{x} = p_x/m +\Omega(t)y, \quad
 \dot{y} = p_y/m ,
 \qquad
\dot{p}_x =0, \quad
\dot{p}_y = -\Omega(t)p_x -m\Omega^2(t) y, 
\]
where $\Omega(t)$ is the {\em cyclotron frequency}. Hence, $p_x = const$, and we arrive at the 
inhomogeneous equation
\be
\ddot{y} + \Omega^2(t) y = -\Omega(t)p_x/m.
\label{inhom}
\ee
 Therefore, all solutions can be expressed in terms of complex functions $\vep(t)$ and 
$\vep^*(t)$, satisfying Equation (\ref{eqvep}) with $\alpha=1$ and the condition (\ref{incondvep}). 
However, due to the presence of function $\Omega(t)$ in the right-hand side of Equation (\ref{inhom}),
the solutions to the complete set of equations contain three additional functions \cite{DMM72}:
\be
\sigma(t) = \int_{t_0}^t\Omega(\tau)\vep(\tau)d\tau = -\int_{t_0}^t\ddot\vep(\tau)/[\Omega(\tau)]d\tau 
= -\,\frac{\dot\vep(t)}{\Omega(t)} + \frac{i}{\sqrt{\Omega_i}} 
-\int_{t_0}^t \frac{\dot\vep(\tau)\dot\Omega(\tau)}{\Omega^2(\tau)}\,d\tau,
\label{defsig1}
\ee
\be
 S(t) = \mbox{Im}(\vep\sigma^*), \quad
\chi(t) = \int_{t_0}^t[1- \Omega(\tau)S(\tau)]d\tau ,
\label{defskap}
\ee
where $t_0$ is the time instant when the frequency $\Omega$ starts to vary (so that $\Omega(t) \equiv \Omega_i$ for $t \le t_0$).
Functions $\vep(t)$ and $\sigma(t)$ are complex, whereas functions $S(t)$ and $\chi(t)$ are real.
After some straightforward
algebra, one can obtain the following form of matrix $\Lambda_Q$ in Equation (\ref{LamQ}):
%
\be
\Lambda_Q(t) = \left\Vert
\begin{array}{cccc}
1 & \sqrt{\Omega_i}\,\mbox{Re}(\sigma) & {\chi}/{m} & {\mbox{Im}(\sigma)}/(m\sqrt{\Omega_i})
\\ \vspace{1mm}
0 & \sqrt{\Omega_i}\,\mbox{Re}(\vep) & -{S}/{m} & {\mbox{Im}(\vep)}/(m\sqrt{\Omega_i})
\\ \vspace{1mm}
0 & 0 & 1 & 0
\\ \vspace{1mm}
0 & m\sqrt{\Omega_i}\,\mbox{Re}(\dot\vep) & -\dot{S} & {\mbox{Im}(\dot\vep)}/\sqrt{\Omega_i}
\end{array} \right\Vert.
\ee
The transformation (\ref{LamUQU}) yields the final matrix $\Lambda_q(t)$. Writing it in the same form as in Equation (\ref{Lamq-0}),
 we find the following expressions for the $2\times2$ blocks:
\be
\lambda_1 = \frac{\sqrt{\Omega_i}}{\Omega(t)} \left\Vert
\begin{array}{cc}
\mbox{Im}(\dot\vep) & - \mbox{Re}(\dot\vep)
\\
-\Omega(t)\mbox{Im}(\vep) & \Omega(t) \mbox{Re}(\vep)
\end{array} \right\Vert,
\quad
\lambda_2 = \frac{\sqrt{\Omega_i}}{\Omega(t)}\left\Vert
\begin{array}{cc}
0 & - \mbox{Re}\left(\dot\vep +\sqrt{\Omega_i}\,\dot{S}\right)
\\
0 &  \mbox{Re}\left(\Omega\vep - \sqrt{\Omega_i}\,\dot{\chi}\right)
\end{array} \right\Vert,
\label{lam2Land}
\ee
\be
\lambda_3 = \left\Vert
\begin{array}{cc}
1 - \displaystyle{\frac{\sqrt{\Omega_i}}{\Omega(t)}}\mbox{Im}(\Omega\sigma + \dot\vep) & 
\displaystyle{\frac{\sqrt{\Omega_i}}{\Omega(t)}}\mbox{Re}(\Omega\sigma + \dot\vep)
\\
0 &  0
\end{array} \right\Vert,
\label{lam3Land}
\ee
\be
\lambda_4 = \left\Vert
\begin{array}{cc}
1  & \sqrt{\Omega_i}\,\mbox{Re}(\sigma + \dot\vep/\Omega) -\Omega_i (\chi - \dot{S}/\Omega)
\\
0 &  \Omega_i/\Omega
\end{array} \right\Vert.
\label{lam4Land}
\ee

\subsection{Mean energy}

The mean energy can be written as follows.
\be
{\cal E}(t) = {\cal E}_i/(2\Omega_i)\left[K_{\Omega}(t) + s^{-1}\Upsilon K_{Y}(t) - 2\rho K_{\rho}(t)\right], \quad
{\cal E}_i = m\Omega_i^2 G,
\label{EqfinLand}
\ee
\be
K_{\Omega}(t) = |\dot\vep|^2 + \Omega^2(t)|\vep|^2, \quad
K_{Y}(t) = U^2(t) + V^2(t), \quad
K_{\rho}(t) = \mbox{Re}(\dot\vep) U(t) + \Omega(t) \mbox{Re}(\vep) V(t),
\label{KY}
\ee
\be
V(t) = \Omega\mbox{Re}(\vep) -\sqrt{\Omega_i}\,\dot\chi , \quad
U(t) = \mbox{Re}(\dot\vep) +\sqrt{\Omega_i}\,\dot{S} .
\label{UV}
\ee

\subsubsection{Adiabatic evolution}

In the adiabatic approximation, one can use the  solution
\be
\vep(t) \approx [\Omega(t)]^{-1/2}\exp[i\phi(t)], \quad \dot\vep(t) \approx i \Omega(t)\vep(t), \quad
\phi(t) = \int_0^t \Omega(\tau)d\tau
\label{vepad}
\ee
Then, neglecting the derivative $d\Omega/dt$  in (\ref{defsig1}) and other formulas, one can write
\[
\sigma(t) \approx -i\vep(t) + \frac{i}{\sqrt{\Omega_i}}, \qquad
S(t) \approx \frac1{\Omega(t)} -\frac{\cos(\phi)}{\sqrt{\Omega(t)\Omega_i}}, 
\]
\[
\chi(t) \approx \frac{\sin(\phi)}{\sqrt{\Omega(t)\Omega_i}}, \qquad
\left(
\begin{array}{c}
\dot{S} \\
\dot\chi
\end{array} \right)
 \approx 
 \sqrt{\frac{\Omega(t)}{\Omega_i}}
 \left(
\begin{array}{c}
 \sin(\phi) \\
 \cos(\phi) 
 \end{array}
 \right),
\]
 so that $U(t)= V(t) =0$. Hence, $K_{Y}(t) = K_{\rho}(t) =0$  and ${\cal E}(t) ={\cal E}_i \Omega(t)/\Omega_i$. 
 This means that  the energy variation does not depend on the choice 
 of the gauge in the adiabatic approximation, provided the frequency $\Omega(t)$ does not pass through zero value, when the
 approximation (\ref{vepad}) fails. 

\subsubsection{Non-adiabatic evolution}
 
 But the results in the cases of $\alpha=0$ and $\alpha=1$ are different for non-adiabatic variations of $\Omega(t)$.
One of the reasons is the necessity to know, in the asymptotic regime $t>T$, in addition to two complex dimensionless coefficients
$u_{\pm}$ in Equation (\ref{uvsol}) (where  $\omega_{f}$ must be replaced with $\Omega_f$), 
the third  complex constant dimensionless coefficient $u_{\sigma}$, describing the behavior of the function $\sigma(t)$ for $t>T$:
\be
\sigma(t) = -\,\frac{\dot\vep(t)}{\Omega_f} + \frac{u_{\sigma}}{\sqrt{\Omega_i}}, 
\quad u_{\sigma}= {i} - \sqrt{\Omega_i} \int_{t_0}^T \frac{\dot\vep(\tau)\dot\Omega(\tau)}{\Omega^2(\tau)}\,d\tau.
\label{assig}
\ee
Then, 
\[S(t) = \Omega_f^{-1} + \mbox{Im}[u_{\sigma}^*\vep(t)]/\sqrt{\Omega_i}, \qquad 
\dot\chi(t) = -\Omega_f \mbox{Im}[u_{\sigma}^*\vep(t)]/\sqrt{\Omega_i}, \qquad t >T. 
\]
The final mean energy ratio equals
\be
{\cal E}_f/{\cal E}_i = (|\Omega_f|/\Omega_i)\left\{1 + 2|u_{-}|^2
+s^{-1}\Upsilon \left[ \left(|a|^2 + |b|^2\right)/2 +  \mbox{Im}(ba^*)\right]
-\rho \left[ |a|^2  +  \mbox{Im}(ba^*)\right] \right\},
\label{asskap}
\ee
where
\be
a= u_{+} + u_{-}^*, \quad b = u_{+}u_{\sigma}^* - u_{-}^* u_{\sigma}.
\label{ab}
\ee

\subsubsection{Sudden jump}
\label{sec-sudj}

In the case of instantaneous jump, the coefficients $u_{\pm}$ are real: see Equation (\ref{upmjump}). 
Calculating  the first integral in (\ref{defsig1}) 
with $\vep(t)$ given by (\ref{uvsol}), we obtain the pure imaginary coefficient
$u_{\sigma}=i\Omega_i/\Omega_f$ (note that its sign depends on the sign
of magnetic field). Then, formula (\ref{asskap})  results in the relation
(which holds for positive and negative values of the final frequency $\Omega_f$)
\be
{\cal E}_f/{\cal E}_{i} =  \left[ \Omega_i^2 + \Omega_f^2 + s^{-1}\Upsilon(\Omega_i -\Omega_f)^2
+ 2\rho\Omega_f(\Omega_i -\Omega_f)\right]/(2\Omega_i^2).
\label{EqfLand}
\ee
It differs from Equation (\ref{Eqfcirc2}) for the circular gauge. 
In particular, ${\cal E}_f/{\cal E}_{i} = (1+ s^{-1}\Upsilon)/2$ when
$\Omega_f \to 0$. 
The same result can be obtained directly from formulas (\ref{EqfinLand})-(\ref{UV}), if one uses solution
(\ref{vep-om0}) (with $\Omega$ instead of $\omega$) and its consequencies: $K_{\Omega} =V^2 =\Omega_i$ and $U=0$.
This means that the energy does not change if  $s^{-1}\Upsilon =1$ and $\Omega_f = 0$
(for any value of parameter $\rho$).
 The final energy for $\Omega_f = -\Omega_i$ is also  different from the case of
$\alpha=0$: ${\cal E}_{f} /{\cal E}_i = 1 + 2s^{-1}\Upsilon -2\rho$.

\subsubsection{Parametric resonance}

The parametric resonance occurs now at the {\em twice cyclotron frequency\/} $2\Omega_i$.
Therefore, one should replace $\omega \to \Omega$ in Equation (\ref{upm-param}) and
 calculate the functions $\dot\vep$, $\sigma$, $S$ and $\chi$, assuming $u_{\pm}$ as constant
coefficients (but remembering that $u_{-}$ is pure imaginary now). Then, $u_{\sigma}=i(u_{+} -u_{-})$ and $b=-i$.
In this case, Equation (\ref{asskap}) leads to the formula
\be
{\cal E}(t)/{\cal E}_i = \cosh(2\Omega_{i}\gamma t) + s^{-1}\Upsilon\left[\cosh^2(\Omega_{i}\gamma t) -\cosh(\Omega_{i}\gamma t)\right]
-\rho \left[\cosh(2\Omega_{i}\gamma t) -\cosh(\Omega_{i}\gamma t)\right]. 
\label{EqresLan}
\ee
This formula is different from (\ref{Eqres0}), even in the low-temperature case. 

\subsection{Mean magnetic moment}

Using Equations (\ref{Lkin}), (\ref{M2}), (\ref{sigr}), (\ref{sigrc}), (\ref{lam2Land})--(\ref{lam4Land}),
we can write the mean magnetic moment as
\be
{\cal M}(t)= -\,\frac{\mu_B{\cal C}}{2\sqrt{\Omega_i}}\PG{S_{\Omega}(t)+ s^{-1}\Upsilon\sqrt{\Omega_i}S_{Y}(t)-\rho S_{\rho}(t)},
\label{mean-M-Land}
\ee
where
\be
S_{\Omega}(t)=\mbox{Im}(\dot\vep)+\sqrt{\Omega_i}\PG{\Omega(t) |\vep|^2-\mbox{Re}(\dot\vep\sigma^*)}, 
\label{SOm-gen}
\ee
\be
S_{Y}(t)= \sqrt{\Omega_i}N(t)+M(t) +\Omega_i \PC{ \chi \dot{S} - S\dot{\chi}}, \qquad
S_{\rho}(t)=\mbox{Im}(\dot\vep)+\Omega_i N(t)+2\sqrt{\Omega_i}M(t),
\ee
\be
N(t)=\PC{1-2\dot\chi}\mbox{Re}(\vep)+\chi \mbox{Re}(\dot\vep)-\dot{S}\mbox{Re}(\sigma), \qquad
M(t)=\Omega(t)\mbox{Re}^2(\vep)-\mbox{Re}(\dot\vep)\mbox{Re}(\sigma).
\label{NM-gen}
\ee
The approximate solution (\ref{vepad}) results in the following formula in the adiabatic case [when $\Omega(t)>0$]: 
\be
{\cal M}_{ad}(t)=\mu_B  {\cal C}\PG{\frac{\rho[\Omega(t)+\Omega_i]\cos(\varphi)}{2\sqrt{\Omega(t)\Omega_i}}-1}.
\label{MadiabLand}
\ee
It is different from (\ref{Madiab}), because it means the divergent magnetic moment when $\Omega(t) \to 0$:
\be
{\cal M}_{ad}(t) \approx \mu_B  {\cal C}\PG{\frac{\rho\Omega_i\cos(\varphi)}{2\sqrt{\Omega(t)\Omega_i}}-1}.
\label{MadiabLand-0}
\ee
One can doubt in formula (\ref{MadiabLand-0}), because solution (\ref{vepad}) is not justified when $\Omega(t) \approx 0$.
However, the exact solution in the inverse linear decay case (Section \ref{sec-slow-Land-invlin}) 
leads to formula (\ref{M-ad-Land-invlin}) coinciding with (\ref{MadiabLand-0}).

The explicit form of coefficients (\ref{SOm-gen})-(\ref{NM-gen}) is given in Appendix \ref{ap-SOm-as}.
They lead to the following simple formula in the case of sudden jump of magnetic field
(note that it is valid for positive as well as negative values of $\Omega_f$):
\beqn
{\cal M}(t) &=& \frac{(-\mu_B  {\cal C})}{2\Omega_f\Omega_i}
\left\{\Omega_f^2+\Omega_i^2+s^{-1}\Upsilon\PC{\Omega_f \!-\! \Omega_i}^2-2\rho\Omega_f\PC{\Omega_f \!-\! \Omega_i}
\right. \nonumber \\  && \left.
+\Omega_i\cos(\Omega_f t)\PG{(\Omega_f \!-\! \Omega_i)(1 \!+\! s^{-1}\Upsilon)-2\rho\Omega_f} \right\}.
\eeqn
For $\Omega_f=0$ and isotropic initial traps ($s=1$) the result coincides with the circular gauge formula (\ref{Mjump-om0}):
\be
{\cal M}_f = \mu_B  {\cal C} \left(s^{-1}\Upsilon -1 \right)/2.
\ee
Here, we see an important role of the asymmetry parameter $s$. If $s\le 1$, them ${\cal M}_f$ is always positive. 
However, ${\cal M}_f$ can be negative if $s\gg 1$, even in the high-temperature case.

The behavior of the mean magnetic moment after the sudden inversion of magnetic field is quite different now from
that given by formula (\ref{M-circ-sudinv}) for the circular gauge:
\be
{\cal M}(t)= \mu_B  {\cal C}\left[s^{-1}\Upsilon + 2\left(1+s^{-1}\Upsilon -\rho\right)\sin^2(\Omega_i t)\right].
\label{M-Land-sudinv}
\ee
In particular, the ratio $R \equiv {|\widetilde{\Delta {\cal M}}|}/{|\langle\langle {\cal M}\rangle\rangle|}$ varies 
between $2/3$ at zero temperature and $1/2$ in the high-temperature regime (if $s=1$).

In the case of parametric resonance, we have the following explicit expressions for the functions  determining the
evolution of mean magnetic moment:
\beqnn
{\sqrt{\Omega_i}}\sigma &=& \PG{\cos(\Omega_i t)-1}\sinh(\Omega_i \gamma t)+\sin(\Omega_i t)\cosh(\Omega_i \gamma t)
\\ &&
-i\left\{\PG{\cos(\Omega_i t)-1}\cosh(\Omega_i \gamma t)+\sin(\Omega_i t)\sinh(\Omega_i \gamma t)\right\},
\eeqnn
\[
S=[1-\cos(\Omega_i t)]/{\Omega_i}, \qquad \dot{S}=\sin(\omega_i t), \qquad \chi= \sin(\Omega_i t)/{\Omega_i}, 
\qquad \dot{\chi}=\cos(\Omega_i t),
\]
\[
S_\Omega=-\sqrt{\Omega_i}\PH{\sin(\Omega_it)S_2(t) +\cos(\Omega_it)C_2(t)-2\cosh(2\Omega_i \gamma t)},
\]
\[
S_Y=\PG{2-\cos(\Omega_i t)}C_1(t)-\sin(\Omega_i t)\sinh(2\Omega_i \gamma t)/2,
\]
\[
S_\rho=-\sqrt{\Omega_i}\PH{\sin(\Omega_it)S_2(t) +2\cos(\Omega_i t)\PG{C_1(t) -1} -2C_2(t)},
\]
where
\[
S_2(t) = \sinh(2\Omega_i \gamma t)-\sinh(\Omega_i \gamma t), \quad
C_2(t) = \cosh(2\Omega_i \gamma t)-\cosh(\Omega_i \gamma t), 
\]
\[
C_1(t) = \cosh^2(\Omega_i \gamma t)-\cosh(\Omega_i \gamma t).
\]

Formulas describing quantum fluctuations of the energy and magnetic moment are very cumbersome for the Landau gauge. For this reason
we do not bring them here.

\section{Landau gauge: explicit examples}
\label{sec-Landexp}

\subsection{Inverse linear decrease of magnetic field}
\label{sec-1t-Land}

Explicit expressions for the functions $\vep(t), \sigma(t), S(t)$ and $\chi(t)$ can be obtained for $\Omega(t)=\Omega_0/\tau$, where
the same notation is used as in Section \ref{sec-om1t}, with the replacement $\omega \to \Omega$. 
Using the solution (\ref{vep-tau}), one can obtain the following 
explicit expression for the function $\sigma(t)$: 
\be
\sigma(t)=\frac{\sqrt{\tau}\PG{\tau^{-r}\PC{2r+1}^2-\tau^r\PC{2r-1}^2}-8r}{8ur\sqrt{\Omega_0}}
+i\frac{4r-\sqrt{\tau}\PG{\tau^{-r}\PC{2r+1}+\tau^r\PC{2r-1}}}{4r\sqrt{\Omega_0}}.
\ee
The expressions for functions $S(t)$ and $\chi(t)$ are different for real and imaginary values of coefficient 
$r=\sqrt{1/4 -u^2} $. 

\subsubsection{Fast variations}

In the case of $u<1/2$ we have
\[
S(t) = \frac{\sqrt{\tau}}{4r\Omega_0}\PG{4r\sqrt{\tau}-\tau^{-r}\PC{2r-1}-\tau^r\PC{2r+1}}, 
\quad
\dot{S}(t)=\frac{8r\sqrt{\tau}+\tau^{-r}\PC{2r-1}^2-\tau^r\PC{2r+1}^2}{8ur\sqrt{\tau}},
\]
\[
\chi(t) =\frac{t_0\sqrt{\tau}}{2r}\PC{\tau^r-\tau^{-r}},
\qquad
\dot{\chi}(t)=\frac{\tau^{-r}\PC{2r-1}+\tau^r\PC{2r+1}}{4r\sqrt{\tau}}.
\]
Then, the following explicit expressions for the functions entering Equations (\ref{EqfinLand})-(\ref{UV})
can be obtained:
\[
K_{\Omega}(t) = \frac{\Omega_0}{4r^2\tau}\left[\left(\tau^{-r}-\tau^r\right)^2 + 8r^2\right],  \quad
K_{\rho} = \frac{\Omega_0 \left(\tau^{-r}-\tau^r\right)}{4r^2\tau}\left[\tau^{-r}-\tau^r + 2r\sqrt{\tau}\right],
\]
\[
U(t) = \frac{\sqrt{\Omega_0}}{u} + \frac{\sqrt{\Omega_0}}{4ur\sqrt{\tau}}\PG{\tau^{-r}\PC{1-2r}-\tau^r\PC{2r+1}}, \quad
V(t) = \frac{\sqrt{\Omega_0}}{2r\sqrt{\tau}}\PC{\tau^{-r}-\tau^r}.
\]
The leading terms of these expressions for $\tau \gg 1$ result in the following coefficients of Equation (\ref{KY}):
\[
K_{\Omega}(\tau) \approx \frac{\Omega_0}{4r^2}\tau^{-\delta}, \quad 
K_{Y}(\tau) \approx \frac{\Omega_0}{4r^2}\tau^{-\delta} + \frac{\Omega_0}{u^2}\left[
1- \frac{2r+1}{4r}\tau^{-\delta/2}\right]^2,
\quad
K_{\rho}(\tau) \approx \frac{\Omega_0}{4r^2}\left[ \tau^{-\delta} - 2r \tau^{-\delta/2}\right], 
\]
where $\delta = 1- 2r$.
If $u \ll 1$, then $r \approx 1/2$  and $\delta \approx 2u^2$. If the time variable $\tau$ is not extremely big, so that 
$\tau^{-\delta} \approx 1$, 
we arrive at the formula ${\cal E}(t)/{\cal E}_{i} = (1+ s^{-1}\Upsilon)/2$,
coinciding with the sudden jump approximation formula with $\Omega_f=0$ of Section \ref{sec-sudj}.
On the other hand, 
$K_{\Omega}(\infty)= K_{\rho}(\infty)= 0$, while $K_{Y}(\infty)= \Omega_0/u^2$. This results
 in the nonzero asymptotic ratio 
\be
{\cal E}(\infty)/{\cal E}_{i} = s^{-1}\Upsilon/(2u^2),
\label{Einf-u<12}
\ee
 which can be very high if $u\ll 1$. This is a great difference from the case of circular gauge considered in Section \ref{fastlincirc},
 where the mean energy finally decays to zero value.
However, the asymptotic ratio (\ref{Einf-u<12}) can be achieved for extremely big values of time, 
since the relative corrections are of the order of $\tau^{-u^2}$. 
Consequently, the accuracy of 10\% can be achieved for $\tau \sim 10^{1/u^2}$. For example, taking $u=0.1$,
we need $\tau \sim 10^{100}$.

{ The time-dependent functions determining the evolution of the mean magnetic moment according to Equation (\ref{mean-M-Land}) 
have the following form:}
\be
S_{\Omega}(t)= \frac{\sqrt{\Omega_0}}{2r\tau}\PC{4r\sqrt{\tau}-T_-},
\ee
\be
S_{Y}(t)=\frac{\PC{4r^2\tau+\tau-2}T_{-} - 4r\PC{\tau+1}T_{+}+16r\sqrt{\tau}}{8ru^2\sqrt{\tau}} ,
\ee
\be
S_{\rho}(t)=\frac{\sqrt{\Omega_0}}{16ru^2\sqrt{\tau}}\PH{\PG{\tau-5+4r^2\PC{3\tau-1}}T_{-}-2r\PC{4r^2+3}\PC{\tau+1}T_{+}
+32r\sqrt{\tau}},
\ee
where
$
T_+=\tau^{r}+\tau^{-r}$ and $T_-=\tau^{r}-\tau^{-r}$.

\subsubsection{Slow variations}
\label{sec-slow-Land-invlin}

If $u>1/2$, then, using the notation $\gamma = \sqrt{u^2 -1/4} = ir$ and $\nu = \gamma \ln(\tau)$, we can write
\[
\vep(t) = \frac{\sqrt{\tau}}{2\gamma \sqrt{\Omega_0}}\left[2\gamma\cos(\nu) -\sin(\nu) + 2iu\sin(\nu)\right],
\quad
\dot\vep(t) = \frac{\sqrt{\Omega_0}}{2\gamma \sqrt{\tau}}\left[2i\gamma\cos(\nu) +i\sin(\nu) - 2u\sin(\nu)\right],
\]
\[
K_{\Omega} = 2\Omega_0/\tau + \Omega_0 \sin^2(\nu)/(\gamma^2 \tau),
\]
\[
\sigma(t)=\frac{\sqrt{\tau}\PG{\PC{4\gamma^2-1}\sin(\nu)+4\gamma\cos(\nu)}-4\gamma}{4u\gamma\sqrt{\Omega_0}}
+ i\frac{\sqrt{\tau}\PG{\sin(\nu)-2\gamma\cos(\nu)}+2\gamma}{2\gamma\sqrt{\Omega_0}},
\]
\[
S(t)=\frac{\sqrt{\tau}}{2\gamma\Omega_0}\PG{2\gamma\sqrt{\tau}-2\gamma\cos(\nu)-\sin(\nu)},
\qquad
\dot{S}(t)=\frac{\PC{4\gamma^2-1}\sin(\nu)+4\gamma\PG{\sqrt{\tau}-\cos(\nu)}}{4u\gamma\sqrt{\tau}},
\]
\[
\chi(t)=\frac{t_0\sqrt{\tau}\sin(\nu)}{\gamma},
\qquad
\dot{\chi}(t)=\frac{2\gamma\cos(\nu)+\sin(\nu)}{2\gamma\sqrt{\tau}},
\]
\[
 V(t)= -\frac{\sqrt{\Omega_0}\sin(\nu)}{\gamma\sqrt{\tau}}, \qquad
U(t) = \frac{\sqrt{\Omega_0}}{u}  -\,\frac{\sqrt{\Omega_0}}{2u\gamma\sqrt{\tau}}\PH{2\gamma\cos(\nu) +\sin(\nu)}.
\]
The presence of the constant term $\sqrt{\Omega_0}/{u}$ in the expression for $U(t)$ imposes restrictions on the validity of the
adiabatic approximation for the Landau gauge. If this term were absent, we would have the
relation ${\cal E}(t)/{\cal E}_i \approx \Omega(t)/\Omega_i$ (with oscillating corrections of the order of $u^{-2}$ if $u \gg 1$)
for any value of the time variable $t$, similar to Equation (\ref{E-exact-ubig}). 
However, in the present case, this relation holds only under the condition 
$\tau \ll u^2$, i.e., $t \ll t_0(\Omega_0 t_0)^2$. For bigger values of $t$, the true mean energy goes to the finite asymptotic value
${\cal E}(\infty)= {\cal E}_{i}  s^{-1}\Upsilon/(2u^2)$. While this value is small for $u\gg 1$, it is different from zero.
Certainly, this failure of the adiabatic approximation for very big times is due to the existence of function $\sigma(t)$
in addition to $\vep(t)$ for the Landau gauge.

 The time-dependent functions determining the evolution of the mean magnetic moment according to Equation (\ref{mean-M-Land}) 
have the following form:
\[
S_{\Omega}(t)=\frac{\sqrt{\Omega_0}}{4\gamma^2\sqrt{\tau}}\PH{\sqrt{\tau}\PG{8\gamma^2+\sin^2(\nu)}-4\gamma\sin(\nu)},
\]
\[
S_{Y}(t)=\frac{\cos(\nu)\PG{\sqrt{\tau}\cos(\nu)-16\gamma^2\PC{\tau+1}}-4\gamma\sin(\nu)\PC{4\gamma^2\tau-\tau+3}
+\sqrt{\tau}\PC{32\gamma^2-1}}{16u^2\gamma^2\sqrt{\tau}} , 
\]
\[
\frac{S_{\rho}(t)}{\sqrt{\Omega_0}} = \frac{\cos(\nu)\PG{\sqrt{\tau}\cos(\nu)+8\gamma^2\PC{\tau+1}\PC{2\gamma^2-1}}-8\gamma\sin(\nu)\PG{\gamma^2\PC{3\tau+1}+2}+\sqrt{\tau}\PC{32\gamma^2-1}}{16u^2\gamma^2\sqrt{\tau}}.
\]
If $u \approx \gamma \gg 1$ and $\tau \gg 1$, then, $S_{\Omega}(t) \approx 2\sqrt{\Omega_0}$, 
$S_{\rho}(t) \approx \sqrt{\Omega_0 \tau}\cos(\nu)$ and $S_{Y}(t) \approx 0$ (being of the order of $\sqrt{\tau}/\gamma$).
Hence,
\be
{\cal M}(t) \approx -\mu_B {\cal C}\left[1 - \rho \sqrt{\tau}\cos(\nu)/2 \right],
\label{M-ad-Land-invlin}
\ee
and this formula coincides with (\ref{MadiabLand-0}) for $\Omega(t) = \Omega_0/\tau$.

\subsubsection{Intermediate case}

If $u=1/2$, then,
\[
\vep(t)=\frac{\sqrt{\tau}\PG{2 +(i-1)\ln(\tau)}}{2\sqrt{\Omega_0}}, \qquad 
\dot{\vep}(t)=\frac{\sqrt{\Omega_0}\PG{2i +(i-1)\ln(\tau)}}{2\sqrt{\tau}},
\]
\[
\sigma(t)=\frac{\sqrt{\tau}\PG{4-\ln(\tau)}-4}{2\sqrt{\Omega_0}}+i\frac{\sqrt{\tau}\PG{\ln(\tau)-2}+2}{2\sqrt{\Omega_0}},
\]
\[
S(t) = \frac{\sqrt{\tau}\PG{2\sqrt{\tau}-\ln(\tau)-2}}{2{\Omega_0}}, 
\qquad
\dot{S}(t)=\frac{4\sqrt{\tau}-\ln(\tau)-4}{2\sqrt{\tau}},
\]
\[
\chi(t) =\frac{\sqrt{\tau}\ln(\tau)}{2\Omega_0},
\qquad
\dot{\chi}(t)=\frac{\ln(\tau)+2}{2\sqrt{\tau}},
\]
\[
{\cal E}(\tau) = \frac{{\cal E}_i}{2\tau}\left(\ln^2(\tau)+2 + 2 s^{-1}\Upsilon
\PH{2\PC{\tau+1}+\PG{2+\ln(\tau)}\PG{\ln(\tau)-2\sqrt{\tau}}} -2\rho \ln(\tau)\PG{\ln(\tau)-\sqrt{\tau}}
\right),
\]
with the asymptotic nonzero ratio ${\cal E}(\infty)/{\cal E}_i = 2s^{-1}\Upsilon$.

\begin{figure}[htb]
\includegraphics[height=2.32truein,width=3.0truein,angle=0]{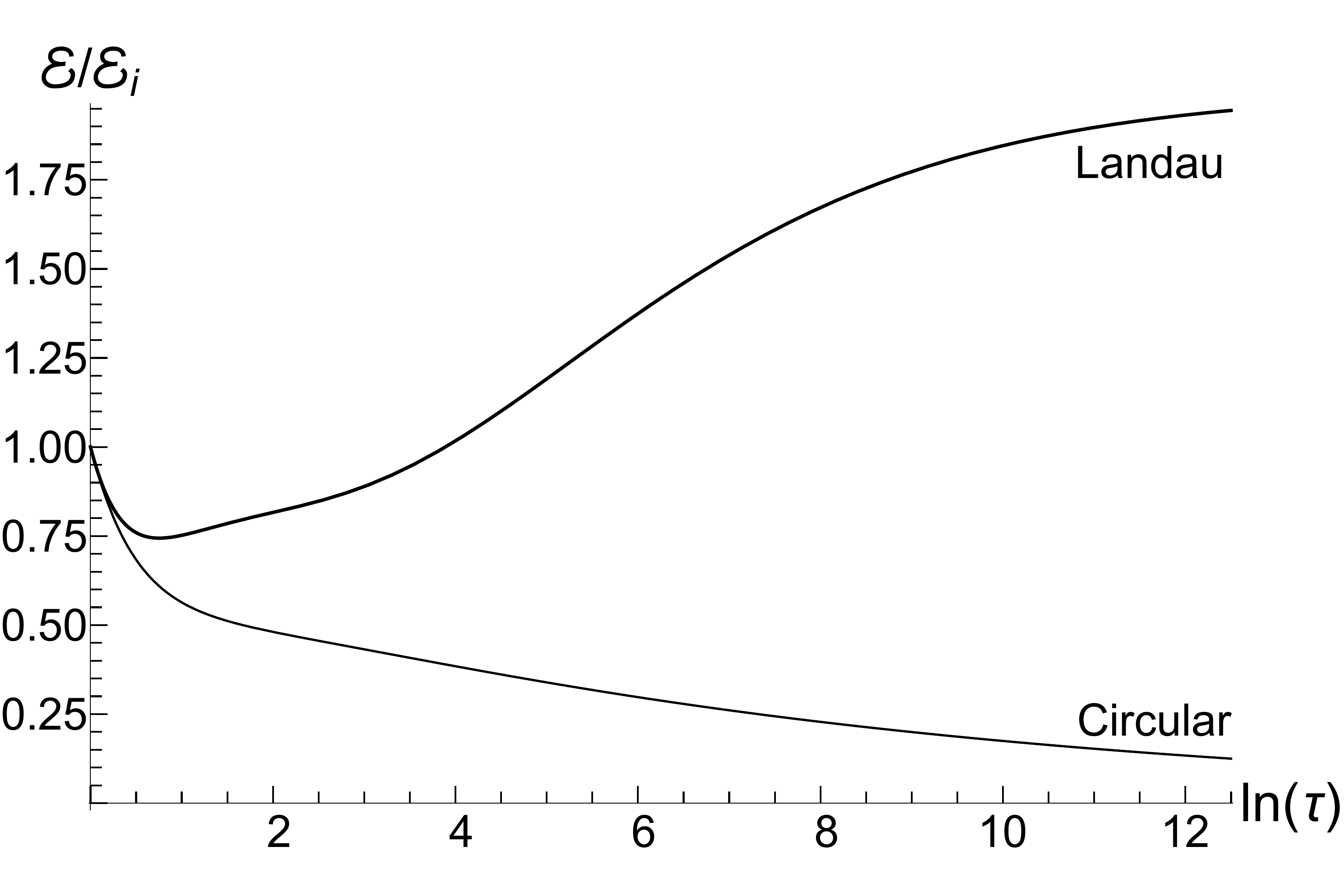}
\includegraphics[height=2.32truein,width=3.0truein,angle=0]{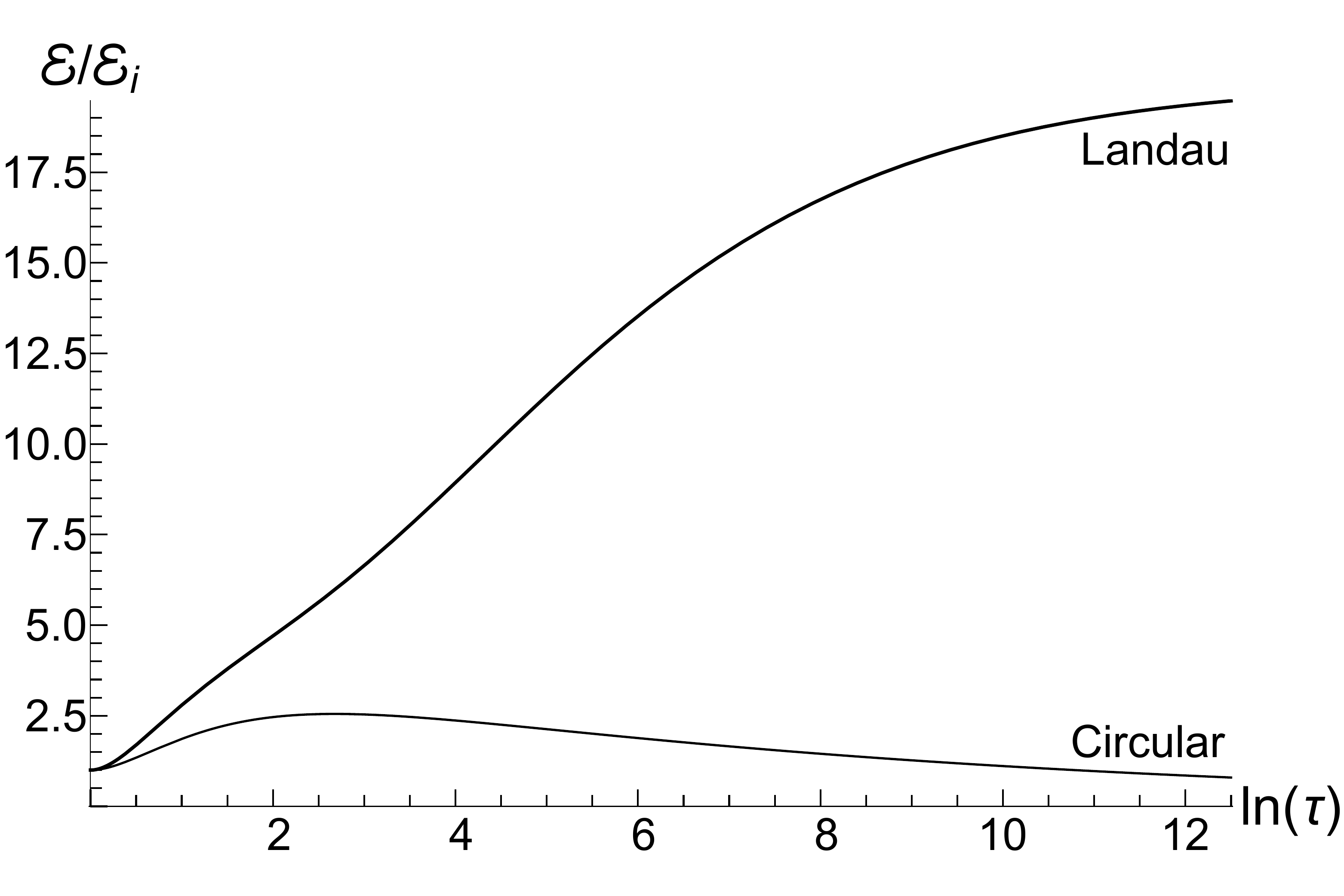}
\caption{\small The ratio ${\cal E}(\tau)/{\cal E}_i$ versus the dimensionless time $\tau = 1 +t/t_0$
for the circular and Landau gauges with the same initial cyclotron frequency $\Omega_0$
and time-scale parameter $t_0$, in the case of inverse linear decay of magnetic field
$B(t)=B_0/(1 + t/t_0)$ with $\Omega_0 t_0 =1/2$.
Left: the low temperature case, $\rho=0$, $s^{-1}\Upsilon=1$. Right: the high temperature case, $\rho=1$, $s^{-1}\Upsilon=10$.
 }
\label{fig-E-circ-Land-invlin}
\end{figure}

The comparison of the functions ${\cal E}(t)/{\cal E}_i$ for the circular and Landau gauges in the case of inverse linear law
of decrease of the magnetic field $B(t)=B_0/(1 + t/t_0)$ is given in Figure \ref{fig-E-circ-Land-invlin}.
The parameters $B_0$ and $t_0$ are assumed the same for the two gauges. The parameter $u=\Omega_0 t_0 =1/2$ is chosen for the
Landau gauge. Hence, the ratio ${\cal E}(t)/{\cal E}_i$ goes asymptotically to the nonzero value 
$2s^{-1}\Upsilon$.
However, since $\Omega=2\omega$, the corresponding value of parameter $u_c = \omega_0 t_0$ for the circular gauge 
is twice smaller: $u_c=1/4$. This means that the evolution in the circular gauge is given by Equation (\ref{E-exact}) with $r=\sqrt{3}/4$.
In this case, the ratio ${\cal E}(t)/{\cal E}_i$ goes asymptotically to zero approximately as $\tau^{-0.14}$.
\begin{figure}[htb]
\includegraphics[height=2.32truein,width=3.0truein,angle=0]{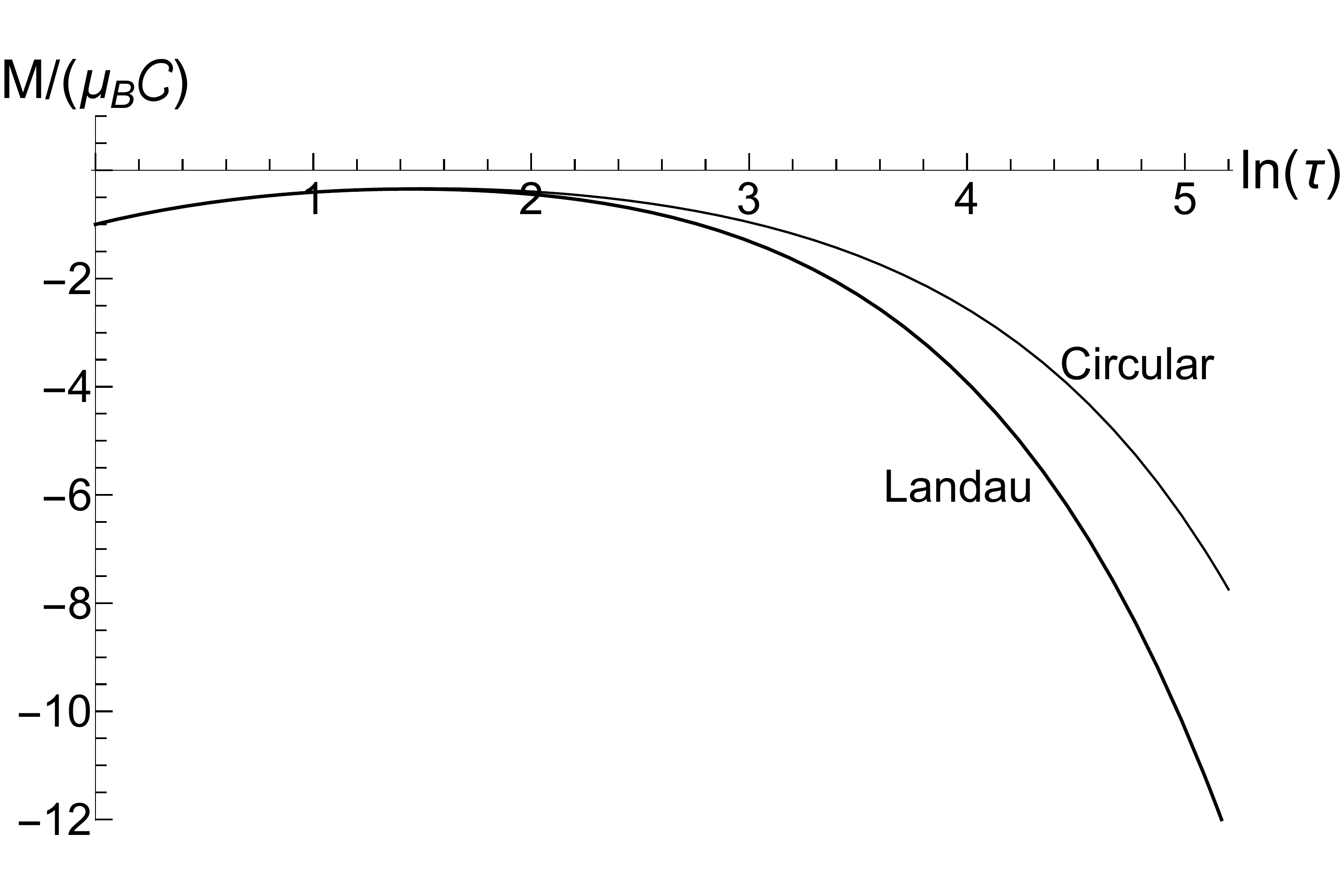}
\includegraphics[height=2.32truein,width=3.0truein,angle=0]{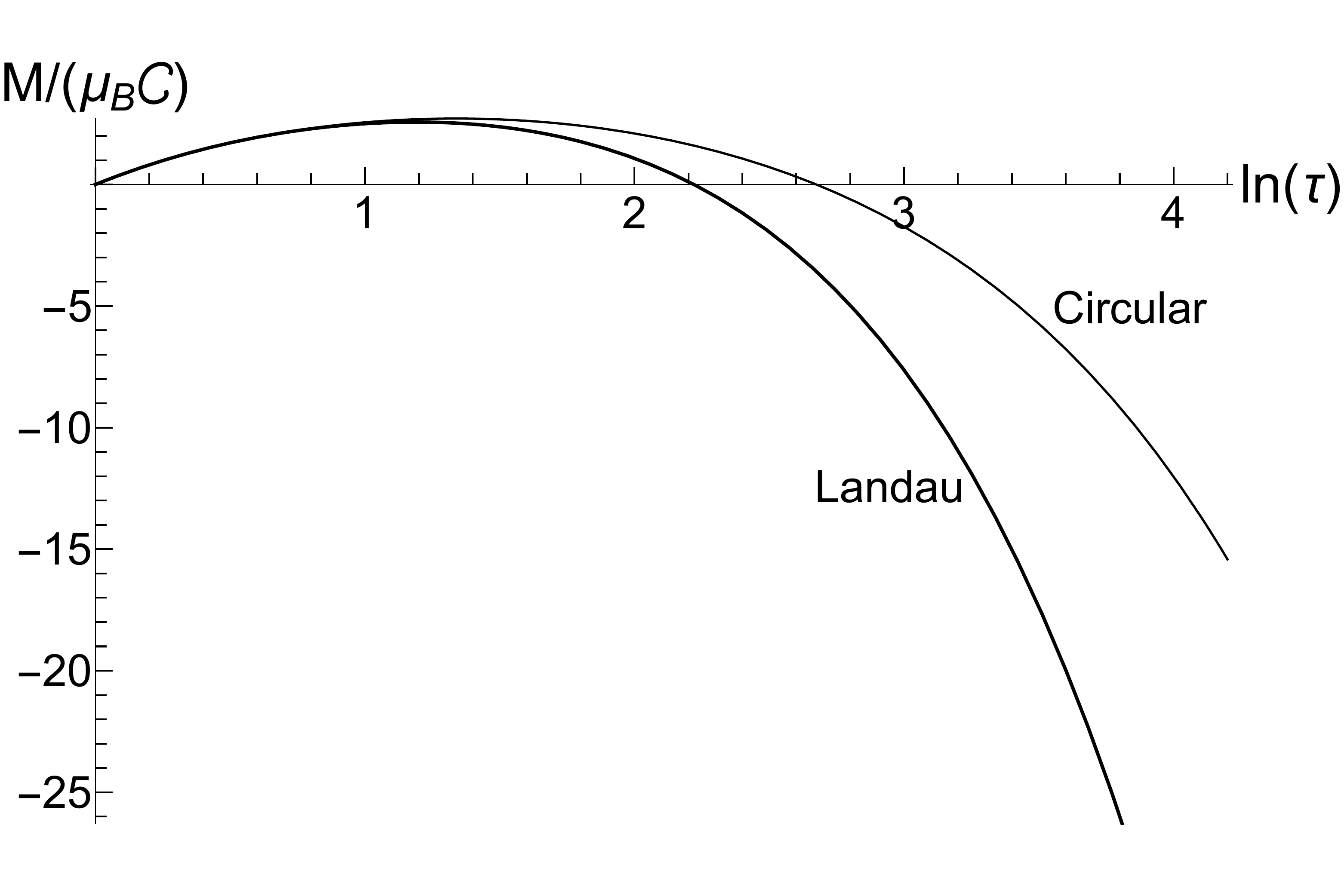}
\caption{\small The mean magnetic moment ${\cal M}(\tau)$ versus the dimensionless time $\tau = 1 +t/t_0$
for the circular and Landau gauges with the same initial cyclotron frequency $\Omega_0$
and time-scale parameter $t_0$, in the case of inverse linear decay of magnetic field
$B(t)=B_0/(1 + t/t_0)$ with $\Omega_0 t_0 =1/2$.
Left: the low temperature case, $\rho=0$, $\Upsilon=1$. Right: the high temperature case, $\rho=1$, $\Upsilon=10$.
 }
\label{fig-M-circ-Land-invlin}
\end{figure}  

Figure \ref{fig-M-circ-Land-invlin}  shows the evolution of the mean magnetic moment under the same conditions.
The mean magnetic moment in the case of Landau gauge with $u=1/2$ behaves as
\beqn 
{\cal M}(\tau)& =& -[\mu_B {\cal C}/(2\sqrt{\tau})]\left\{ 2\sqrt{\tau}-\ln(\tau)
+s^{-1}\Upsilon\PG{\ln(\tau)\PC{\tau-2}
-4\left(\sqrt{\tau} -1\right)^2}
\right. \nonumber \\ && \left.
-(\rho/2)\PG{\ln(\tau)\PC{\tau-5}+16\sqrt{\tau}-6\PC{1+\tau}}
\right\}.
\eeqn
The leading term  for $\tau \gg 1$ is 
${\cal M}(\tau) \approx -\mu_B {\cal C}\sqrt{\tau}\ln(\tau)\left(2s^{-1}\Upsilon - \rho\right)/4$.

\subsection{Exponential-like decrease of the magnetic field}
\label{sec-exp-Land}

Another example of explicit formulas in terms of elementary functions corresponds to the dependence
$\Omega(t) = \Omega_0 \sqrt{2}/\cosh(\Omega_0 t)$. 
In all other cases, we did not succeed to calculate the integral (\ref{defsig1}) analytically.
Using solution (\ref{vep-tanh-om0}) for $\vep(t)$ (with $\tau=\Omega_0 t$), one can find all additional necessary functions.
For the sake of simplicity, we assume here that $\Omega_0=1$. Then,
\be
\vep(\tau) = 2^{-1/4}\left[1 - \tau\tanh(\tau) +i\sqrt{2} \tanh(\tau)\right], \qquad
\sigma(\tau) = 2^{1/4}\left[ \frac{\tau- i\sqrt{2}}{\cosh(\tau)} +i\sqrt{2}\right],
\ee
\be
S(\tau) = \sqrt{2}\left[\frac1{\cosh(\tau)} -1 +\tau\tanh(\tau)\right], \qquad
\chi(\tau) = \frac{2\tau}{\cosh(\tau)}  +\tau -2\tanh(\tau).
\ee

\subsubsection{Evolution of the mean energy}

The mean energy in this case is given by Equation (\ref{EqfinLand}) with the following time dependent coefficients:
\be
K_\Omega=\frac{2+\PG{\tau+\sinh(\tau)\cosh(\tau)}^2}{\sqrt{2}\cosh^4(\tau)}+\frac{\sqrt{2}\PH{2\tanh^2(\tau)+\PG{1-\tau\tanh(\tau)}^2}}{\cosh^2(\tau)},
\label{KOmtau}
\ee
\be
V(\tau)=\frac{2^{1/4}\PG{2+\tau\sinh(\tau)-\cosh^2(\tau)-\cosh(\tau)}}{\cosh^2(\tau)}, \qquad
U(\tau)=\frac{\tau+\sinh(\tau)\PG{\cosh(\tau)-2}}{2^{1/4}\cosh^2(\tau)},
\label{VUtau}
\ee
\be
K_\rho=\frac{2^{1/4}\PG{1-\tau\tanh(\tau)}V(\tau)}{\cosh(\tau)}-\frac{\PG{\tau+\sinh(\tau)\cosh(\tau)}U(\tau)}{2^{1/4}\cosh^2(\tau)}.
\label{Krhotau}
\ee
When $\tau\to\infty$,  
\be
{\cal E}_{Land}(\infty)/{\cal E}_i=(1+3s^{-1}\Upsilon+2\rho)/4.
\label{Einf-Land}
\ee
The right-hand side of this equation is four times higher than for the circular gauge in isotropic traps at
zero temperature  {\em with the same ratio $\omega_i/\Omega_0$}: see Equation (\ref{Einf-circ-exp}).
However, the situation can be inverted for strongly anisotropic initial traps with $s \gg 1$, when 
${\cal E}_{circ}(\infty) \gg {\cal E}_{Land}(\infty)$.

\subsubsection{Evolution of the mean magnetic moment}

The time-dependent coefficients of formula (\ref{mean-M-Land}) for the mean magnetic moment have the following form:
\be
S_\Omega=\frac{2^{1/4}}{\cosh^2(\tau)}\PG{\PC{\tau^2+3}\cosh(\tau)-\tau\sinh(\tau)-1},
\label{SOmtau}
\ee
\be
S_Y=\frac{\sqrt{2}}{\cosh^2(\tau)}\PG{-\cosh^2(\tau)+\PC{\tau^2+3}\cosh(\tau)-3\tau\sinh(\tau)+\tau^2-2},
\label{SYtau}
\ee
\be
S_\rho=\frac{2^{1/4}}{\cosh^2(\tau)}\PG{\cosh^2(\tau)-2\PC{\tau^2+1}\cosh(\tau)+2\tau\sinh(\tau)-\tau^2+3}.
\label{Srhotau}
\ee
The asymptotic value  is positive for any values of parameters in this special case ($\mu_{Land}=\sqrt{2}$):
\be
{\cal M}_{Land}(\infty)={\mu_B{\cal C}\PC{\rho+s^{-1}\Upsilon}}/{2}.
\label{M-infty-Land-exp}
\ee

\subsection{Dynamics of ``fast jump to zero''}

One more simple example is the case of $\Omega(t) = \Omega_i/\cosh(\kappa t)$ with $\kappa \gg \Omega_i$. As was shown in
Section \ref{sec-MMM-mu0}, function $\vep(t)$ in this case can be chosen as $\vep(t) = \Omega_i^{-1/2}\left( 1 + i\Omega_i t \right)$.
However, calculating the function $\sigma(t)$, one can neglect the term $i\Omega_i t$, since function $\Omega(t)$ goes to zero 
exponentially at $t \sim \kappa^{-1}$, when $\Omega_i t \sim \Omega_i /\kappa \ll 1$. Consequently, the function $\sigma(t)$ 
is {\em real\/} in this approximation. Hence, $S(t) \equiv 0$ and $\chi(t) =t$. Then, Equations (\ref{EqfinLand})-(\ref{UV})
result in the formula
\be
{\cal E}(t)/{\cal E}_i = \frac12\left\{1 + \frac{1}{\cosh^2(\tau)}  + s^{-1}\Upsilon \left[1 - \frac{1}{\cosh(\tau)}\right]^2
+  \frac{2\rho}{\cosh(\tau)}\left[1 - \frac{1}{\cosh(\tau)}\right] \right\}. 
\label{ELand-tau0}
\ee
Equations (\ref{mean-M-Land})-(\ref{NM-gen}) lead to the following expression for the time-dependent
 mean magnetic moment:
\be
{\cal M}(t)= -\,\frac{\mu_B{\cal C}}{2} \left\{1 + \frac{1}{\cosh(\tau)}  - s^{-1}\Upsilon \left[1 - \frac{1}{\cosh(\tau)}\right]
-  \frac{2\rho}{\cosh(\tau)} \right\}, 
\label{mean-M-Land-mild0}
\ee
with ${\cal M}(\infty)= -\mu_B{\cal C}\left( 1- s^{-1}\Upsilon \right)/2$.
Formula (\ref{mean-M-Land-mild0}) coincides with (\ref{meanmag-mild}) for isotropic traps ($s=1$). 
However, the behavior is different if $s\neq 1$.
In particular, ${\cal M}_{Land}(\infty)$ is positive for $s^{-1}\Upsilon > 1$ and negative for $s^{-1}\Upsilon < 1$, whereas 
${\cal M}_{circ}(\infty)$ is positive for any values of  $s$ and $\Upsilon$ (unless $s=\Upsilon=1$).

\begin{figure}[htb]
\includegraphics[height=2.12truein,width=3.0truein,angle=0]{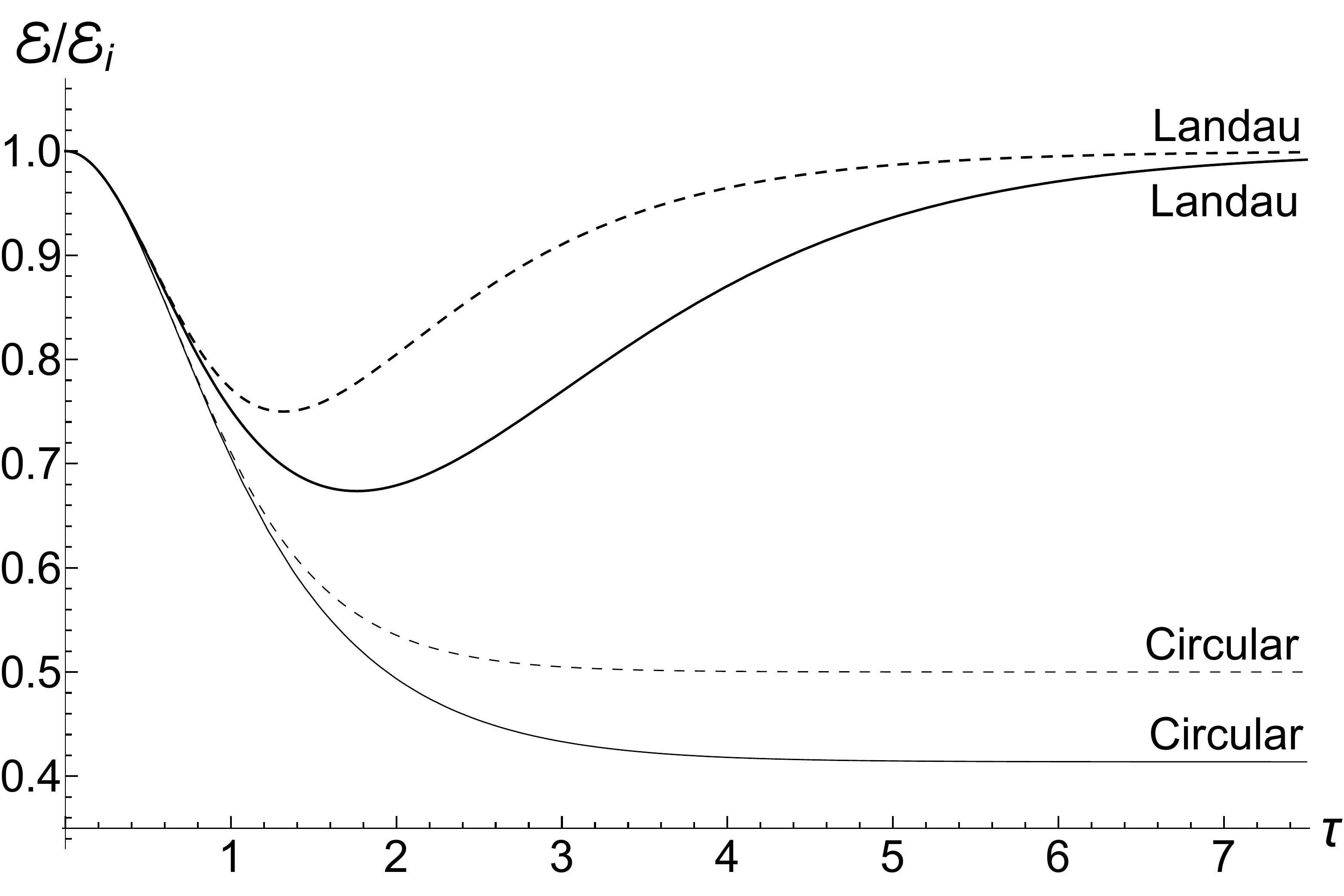}
\includegraphics[height=2.12truein,width=3.0truein,angle=0]{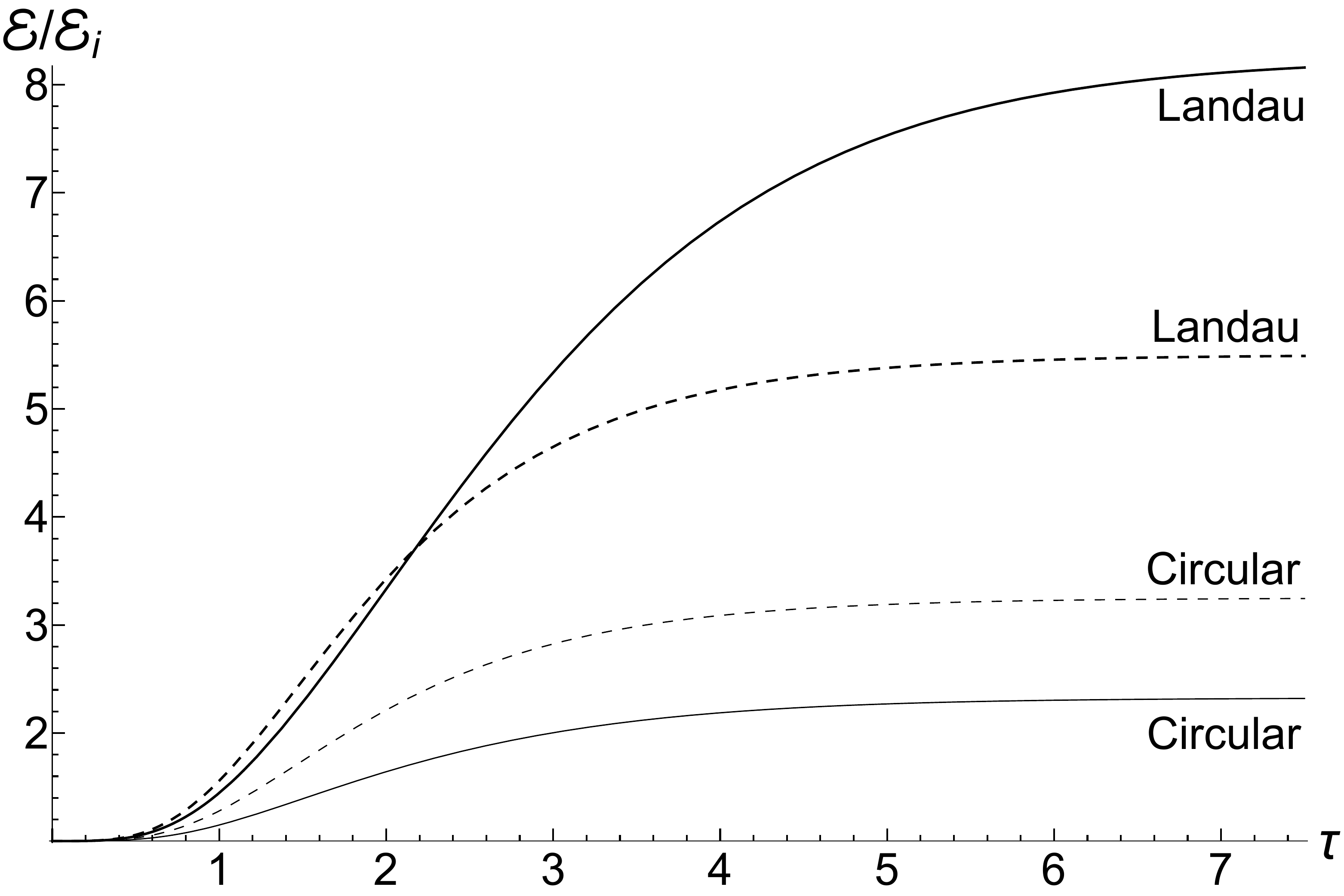}
\caption{\small The ratios ${\cal E}(\tau)/{\cal E}_i$ in the isotropic traps ($s=1$)
versus the dimensionless time $\tau = \kappa t$
for the circular and Landau gauges with $B(t) = B_0/\cosh(\kappa t)$. 
Solid lines: $\mu_{Land} = \sqrt{2}$ and $\mu_{circ}= \sqrt{2}/2$. Dashed lines: $\mu_{Land} \ll 1$ and $\mu_{circ} \ll 1$.
Left: the low temperature case, $\rho=0$, $\Upsilon=1$. Right: the high temperature case, $\rho=1$, $\Upsilon=10$. 
 }
\label{fig-E-circ-Land-exp}
\end{figure}

\begin{figure}[htb]
\includegraphics[height=2.12truein,width=3.0truein,angle=0]{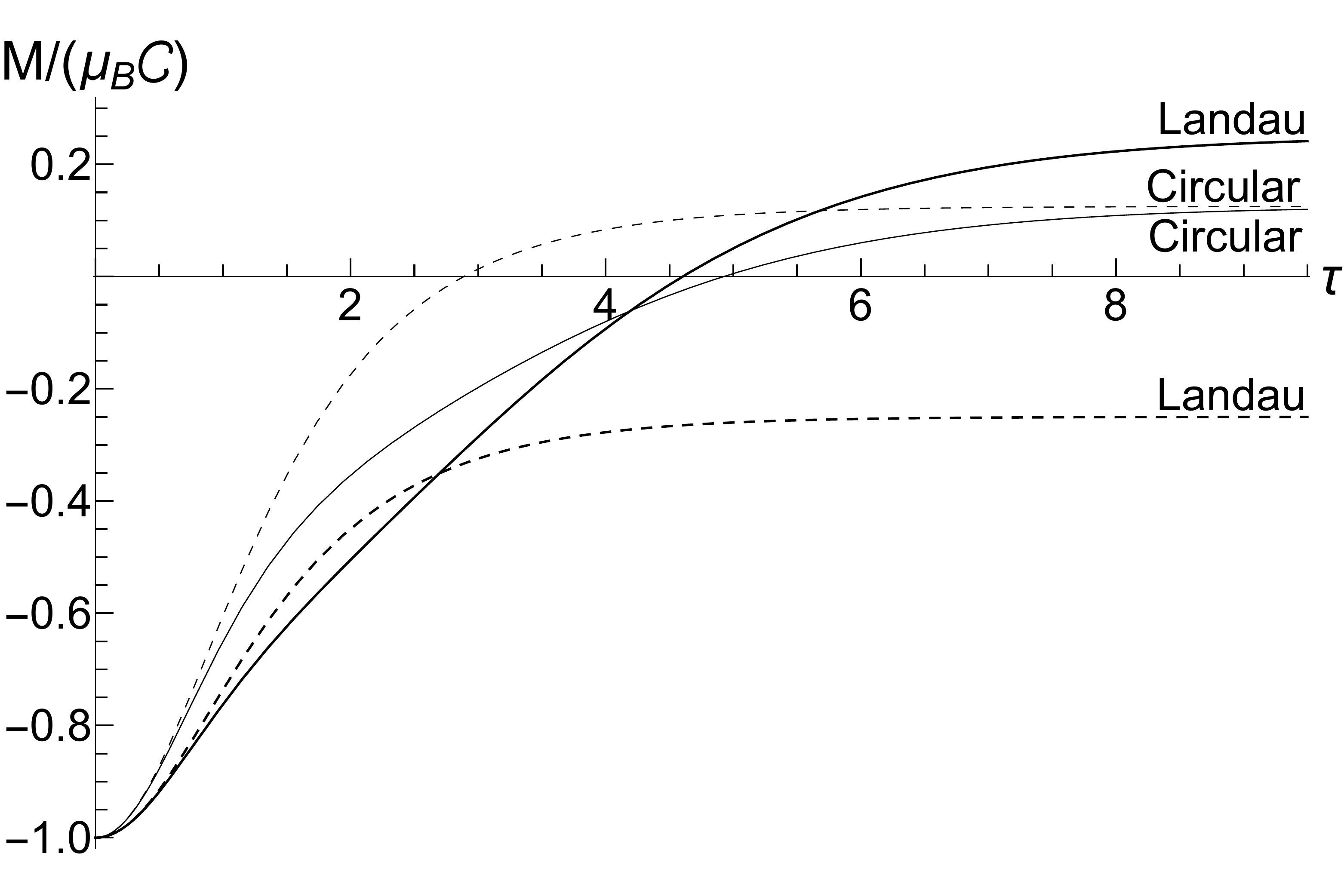}
\includegraphics[height=2.12truein,width=3.0truein,angle=0]{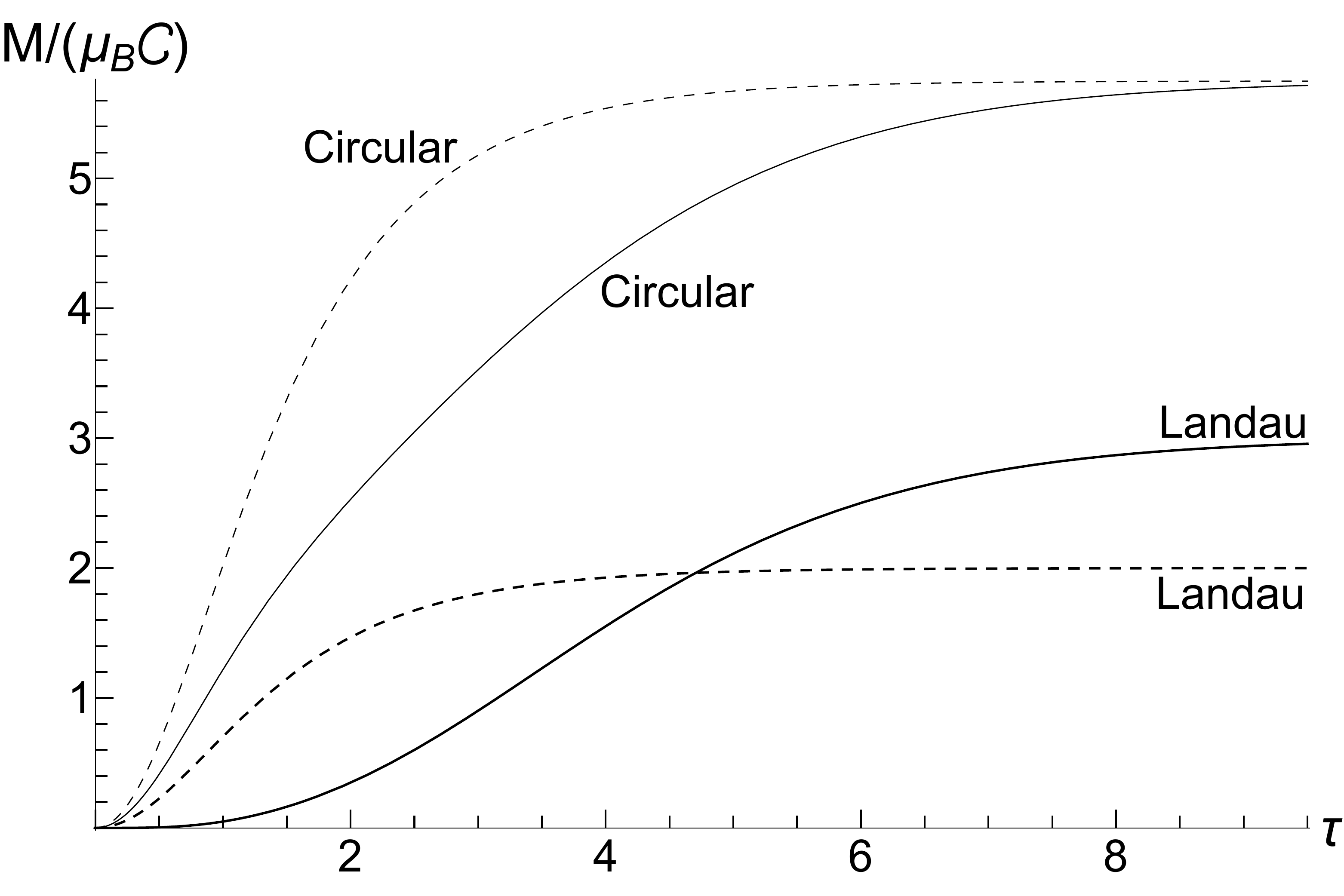}
\caption{\small The mean magnetic moment ${\cal M}(\tau)$ in the anisotropic traps with $s=2$
versus the dimensionless time $\tau = \kappa t$
for the circular and Landau gauges with with $B(t) = B_0/\cosh(\kappa t)$. 
Solid lines: $\mu_{Land} = \sqrt{2}$ and $\mu_{circ}= \sqrt{2}/2$. Dashed lines: $\mu_{Land} \ll 1$ and $\mu_{circ} \ll 1$.
Left: the low temperature case, $\rho=0$, $\Upsilon=1$. Right: the high temperature case, $\rho=1$, $\Upsilon=10$. 
 }
\label{fig-M-circ-Land-exp}
\end{figure}

The comparison of functions ${\cal E}(\tau)$ and ${\cal M}(\tau)$ for the circular and Landau gauges in the case of the ``mild''
exponential decrease of the magnetic field $B(t) = B_0/\cosh(\kappa t)$ is made in Figures \ref{fig-E-circ-Land-exp} and
\ref{fig-M-circ-Land-exp}. We consider two values of the ratio $\mu_{Land}=\Omega_i/\kappa$ (normalized by the {\em cyclotron frequency}): 
$\mu_{Land} = \sqrt{2}$ and $\mu_{Land} \ll 1$. In the first case, we use functions (\ref{EqfinLand}) and 
(\ref{mean-M-Land}) with coefficients (\ref{KOmtau})-(\ref{Krhotau}) and 
(\ref{SOmtau})-(\ref{Srhotau}), respectively, for the Landau gauge.
However, since  $\Omega(t)=2\omega(t)$ for the same magnetic field, 
 one should remember that $\mu_{circ}=\mu_{Land}/2$. For this reason, plots for the circular gauge are made using formulas from 
 Section \ref{sec-mild} with  $\mu_{circ}= \sqrt{2}/2$. Hence, the asymptotic value (\ref{Einf-Land}) should be compared with
the value given by Equations (\ref{DqGam}) and (\ref{Ff-mild}) for $\nu = (\sqrt{3}-1)/2$.
In this case, $D_q \approx (-0.43+0.48 {i}) /\sqrt{\omega_i}$ and
${\cal E}_{circ}(\infty)/{\cal E}_i \approx 0.21 \left( 1 + s_0\Upsilon\right) + 0.05\rho $.

When $\mu_{Land}\ll 1$, this parameter does not enter formulas for ${\cal E}(\tau)$ and ${\cal M}(\tau)$. In this case, we used
Equations (\ref{ELand-tau0}) and (\ref{mean-M-Land-mild0}) for the Landau gauge. The equations used for the circular gauge were
(\ref{Efratio-tau-0}) and (\ref{meanmag-mild}).
The coincidence of the ratios ${\cal E}_{Land}(\infty)/{\cal E}_i$ for two values of parameter $\mu$ in the low-temperature case
is accidental: these ratios are different if $s^{-1}\Upsilon \neq 1$. 

\section{Discussion}
\label{sec-fin}

We have obtained several exact results describing the dynamics governed by Hamiltonian (\ref{Ham}) with two gauges: 
the circular and Landau ones.
The dynamics is quite reach, depending on the concrete time dependence of the magnetic field $B(t)$.
All explicit analytic examples and figures clearly show that the dynamics can be quite different for the two gauges of the same
{\em time-dependent\/} magnetic field. The only exception is the case of the adiabatic variation of the magnetic field,  provided
the ratio of the final and initial frequencies is not too small, so that the simple adiabatic solution (\ref{vepad}) 
to Equation (\ref{eqvep}) can be justified.
In all the cases, fluctuations of the magnetic moment turn out extremely strong. Our results show that the time-dependent variance 
of the magnetic moment can be much higher than the square of its mean value. This is a generalization of the result found in 
Ref. \cite{DDmag} for the equilibrium state.

Important consequences of numerous examples are the conditions of validity of two frequently used approximations: the ``sudden jump'' and 
adiabatic ones. For the {\em monotonous\/} variations of the cyclotron frequency $\Omega(t)$, a simple parameter distinguishing
between the two extreme cases is the ratio $\mu=\Omega_i/\kappa$, where $\kappa^{-1}$ is some characteristic time of the transition 
from the initial frequency $\Omega_i$ to the final $\Omega_f$. Formally, the ``sudden jump'' corresponds to $\mu \ll 1$, while
the adiabatic approximation corresponds to $\mu \gg 1$. However, our examples show that in many cases a reasonable accuracy of the
approximations can be achieved when $\mu$ is a few times smaller or bigger than unity. Practically, the values $\mu=0.1$ and $\mu=10$
can be quite sufficient.
This result is important, because it justifies the reasonableness of the ``sudden jump'' approach in numerous applications, in particular,
in our papers \cite{DH18,DH-jump,DH19}. However, such justifications are not universal: they work well if only the ``transition time''
is well defined, as in the cases of exponential-like decay. For more slow frequency evolution laws the situation can be more
complicated: see Sections \ref{sec-om1t}, \ref{sec-om2t} and \ref{sec-1t-Land}.

An interesting exceptional case is $\Omega_f=0$. It has been known for a long time (starting, perhaps, from Ref. \cite{Varro})
that the description of the limit transition from a nonzero magnetic field to the free motion is a nontrivial problem
(for a similar problem for the harmonic oscillator with a time-dependent frequency see, for example, paper \cite{Tiba21}).
Our results show that the mean values of the energy and magnetic moment tend to some constant values, which are different for the
Landau and circular gauges. Moreover, these constant values are sensitive to the concrete forms of the time-dependent frequency
$\Omega(t)$. For example, the values of ${\cal E}_{circ}(\infty)$ and ${\cal M}_{circ}(\infty)$ do not depend on the speed of the
frequency decay for the ``mild exponential decay'' $\Omega(t) = \Omega_i/\cosh(\kappa t)$. On the other hand, analogous final
values for the Landau gauge strongly depend on the parameter $\kappa$ in the anisotropic case: even signs of the final magnetic
moments can be opposite. 
Quite different pictures are observed when the asymptotic forms of the function $\Omega(t)$ are non-exponential, e.g.,  
inverse power laws $\Omega(t) \sim t^{-b}$ with $b>0$. If $b=2$, Figures \ref{fig-E-t2-var-u} and \ref{fig-M-t2-var-u} still
show the existence of finite values ${\cal E}_{circ}(\infty)$ and ${\cal M}_{circ}(\infty)$, which are well different from
the case of exponential decay.
On the other hand, ${\cal M}_{circ}(t)$ and ${\cal M}_{Land}(t)$ can grow unlimitedly when $t\to\infty$ if $b=1$ and the
characteristic time scale $t_0$ is relatively small: see Figures \ref{fig-M-logtau-var-u}
and \ref{fig-M-circ-Land-invlin}.
Another intriguing feature of the special case of $b=1$ is that
neither adiabatic nor sudden jump approximations work in the whole time axis, although both approximations 
can have sense inside some limited time intervals for appropriate values of parameters. 
In particular,  under the condition $\Omega_0 t_0 \ll 1$,
the mean energy and magnetic moment rapidly attain the values predicted by the sudden jump approximation formulas,
as one can expect. A totally unexpected result is that
 after very long time intervals the functions ${\cal E}(t)$ and ${\cal M}(t)$ go to the final values which are very
different from the sudden jump predictions (and different for the circular and Landau gauges). Perhaps, this is a consequence
of the absence of a well defined value of the ``transition time'' for this kind of evolution with a very long 
non-exponential ``tail''.
Probably, a study of a more general situation, with an arbitrary value of parameter $b$, could be interesting. 
However, we leave this problem for another publication.

While the choice of $\Omega_f=0$ enables us to find several simple exact solutions to Equation (\ref{eqvep}), it is necessary
to remember that this limit in Hamiltonian $\hat{H}_0$, given by Equation (\ref{Ham}), can be doubtful from the point of view 
of description of real
physical situations, where a quantum particle is always confined within some container or trap. Probably, a more adequate
Hamiltonian in this case could be
\be
\hat{H}_g = \hat{H}_0 + M\left(g_1^2\hat{x}^2 + g_2^2\hat{y}^2\right)/2.
\label{Hg}
\ee
A preliminary investigation in this direction for the circular gauge and $g_1=g_2$ was performed recently in paper  \cite{D21}.
A general case with $g_1 \neq g_2$ seems worth studying, especially in connection with the Landau gauge.

Some results, especially related to the behavior of the magnetic moment, seem paradoxical. Indeed, the nonzero value of parameter $\rho$
in the initial state is necessary to have the correct Landau--Darwin value (\ref{EiMi}) of the mean magnetic moment in the equilibrium state
of a free charged particle in a uniform magnetic field.
However, formula (\ref{Madiab}) gives an oscillating mean magnetic moment even in the case of constant frequency $\omega$
(when this formula is {\em exact}). On the other hand, all mean values cannot depend on time in any equilibrium state
described by the density operator $\hat\rho =\exp(-\beta\hat{H})$, if $\hat{H}$ is time independent...
A possible explanation of this controversy is that the covariance matrix (\ref{sig0}) corresponds, strictly speaking, to the equilibrium 
state of the system, described not by the free Hamiltonian $\hat{H}_0$, given by Equation (\ref{Ham}), but by the Hamiltonian (\ref{Hg})
 with $g_k\ll \omega_i$.
It seems that the abrupt switching off the confining parabolic potential at $t=0$ transforms the equilibrium state of Hamiltonian
$\hat{H}_g$ into the non-equilibrium state of Hamiltonian $\hat{H}_0$, so the further evolution of some quantities becomes
time-dependent. The Hamiltonian (\ref{Ham}) possesses many attractive features, related to the existence of constants of motion
$x_c$ and $y_c$. On the other hand, probably, it is oversimplified in some respects, because, for example, the formal
equilibrium density operator $\exp(-\beta\hat{H}_0)$ cannot be normalized: its trace equals infinity.
This issue needs a more detailed study. 

One more intriguing problem is related to the case of very slow variation of the cyclotron frequency $\Omega(t)$.
At first glance, it is sufficient to use a simple solution (\ref{vepad}) to calculate all mean values and probabilities
\cite{MMT70,MMT73}. An immediate consequence is the linear dependence ${\cal E}_f/{\cal E}_i = \omega_f/\omega_i$
(the well known adiabatic invariant), clearly seen in Figures \ref{fig-EfEi-wf} and \ref{fig-EfEi-wfEps} 
 for any values of parameters $s_0\Upsilon$ and $\rho$.  
This linear dependence holds for the circular gauge as well as for the Landau gauge. Probably, such a gauge independence
can be explained by the extremely small values of the induced electric fields, when the difference between 
their geometries becomes insignificant.
However, the solution (\ref{vepad}) is not valid when the frequency becomes close to zero, especially when it passes through zero value
and becomes negative. Two examples {\em for the circular gauge\/} in Sections \ref{E-semiaxis} and \ref{sec-EEfullline} show that
the ratio ${\cal E}_f/{\cal E}_i$ as function of $|\omega_f|/\omega_i$ is again a straight line when $\omega_f <0$, 
but the proportionality
coefficient is bigger than unity: it equals $3$ in the low-temperature regime, while it can be even much bigger in the
 high-temperature regime. Unfortunately, we do not know, what can happen {\em for the Landau gauge}, since we did not succeed
 to find explicit solutions when $\omega(t) <0$ for this gauge (except for the sudden jump approximation). 
 This is a challenge for further studies, as well as the general
adiabatic case with an arbitrary gauge parameter $\alpha$ and negative final frequency $\omega_f$.
Another challenge is the case of slow variation of the gauge parameter itself, when $\alpha=\alpha(t)$. Physically,
it means a slow change of the {\em shape of solenoid} without any change of the magnetic field inside.

It was shown that the dynamics of the initial {\em high-temperature\/} equilibrium states can be quite different from the
evolution of the initial low-temperature states. In particular, the initial small mean magnetic moment can be strongly amplified
(by the factor of the order of $s_0\Upsilon \gg 1$)
when the magnetic field depends on time (and the mean energy can be strongly amplified, as well), even if the magnetic field
{\em decreases}. 
The reason is that fluctuations of the guiding center coordinates are much
stronger than fluctuations of the relative coordinates in the high-temperature equilibrium state, according to
Equations (\ref{sig0}) and (\ref{Geq}). Due to the dynamical coupling between the guiding center and relative coordinates
in the time-dependent magnetic field, the fragile statistical balance between the initial equilibrium fluctuations 
of the relative and guiding center coordinates is broken in
the process of evolution, so that the contribution of strong guiding center fluctuations to the mean magnetic moment 
 becomes dominant.
 
 In order to avoid possible misunderstandings, we stress that the interaction
with any reservoir during the evolution from the fixed initial state is neglected in our paper. 
From the physical point of view,  the interaction with a thermal reservoir can be important, especially in the case
of very slow (adiabatic) evolution, when various relaxation times can enter the game. However, this problem needs a separate
study, because the problem of relaxation in the presence of a magnetic field is nontrivial even for the constant field 
\cite{DOM85,Li90,Li96,Smirnov98,Schuch03,Genkin09,Band10,Kalan19},
and it can be more complicated for time-dependent fields \cite{Pachon14}.

It is worth mentioning that the linear vector potential (\ref{pv}) with any time-dependent function $B(t)$ is,
as a matter of fact, {\em an approximation\/} in the absence of distributed external currents. However, this approximation
is quite good in the non-relativistic case, because the spatial inhomogeneity scale of the electromagnetic field
is proportional to the light velocity $c$, whereas the cyclotron radius of a charged particle (defining the admissible
inhomogeneity scale of the magnetic field) is proportional to the particle velocity $v \ll c$. For more details
one can consult Refs. \cite{DelMi,DH19}.

Note that the evolution of the covariance matrix and combinations of its components,
 related to the mean energy and magnetic moment, strongly depends on the choice of the initial conditions. 
We considered the case which seems ``the most natural'' -- the initial ``equilibrium'' state described by means
of four parameters. But the covariance matrix is determined by {\em  ten\/} parameters in the most general case.
Hence, the dynamics corresponding to other initial conditions (e.g., some kinds of ``cat'' states) can be even more
fascinating (despite that such choices could seem rather artificial).

\section*{Acknowledgments} 
The authors are grateful to the anonymous referees for the important comments and suggestions. 
We thank  C. Farina, S.S. Mizrahi and A.E. Santana for the interest to our work and 
useful discussions.
V.V.D. acknowledges the partial support of the Brazilian funding agency 
Conselho Nacional de Desenvolvimento Cient\'{\i}fico e Tecnol\'ogico (CNPq).

\appendix

\section{Details of general solutions for the circular gauge}
\label{ap-detcirc}

It is convenient to introduce the $2\times2$ rotation matrix
\be
{\cal R} = \left\Vert
\begin{array}{c c }
\cos\vf	 &	\sin\vf	\\ 
-\sin\vf & \cos\vf 
\end{array}
\right\Vert ,
\quad \vf(t) = \int_0^t \omega(\tau)d\tau.
\label{vf}
\ee
Then, the $4\times4$ matrix $\Lambda_Q$ of Equation (\ref{LamQ}) can be written as
\be
\mat{\Lambda}_Q = \omega_{i}^{1/2} \left\Vert
\begin{array}{c c}
 \mbox{Re}(\vep) {\cal R} & \mbox{Im}(\vep) {\cal R}/(m\omega_{i}) \\
 m\mbox{Re}(\dot\vep) {\cal R} & \mbox{Im}(\dot\vep) {\cal R}/\omega_{i} 
\end{array}
\right\Vert.
\ee
Calculating the matrix product (\ref{LamUQU}), we obtain matrix $\mat{\Lambda}_q(t)$.
Its $2\times2$ blocks have a similar structure,
\be
\lambda_j(t) = \frac{\sqrt{\omega_{i}}}{2\omega(t)} \left\Vert
\begin{array}{c c}
c_j(t) & s_j(t) \\
-s_j(t) & c_j(t)
\end{array}
\right\Vert,
\label{lamjt}
\ee
with the following coefficients:
\beqnn
&&c_1 = a_{+}\cos\vf - b_{-}\sin\vf, \quad s_1 =  a_{+}\sin\vf +b_{-}\cos\vf , 
\\
&&c_2 = a_{-}\cos\vf + b_{+}\sin\vf, \quad s_2 =  a_{-}\sin\vf -b_{+}\cos\vf ,
\\
&&c_3 = a_{-}\cos\vf - b_{+}\sin\vf, \quad s_3 =  a_{-}\sin\vf +b_{+}\cos\vf ,
\\
&&c_4 = a_{+}\cos\vf + b_{-}\sin\vf, \quad s_4 =  a_{+}\sin\vf -b_{-}\cos\vf ,
\eeqnn
where
\[
a_{\pm}(t) = \omega(t)\mbox{Re}(\vep) \pm \mbox{Im}(\dot\vep), \quad
b_{\pm}(t) = \omega(t)\mbox{Im}(\vep) \pm \mbox{Re}(\dot\vep).
\]
Then, Equations (\ref{sigr}), (\ref{sigrc}) and (\ref{lamjt}) result in the following expressions for blocks of matrix $\mat{\sigma}_q(t)$:
\be
G\sigma_r = \frac{G\omega_i}{4\omega^2(t)}
\left\Vert
\begin{array}{c c}
c_1^2 + s_1^2 + \Upsilon C_2 -2\rho\left(c_1c_2 +s_1s_2\right) & 
\Upsilon s_2c_2\left( s^{-1} -s\right)
 \\
 \Upsilon s_2c_2\left( s^{-1} -s\right) &
c_1^2 + s_1^2 + \Upsilon S_2 -2\rho\left(c_1c_2 +s_1s_2\right)
\end{array}
\right\Vert,
\label{sigrcirc}
\ee
\be
G\sigma_c = \frac{G\omega_i}{4\omega^2(t)}
\left\Vert
\begin{array}{c c}
c_3^2 + s_3^2 + \Upsilon C_4 -2\rho\left(c_3c_4 +s_3s_4\right) & 
\Upsilon s_4c_4\left( s^{-1} -s\right)
 \\
 \Upsilon s_4c_4\left( s^{-1} -s\right) &
c_3^2 + s_3^2 + \Upsilon S_4 -2\rho\left(c_3c_4 +s_3s_4\right)
\end{array}
\right\Vert,
\label{sigccirc}
\ee
where $C_k = sc_k^2 + s^{-1}s_k^2$ and $S_k = ss_k^2 + s^{-1}c_k^2$ for $k=2,4$,
\be
G\sigma_{rc} = \frac{G\omega_i}{4\omega^2(t)}\left(\Sigma_0 + \Upsilon\Sigma_{\Upsilon} -\rho\Sigma_{\rho}\right),
\label{sigrccirc}
\ee
\be
\Sigma_0 =
\left\Vert
\begin{array}{c c}
c_1c_3 + s_1s_3  & s_1c_3 - c_1s_3
 \\
 c_1s_3 - s_1c_3  & c_1c_3 + s_1s_3 
\end{array}
\right\Vert,
\qquad
\Sigma_{\Upsilon} =
\left\Vert
\begin{array}{c c}
s c_2c_4 + s^{-1} s_2s_4  & s^{-1} s_2c_4 - s c_2s_4
 \\
s^{-1} c_2s_4 - s s_2c_4  & s s_2s_4 + s^{-1} c_2c_4
\end{array}
\right\Vert,
\ee
\be
\Sigma_{\rho} =
\left\Vert
\begin{array}{c c}
c_1c_4 + s_1s_4 + c_2c_3 + s_2s_3  & s_1c_4 + s_2c_3 - c_1s_4 - c_2s_3
 \\
c_1s_4 + c_2s_3 - s_1c_4 - s_2c_3 & c_1c_4 + s_1s_4 + c_2c_3 + s_2s_3
\end{array}
\right\Vert.
\ee

We see  that the trap anisotropy ($s\neq 1$) complicates significantly all formulas. 
For this reason, we studied the fluctuations of the energy and magnetic moment in
 the simplest case of $s=1$, when
 matrices $\sigma_r$ and $\sigma_c$ are proportional to the unit $2\times2$ matrix $I_2$:
\be
G\sigma_r = \frac{G\omega_i}{4\omega^2(t)}\left[ |F_-|^2 + \Upsilon |F_+|^2 - 2\rho \mbox{Re}(F_- F_+) \right] I_2, 
\label{sigr-sigc}
\ee
\be
G\sigma_c = \frac{G\omega_i}{4\omega^2(t)}\left[ |F_+|^2 + \Upsilon |F_-|^2 - 2\rho \mbox{Re}(F_- F_+) \right] I_2.
\label{sigr-sigc-c}
\ee
The only matrix $\sigma_{rc}$ is not diagonal for $s=1$:
\be
G\sigma_{rc} = \frac{G\omega_i}{4\omega^2(t)}
\left\Vert
\begin{array}{c c}
(1 + \Upsilon)\mbox{Re}\left(F_{-} F_{+}^*\right) -\rho \mbox{Re}\left(F_{-}^2 + F_{+}^2\right)
 & -2\omega(1 + \Upsilon)\mbox{Re}\left(\dot\vep \vep^*\right)
 \\
2\omega(1 + \Upsilon)\mbox{Re}\left(\dot\vep \vep^*\right) 
& (1 + \Upsilon)\mbox{Re}\left(F_{-} F_{+}^*\right) -\rho \mbox{Re}\left(F_{-}^2 + F_{+}^2\right)
\end{array}
\right\Vert.
\label{sigrc-s1}
\ee

\section{The fourth-order moments in terms of the second-order ones for the Gaussian states}
\label{ap-42}

The following relations hold in the Gaussian states, as special cases of the general formula (\ref{basic}):
\[
\overline{x_r^4} =3\left(\overline{x_r^2}\right)^2, \quad 
\overline{y_r^4} =3\left(\overline{y_r^2}\right)^2,
\]
\beqnn
\overline{x_r^2 y_r^2} &=& \langle \hat{x}_r^2\hat{y}_r^2 + \hat{y}_r^2\hat{x}_r^2 
+ \hat{x}_r\hat{y}_r \hat{x}_r\hat{y}_r +  \hat{y}_r\hat{x}_r \hat{y}_r\hat{x}_r 
+ \hat{x}_r\hat{y}_r^2 \hat{x}_r +  \hat{y}_r\hat{x}_r^2 \hat{y}_r \rangle/6
\\
&=& \langle  \hat{x}_r^2\hat{y}_r^2 + \hat{y}_r^2\hat{x}_r^2 \rangle/2 - [\hat{x}_r,\hat{y}_r]^2/2
= \overline{x_r^2} \cdot \overline{y_r^2} +2 \left(\overline{x_ry_r}\right)^2.
\eeqnn
 Therefore, 
\[
\langle  \hat{x}_r^2\hat{y}_r^2 + \hat{y}_r^2\hat{x}_r^2 \rangle = 2\overline{x_r^2} \cdot \overline{y_r^2}
+4 \left(\overline{x_ry_r}\right)^2
+ [\hat{x}_r,\hat{y}_r]^2 .
\]
As $\PG{\h{x_{r}},\h{x}_c}=\PG{\h{y_{r}},\h{y}_c}=0$, then,
\[
\langle \hat{x}_c^2\hat{x}_r^2 + \hat{y}_c^2\hat{y}_r^2 \rangle = \overline{x_{c}^2 x_{r}^2} + \overline{y_{c}^2 y_{r}^2}
=\overline{x_c^2} \cdot \overline{x_r^2}+\overline{y_c^2} \cdot \overline{y_r^2}
+2\PG{\PC{\overline{x_c x_r}}^2+\PC{\overline{y_c y_r}}^2}.
\]
In the case of four different operators we have
\beqnn
\overline{x_c y_c x_r y_r} &=& \langle\hat{x}_c\hat{y}_c \hat{x}_r \hat{y}_r \!+\! \hat{y}_c\hat{x}_c  \hat{y}_r \hat{x}_r 
\!+\! \hat{x}_c\hat{y}_c \hat{y}_r \hat{x}_r \!+\! \hat{y}_c\hat{x}_c \hat{x}_r \hat{y}_r  \rangle /4
\\
&=&\langle\hat{x}_c\hat{y}_c \hat{x}_r \hat{y}_r + \hat{y}_c\hat{x}_c  \hat{y}_r \hat{x}_r \rangle/2 
+\PG{\h{x_{r}},\h{y}_r}\PG{\h{y_{c}},\h{x}_c}/4
\\&=& \overline{x_c y_c} \cdot \overline{x_r y_r}+ \overline{x_c x_r} \cdot \overline{y_c y_r}+ \overline{x_c y_r} \cdot \overline{y_c x_r}.
\eeqnn
Other useful relations are
\[
\overline{x_c y_r^2 x_r} = \langle\hat{x}_c \left( \hat{y}_r^2 \hat{x}_r 
+ \hat{x}_r\hat{y}_r^2+ \hat{y}_r \hat{x}_r \hat{y}_r\right)\rangle/3, \qquad
\BK{\hat{y}_r \hat{x}_r \hat{y}_r}=\BK{\hat{y}_r^2 \hat{x}_r + \hat{x}_r\hat{y}_r^2}/2.
\]
 Therefore,
\[
\overline{x_c y_r^2 x_r} = \langle\hat{x}_c \left( \hat{y}_r^2 \hat{x}_r \!+\! \hat{x}_r\hat{y}_r^2\right)\rangle/2
= 2\overline{x_c y_r} \cdot \overline{y_r x_r}+ \overline{x_c x_r} \cdot \overline{y_r^2},
\]
so that
\[
\langle\hat{x}_c \left( \hat{y}_r^2 \hat{x}_r + \hat{x}_r\hat{y}_r^2\right) \rangle =4\overline{x_c y_r} \cdot \overline{y_r x_r}
+2 \overline{x_c x_r} \cdot \overline{y_r^2},
\quad
\langle\hat{y}_c \left( \hat{x}_r^2 \hat{y}_r + \hat{y}_r\hat{x}_r^2\right) \rangle =4\overline{y_c x_r} \cdot \overline{x_r y_r}
+2 \overline{y_c y_r} \cdot \overline{x_r^2}.
\]
Similarly,
\[
\langle\hat{x}_r^3 \hat{x}_c \rangle = 3\overline{x_r x_c} \cdot \overline{x_r^2}, \qquad \langle\hat{y}_r^3 \hat{y}_c \rangle = 3\overline{y_r y_c} \cdot \overline{y_r^2}.
\]

\section{Asymptotic formulas for $\kappa \ll \omega_{i,f}$
in the case of exponentially varying frequency on the time semi-axis  }
\label{sec-asymp}

If $\kappa \ll \omega_{i}$ and $\kappa \ll \omega_{f}$, then $|\mu| \gg 1$ and $|\gamma| \gg 1$,
and we need asymptotic formulas for the confluent hypergeometric function $\Phi(a;c;x)$ with big absolute values of the argument $x$ 
and the second parameter $c$. The simplest formula can be found, if one writes $c(c+1)\ldots(c+n-1) \approx c^n$ in Equation (\ref{def-Phi}).
Then $\Phi(a;c;x) \approx (1-x/c)^{-a}$. Using this approximation, we have 
\[
\Phi(1/2; 1- 2i\gamma; 2i\mu) \approx \sqrt{\frac{\omega_{f}}{\omega_{i}}},
 \qquad
\lambda \approx \frac{i\kappa}{\omega_{i}^2}\left(\omega_{i} -\omega_{f}\right),
\qquad
K \approx 2\omega_{f} \left[ 1 + \kappa^2\left(\omega_{i} -\omega_{f}\right)^2/\left(2\omega_{i}^4\right)\right].
\]
However, such a simple result can be justified under the condition $|x/c| <1$ only \cite{BE}, which is equivalent to
the inequality $\omega_{f} > \omega_{i}/2$.
The case of negative final frequency $\omega_f$ can be studied with the aid of a more complicated asymptotic formula 
 [see \cite{BE}, Equation 6.13(2)],
\be
\Phi(a;c;iz) \approx \frac{\Gamma(c)}{\Gamma(c-a)}\left(\frac{e^{i\pi/2}}{z}\right)^a
+ \frac{\Gamma(c)}{\Gamma(a)}e^{iz}(iz)^{a-c}.
\label{asymp2}
\ee
It holds for  $z \to \infty$, provided $z>0$ and $|c| \ll z$.
The last condition does not exclude the possibility that $|c| \gg 1$, if $\kappa \ll |\omega_{f}| \ll \omega_{i}$.
Then we can use also the asymptotic Stirling formula for the Gamma function 
\be
\Gamma(z) \approx \sqrt{2\pi}\exp\left[(z-1/2)\ln(z) -z\right], \quad |z| \gg 1.
\label{Stirling}
\ee
If $c= x_r -2i\gamma$, then 
\[
\ln(x_r-2i\gamma) =\ln(-2i\gamma) +\ln[1 + ix_r/(2\gamma)].
\]
And here a big difference between the cases of positive and negative values of coefficient $\gamma$ arises.
If $\gamma >0$, then $\ln(-2i\gamma)=\ln(2\gamma) -i\pi/2$, so that the product $z\ln(z)$ contains the real part
$-\pi\gamma$. On the other hand, if $\gamma <0$,
\[
\ln(-2i\gamma)= \ln(2i|\gamma|) =\ln(2|\gamma|) +i\pi/2,
\]
and the product $z\ln(z)$ contains the real part $+\pi\gamma$. At the same time, the term 
$i^{a-c}= \exp[i\pi(a-c)/2]$ in Equation (\ref{asymp2}) always has the real part $\exp(-\pi\gamma)$.
Hence, the product $\Gamma(c)(iz)^{a-c}$ is proportional to $\exp(-2\pi\gamma)$ for $\gamma >0$, meaning that the
second term in Equation (\ref{asymp2}) can be neglected if $\gamma \gg 1$. As a result, we have the following asymptotic
expressions in the case of $\kappa \ll \omega_{f} \ll \omega_{i}$:
\[
\Phi(1/2; 1- 2i\gamma; 2i\mu) \approx 
\left(\omega_{f}/\omega_{i}\right)^{1/2}, \qquad
\lambda \approx i\kappa/\omega_{i},
\qquad
K(\infty) \approx 2\omega_{f} \left[ 1 + \kappa^2/\left(2\omega_{i}^2\right)\right].
\]
Consequently, ${\cal E}_f/{\cal E}_i \approx \omega_{f}/\omega_{i}$ in both the cases of $\omega_{f} >0$, 
in accordance with Figure \ref{fig-EfEi-wf}.

On the other hand, the second term in Equation (\ref{asymp2}) cannot be neglected for $\gamma < 0$, since two exponential terms,
$\exp(\pi\gamma)$ and $\exp(-\pi\gamma)$, mutually eliminate each other.
 Moreover, this term
is much bigger than the first one in the case of function $\Phi(3/2; 2-2i\gamma; 2i\mu)$ with $\mu \gg |\gamma|$.
After some algebra, one can obtain the following expressions
[here $\Phi \equiv \Phi(1/2; 1- 2i\gamma; 2i\mu)$]:
\[
\Phi \approx \sqrt{\frac{|\gamma|}{\mu}}\left(i + \sqrt{2} e^{i\rho}\right),  \quad
|\Phi|^2 \approx \frac{|\gamma|}{\mu}\left(3 + 2\sqrt{2} \sin\rho\right), \qquad
\rho = 2(\mu +\gamma) + 2|\gamma|\ln(|\gamma|/\mu),
\]
\[
\lambda \approx \frac{2\sqrt{2} e^{i\rho}}{i + \sqrt{2} e^{i\rho}}, \quad
\mbox{Re}\lambda \approx \frac{4 +2\sqrt{2} \sin\rho}{3 + 2\sqrt{2} \sin\rho}, \quad
1-\mbox{Re}\lambda \approx -\left(3 + 2\sqrt{2} \sin\rho\right)^{-1}.
\]
Then, one can check that identity (\ref{ident}) is fulfilled exactly, so Equation (\ref{Omlam1}) yields
$K(\infty) = 6|\omega_{f}|$ and ${\cal E}_f/{\cal E}_i \approx 3|\omega_{f}|/\omega_{i}$ for $\omega_{f} <0$ and
$\kappa \ll |\omega_{f}| \ll \omega_{i}$, again in accordance with Figure \ref{fig-EfEi-wf} in the zero-temperature case.
Actually, this simple nice result corresponds to the leading terms of the asymptotic expansions of the confluent
hypergeometric function. The nonzero ratio $\omega_{f}/\omega_{i}$ results in some oscillations around the average value,
which are clearly seen in Figure \ref{fig-EfEi-k1-10}. The frequency of these oscillations increases with decrease of
parameter $\kappa$, as can be seen in the formula for coefficient $\rho$.

\section{Analytic corrections to the sudden jump approximation for $\kappa \gg \omega_{i,f}$ 
in the case of exponentially varying frequency on the time semi-axis }
\label{sec-ap-jump}
 
To find corrections to the sudden jump approximation in the case of big (but finite) values of $\kappa$, one needs the
expansion of the confluent hypergeometric function with respect to its argument, up to terms of the
order of $\kappa^{-2}$: 
\[
\Phi(1/2; 1-2i\gamma; 2i\mu) \approx 1 + i\mu -2\mu\gamma -3\mu^2/4, 
\qquad
\lambda \approx  \frac{\omega_{i} -\omega_{f} }{\omega_{i} }\left(1 +2i\gamma + i\mu/2 
 -\mu\gamma/2 -4\gamma^2   \right).
\]
Then, one can verify that the right-hand side of (\ref{ident}) equals unity up to the terms of the order of $\kappa^{-2}$,
confirming the identity (\ref{uvcond}).
 Using (\ref{Omlam1}), we obtain the following expression for
the correction to the final energy (\ref{Eqfcirc2}) due to the finite duration of the ``jump'' 
for the circular gauge (we confine ourselves with the low-temperature case here):
\be
\delta{\cal E} = -\frac{m G}{2\kappa^2} (\omega_{f} -\omega_{i})^2\left[5\left(\omega_{f} +\omega_{i}\right)^2
-4\omega_{i}^2 \right].
\label{deltaE-bigkap}
\ee
Consequently, the sudden jump approximation can be well justified in fact under the condition 
$\kappa \gg |\omega_{f} -\omega_{i}|$. 
The correction (\ref{deltaE-bigkap}) can be positive for negative values of  $\omega_{f}$, 
belonging to the interval $|\omega_{f} +\omega_{i}| < 2\omega_{i}/\sqrt{5}$.
Otherwise, the correction is negative.

\section{Solution for the Epstein--Eckart profile  on the whole line}
\label{sec-appEE-1}

Introducing the  variable $\zeta = \exp(\kappa t)$, one can transform Equation (\ref{eqvep})
with function (\ref{omtanh}) to the form
\be
(\kappa\zeta)^2 \vep^{\prime\prime} +\kappa^2 \zeta  \vep^{\prime} + 
\frac{\left(\omega_f \zeta +\omega_i\right)^2}{(\zeta +1)^2}  \vep =0,
\label{derxi}
\ee
where the prime means the derivative with respect to $\zeta$.
Writing $\vep(\zeta) = \zeta^{\lambda} (1 +\zeta)^d f(\zeta)$, one arrives at the equation
\beqn
&& (\kappa\zeta)^2 (\zeta +1)^2 f^{\prime\prime} 
+ \kappa^2 \zeta (\zeta +1) \left[ 1+2\lambda +\zeta(1 + 2\lambda +2d)\right] f^{\prime}
\nonumber 
\\
&& + \left\{ \kappa^2 \left[ (1+\zeta)^2 \lambda^2 + d(d-1)\zeta^2 + d\zeta (1+\zeta)(2\lambda +1)\right]
 + \left(\omega_f \zeta +\omega_i\right)^2 \right\} f =0.
\label{flamd}
\eeqn
The next step is to find the values of parameters $\lambda$ and $d$,
which transform Equation (\ref{flamd}) to the  form
\be
\zeta(1+\zeta)f^{\prime\prime} + (A +B\zeta) f^{\prime} +Cf=0,
\label{ABC}
\ee
with some constant coefficients $A,B,C$.
The necessary condition is the possibility to divide the coefficient at $f$ by $\zeta(1+\zeta)$.
This means that this coefficient must go to zero for $\zeta=0$ and $\zeta= -1$. Taking $\zeta =0$,
one arrives at the equation $\kappa^2\lambda^2 + \omega_i^2 =0$.
We choose the solution $\lambda = +i\omega_i/\kappa$, because it leads to the correct expression 
$\vep(t) \sim \exp(+i\omega_i t)$ as $t \to -\infty$.
Taking $\zeta = -1$, one arrives at the equation
$\kappa^2 d(d-1) +\left(\omega_i -\omega_f\right)^2 =0$.
 Then, comparing equation (\ref{ABC}) with (\ref{eqF}), one can obtain 
 the solution (\ref{sol-F}) with coefficients  (\ref{d-}).
The choice of parameter $d$ with the negative sign before the square root in Equation (\ref{d-}) 
guarantees that $\vep(t) \sim \exp(+i\omega_i t)$ for all times if $\omega_i =\omega_f$.

\section{Asymptotic functions describing the magnetic moment in the Landau gauge}
\label{ap-SOm-as}

The time dependence of the mean magnetic moment  is contained in functions 
$S_\Omega(t)$, $S_Y (t) $ and $S_{\rho}(t)$. In the asymptotic regime, these functions are determined by the constant coefficients
 $u_{\sigma}$, $u_{\pm}$ and their combinations  
\[
a= u_{+} + u_{-}^*, \qquad b = u_{+}u_{\sigma}^* - u_{-}^* u_{\sigma}, \qquad
 a_- = u_{+} - u_{-}^*, \qquad  b_- = u_{+}u_{\sigma}^* + u_{-}^* u_{\sigma}.
 \]
To simplify formulas, it is helpful to introduce four functions oscillating as $\exp\left(i|\omega_{f}|t\right)$:
\[
A=ae^{i|\omega_{f}|t}, \qquad B=be^{i|\omega_{f}|t}, \qquad 
A_-=a_-e^{i|\omega_{f}|t}, \qquad B_-=b_-e^{i|\omega_{f}|t}.
\]
Then,
\[
S_\Omega=\sqrt{|\Omega_f|}\mbox{Re}(A_- -iB_-)+2|\Omega_f|\sqrt{\Omega_i}\PC{|u_+|^2+|u_-|^2}/\Omega_f, 
\]
\beqnn
S_Y &=& \left(|\Omega_f|/{\Omega_f}\right)\PH{\mbox{Re}(b)\mbox{Im}(A-iB)+ |a-ib|^2
}
\\ &&
+\sqrt{\Omega_i/|\Omega_f|}\,\mbox{Re}(A-iB)
- \sqrt{|\Omega_f|/\Omega_i}\,\mbox{Re}(u_{\sigma})\mbox{Im}(A +iB),
\eeqnn
\beqnn
S_\rho &=& \sqrt{|\Omega_f|}\PG{\mbox{Re}(A_-)
-\mbox{Re}(u_\sigma)\mbox{Re}(B)-2\mbox{Re}(u_\sigma)\mbox{Im}(A)}
\\ &&
+ \frac{\Omega_i\mbox{Re}(A)}{|\Omega_f|^{1/2}}
+\frac{2|\Omega_f|\sqrt{\Omega_i}}{\Omega_f}\PG{|a|^2+\mbox{Im}(ba^*)+\mbox{Re}(b)/2}.
\eeqnn

\section{Time-dependent Landau levels}
\label{ap-deg} 

One of several distinguished features of Hamiltonian (\ref{Ham}) with a constant value of the magnetic field
is the infinite degeneracy of its eigenvalues (frequently called as the {\em Landau levels}), 
whose spectrum is equidistant. This degeneracy is due to the 
existence of additional integrals of motion, i.e., the operators commuting with the Hamiltonian:
the guiding center operators $\hat{x}_c$ and $\hat{y}_c$ (\ref{centroXY}) (which are especially useful for the
Landau gauge) and their consequence -- the generalized angular momentum operator $\hat{L}$ (\ref{L3}) 
(which is more useful for the circular gauge).
A natural question is: what happens for time-dependent magnetic fields? Are there some analogs of the Landau levels,
and whether they continue to be infinitely degenerate? The answer is positive. Probably, it was given in the explicit 
form for the first time in study \cite{DMM-75}. Namely, if $\hat{A}$ and $\hat{B}$ are time-independent operators
in the Schr\"odinger picture (including the case when they are integrals of motion), then operators 
$\hat{\cal A}(t) \equiv \hat{U}(t)\hat{A}\hat{U}^{-1}(t)$ and $\hat{\cal B}(t) \equiv \hat{U}(t)\hat{B}\hat{U}^{-1}(t)$
are the {\em integrals of the motion\/} for the nonstationary problem, satisfying the same commutation relations as
$\hat{A}$ and $\hat{B}$. Here $\hat{U}(t)$ is the evolution operator in the Schr\"odinger picture. The 
time-dependent integrals of the motion $\hat{I}(t)$ satisfy the equation 
$i\hbar \partial \hat{I}/\partial t = \left[\hat{H}(t), \hat{I}(t)\right]$, which was used for the first time by
Lewis and Riesenfeld in the paper \cite{LR}, where they ``guessed'' the existence of {\em quadratic\/} time-dependent
integrals of motion, generalizing the Hamiltonian. The linear integrals of motion were constructed for the first time
in papers \cite{MMT69,MMT70,DMM72}, where it was shown how to construct quadratic integrals of motion (including the
LR-invariant as a special case) from the linear ones. If the set $\{\psi_n(x)\}$ describes the stationary ``Landau states'',
satisfying the equation $\hat{H}_0\psi_n(x) = E_n \psi_n(x)$ at $t<0$, when the Hamiltonian $\hat{H}_0$ is assumed to be
time-independent,
then, the set $\{\Psi_n(x,t)\} = \{\hat{U}(t)\psi_n(x)\}$ can be considered as the ``nonstationary Landau states'',
satisfying the equation $\hat{\cal K}(t)\Psi_n(x,t) = E_n \Psi_n(x,t)$, where 
$\hat{\cal K}(t) = \hat{U}(t)\hat{H}_0\hat{U}^{-1}(t)$. This set of states maintains the infinite degeneracy with respect
to eigenvalues of operators $\hat{\cal X}_c(t) = \hat{U}(t)\hat{x}_c \hat{U}^{-1}(t)$ or 
$\hat{\cal L}_c(t) = \hat{U}(t)\hat{L}\hat{U}^{-1}(t)$.
More details can be found in \cite{183vol}. For the most recent trends in this direction one can consult study \cite{invar}.

\section{Non-equvalence of time-dependent gauges}
\label{ap-gauge} 

Probably, the question about the equivalence or non-equivalence of time-dependent vector potentials yielding the same magnetic fields
needs a discussion. It is well known that the given magnetic field ${\bf B}$ and electric field ${\bf E}$ can be derived from different
sets of the vector and scalar potentials,
\[
{\bf B} = \mbox{rot}{\bf A}, \quad {\bf E} = -\nabla \vf - \frac1{c}\frac{\partial A}{\partial t}, \qquad
{\bf B} = \mbox{rot}{\bf A}^{\prime}, \quad {\bf E} = -\nabla \vf^{\prime} - \frac1{c}\frac{\partial A^{\prime}}{\partial t},
\]
provided the two sets are connected as follows,
\[
{\bf A}^{\prime} = {\bf A} +\nabla \chi, \quad \vf^{\prime} = \vf - \frac1{c}\frac{\partial \chi}{\partial t}.
\]
If the magnetic field does not depend on time, the scalar function $\chi$ can be also chosen as time-independent. Then, 
$\vf^{\prime} = \vf$. In particular, if $\vf=0$, then $\vf^{\prime} = 0$ as well. 
In this case, the choice of the vector potential is, indeed, a matter of convenience, since the physical consequences, 
such as the mean energy or the magnetization, do not depend on this choice (although the relations between the transformed 
quantum states can be rather nontrivial \cite{Konst16,Swenson}).

However, the situation is different if vector ${\bf B}$ depends on time. 
Then, two different vectors ${\bf A}^{\prime}$ and ${\bf A}$ must be also time-dependent, as well
as function $\chi$.
Hence,  $\vf^{\prime} \neq 0$ even if $\vf=0$. Since we consider the systems described by means of Hamiltonian (\ref{Ham})
{\em without any scalar potential}, this means that the systems with different values of the gauge parameter $\alpha$ are 
{\em not equivalent\/} if $\partial {\bf B}/\partial t \neq 0$.
Perhaps, the most clear statement could be that Hamiltonian (\ref{Ham}) with a {\em time-dependent\/} vector potential
${\bf A}({\bf r},t)$ describes the motion of a particle not in the magnetic field, but in the {\em eletromagnetic\/} field
with nonzero vectors ${\bf B}$ and ${\bf E}$. Then, different choices of the ``gauge parameter'' $\alpha$ correspond,
as a matter of fact, to {\em different physical systems}.


\end{document}